\documentclass[11pt,english]{article}
\usepackage[T1]{fontenc}
\usepackage[latin9]{inputenc}
\usepackage{geometry}
\usepackage{color}
\usepackage{babel}
\usepackage{array}
\usepackage{float}
\usepackage{amsmath}
\usepackage{amsthm}
\usepackage{amssymb}
\usepackage{graphicx}
\usepackage{rotfloat}
\usepackage{setspace}
\usepackage[authoryear]{natbib}
\usepackage[unicode=true,pdfusetitle,
 bookmarks=true,bookmarksnumbered=false,bookmarksopen=false,
 breaklinks=false,pdfborder={0 0 1},backref=false,colorlinks=true]
 {hyperref}
\hypersetup{
 linkcolor=blue, citecolor=blue, urlcolor=blue}

\makeatletter

\providecommand{\tabularnewline}{\\}

\theoremstyle{plain}
\newtheorem{assumption}{\protect\assumptionname}
\theoremstyle{plain}
\newtheorem{prop}{\protect\propositionname}
\theoremstyle{definition}
 \newtheorem{example}{\protect\examplename}
\theoremstyle{plain}
\newtheorem{cor}{\protect\corollaryname}
\theoremstyle{definition}
\newtheorem{defn}{\protect\definitionname}
\theoremstyle{plain}
\newtheorem{lem}{\protect\lemmaname}

\usepackage{multirow}
\usepackage{tikz}
\usetikzlibrary{patterns,decorations.pathreplacing}
\usetikzlibrary{arrows}
\usepackage{rotating}
\usepackage{pdflscape}
\usepackage{makecell}
\usepackage{pdflscape}
\usepackage{setspace}
\usepackage{chngcntr}

\DeclareMathOperator{\Cov}{Cov}
\DeclareMathOperator{\Var}{Var}
\DeclareMathOperator{\supp}{supp}
\DeclareMathOperator{\E}{E}

\usepackage{cmap}

\@ifundefined{showcaptionsetup}{}{%
 \PassOptionsToPackage{caption=false}{subfig}}
\usepackage{subfig}
\makeatother

\providecommand{\assumptionname}{Assumption}
\providecommand{\corollaryname}{Corollary}
\providecommand{\definitionname}{Definition}
\providecommand{\examplename}{Example}
\providecommand{\lemmaname}{Lemma}
\providecommand{\propositionname}{Proposition}

\begin{document}
\title{\textbf{\Huge{}2SLS with Multiple Treatments}}
\author{\noindent {\small{}Manudeep Bhuller}\thanks{University of Oslo; CEPR; CESifo; and IZA. E-mail: manudeep.bhuller@econ.uio.no}
\and{\small{} Henrik Sigstad}\thanks{Corresponding author. BI, Norwegian Business School. E-mail: henrik.sigstad@bi.no}}
\date{{\normalsize{}First Version: May 16, 2022\\This Version: May 23,
2024\\}}

\maketitle
\begin{singlespace}
\noindent \textbf{Abstract: }We study what two-stage least squares
(2SLS) identifies in models with multiple treatments under treatment
effect heterogeneity. Two conditions are shown to be necessary and
sufficient for the 2SLS to identify positively weighted sums of agent-specific
effects of each treatment: \emph{average conditional monotonicity}
and \emph{no cross effects}. Our identification analysis allows for
any number of treatments, any number of continuous or discrete instruments,
and the inclusion of covariates. We provide testable implications
and present characterizations of choice behavior implied by our identification
conditions.
\end{singlespace}

\noindent \vfill{}

\begin{singlespace}
\end{singlespace}

\noindent \thispagestyle{empty}\pagebreak{}

\noindent {\footnotesize{}\setcounter{page}{1}}{\footnotesize\par}

\section{Introduction\label{sec:Introduction}}

In many settings\textemdash e.g., education, career choices, and migration
decisions\textemdash estimating the causal effects of a series of
different treatments is valuable. For instance, in the context of
criminal justice, one might be interested in separately estimating
the effects of conviction and incarceration on defendant outcomes
\citep{Humphries2022,Kamat2023}. Identifying treatment effects in
settings with multiple treatments using instruments has, however,
proven challenging \citep{HeckmanEtal2008_Annales}. A common approach
in the applied literature is to estimate a ``multivariate'' two-stage
least squares (2SLS) regression with indicators for receiving various
treatments as multiple endogenous variables and at least as many instruments.\footnote{Examples of studies estimating models with multiple treatments using
2SLS\textemdash either as the main specification or as an extension
of a baseline specification with a binary treatment\textemdash include
\citet{persson2004constitutional,acemoglu2005unbundling,rohlfs2006government,angrist2006instrumental,angrist2009incentives,maestas2015does,muellersmith2015,kline2016evaluating,jaeger2018shift,bombardini2020trade,BhullerEtAl2020,norris2021effects,angrist2022methods,Humphries2022,BhullerSigstad2021,Kamat2023,heinesen2022instrumental}.} While such an approach is valid under homogenous treatment effects,
it does not generally identify meaningful treatment effects under
treatment effect heterogeneity.

To fix ideas about the identification problem, consider a case with
three mutually exclusive treatments\textemdash $T\in\left\{ 0,1,2\right\} $\textemdash and
a vector of valid instruments $Z$. Let $\beta_{1i}$ and $\beta_{2i}$
be the causal effects of receiving treatments $1$ and $2$ relative
to receiving treatment $0$ for agent $i$ on an outcome variable.
Then, the 2SLS estimand of the causal effect of receiving treatment
$1$ is, in general, a weighted sum of $\beta_{1i}$ and $\beta_{2i}$
across agents, where the weights can be negative. Thus, the estimated
effect of treatment $1$ can both put \emph{negative }weight on the
effect of treatment $1$ for some agents \emph{and} be contaminated
by the effect of treatment $2$. In severe cases, the estimated effect
of treatment $1$ can be negative even though $\beta_{1i}>0$ for
all agents. The existing literature has, however, not clarified which
conditions are necessary for the 2SLS estimand of the effect of treatment
$1$ to \emph{assign proper weights}\textemdash non-negative weight
on $\beta_{1i}$ and zero weight on $\beta_{2i}$.

In this paper, we present two necessary and sufficient conditions\textemdash besides
the standard rank, exclusion, and exogeneity assumptions\textemdash for
multivariate 2SLS to assign proper weights\emph{: average conditional
monotonicity }and \emph{no cross effects}. Our results apply in the
general case with $n$ treatments and $m\geq n$ continuous or discrete
instruments. We also provide results for settings where exogeneity
holds only conditional on covariates\textemdash a common feature in
applied work that complicates the analysis of 2SLS already in the
binary treatment case \citep{sloczynski2020should,blandhol2022tsls}.
Finally, our results allow the researcher to specify any set of relative
treatment effects, not just effects relative to an excluded treatment
(treatment $0$).

For expositional ease, we continue with the three-treatment example.
When does the 2SLS estimand of receiving treatment $1$ put non-negative
weight on $\beta_{1i}$ and zero weight on $\beta_{2i}$? To develop
intuition about the required conditions, assume we run 2SLS with the
following two instruments: the linear projection of an indicator for
receiving treatment $1$ on the instrument vector $Z$, which we refer
to as\emph{ }``instrument 1'', and the linear projection of an indicator
for receiving treatment $2$ on the instrument vector $Z$, i.e.\emph{,
}``instrument 2''. Using instruments 1 and 2\textemdash the predicted
treatments from the 2SLS first stage\textemdash as instruments is
numerically equivalent to using the full vector $Z$ as instruments.
The first condition\textemdash average conditional monotonicity\textemdash requires
that, conditional on instrument 2, instrument 1 does not, on average,
induce agent $i$ out of treatment $1$.\footnote{This condition generalizes \citet{FrandsenEtAl2019}'s average monotonicity
condition to multiple treatments and is substantially weaker than
the Imbens-Angrist monotonicity condition \citep{imbens1994identification}.} The second condition\textemdash no cross effects\textemdash requires
that, conditional on instrument 1, instrument 2 does not, on average,
induce agent $i$ into or out of treatment $1$. The latter condition
is necessary to ensure zero weight on $\beta_{2i}$ and is particular
to the case with multiple treatments. We show that this condition
is equivalent to assuming homogeneity in agents' relative responses
to the instruments: How instrument $1$ affects treatment $2$ compared
to how instrument $2$ affects treatment $2$ can not differ between
agents.

We derive a set of testable implications of our conditions. First,
we show that when outcome variable transformations interacted with
a treatment $1$ indicator are regressed on the instruments, the coefficient
on instrument $1$ should be non-negative and the coefficient on instrument
$2$ should be zero. Second, we show that the same must hold when
regressing an indicator for treatment $1$ on the instruments in subsamples
that are constructed using \emph{pre-determined }covariates. Using
data from \citet{BhullerEtAl2020}, we show how these tests can be
implemented in practice.

Building upon our general identification results, we consider a prominent
special case: 2SLS with $n$ mutually exclusive treatments and $n-1$
mutually exclusive binary instruments. This case is a natural generalization
of the canonical case with a binary treatment and a binary instrument
to multiple treatments. A typical application is a randomized controlled
trial where each agent is randomly assigned to one of $n$ treatments,
but compliance is imperfect. In this setting, our two conditions require
that each instrument affects exactly one treatment indicator. In particular,
there must be a labeling of the instruments such that instrument $k$
moves agents only from the excluded treatment $0$ into treatment
$k$. This result gives rise to additional testable implications of
our conditions in models with one binary instrument per treatment
indicator.

The requirement that each instrument affects exactly one choice margin
restricts choice behavior in a particular way. In particular, 2SLS
assigns proper weights only when choice behavior can be described
by a selection model where the excluded treatment is always the preferred
alternative or the next-best alternative. It is thus not sufficient
that each instrument influences the utility of only one choice alternative.
For instance, an instrument that affects only the utility of receiving
treatment $1$ could still affect the take-up of treatment $2$ by
inducing agents who would otherwise have selected treatment $2$ to
select treatment $1$. Such cross effects are avoided if the excluded
treatment is always at least the next-best alternative. To apply 2SLS
in this case, the researcher must argue why the excluded treatment
is always the best or the next-best alternative. Our results essentially
imply that unless researchers can infer next-best alternatives\textemdash as
in \citet{kirkeboen2016} and the following literature\textemdash 2SLS
in models with one binary instrument per treatment indicator does
not identify a meaningful causal effect under arbitrary heterogeneous
effects.

Until now, we have considered \emph{unordered} treatment effects\textemdash treatment
effects relative to an excluded treatment. Our results, however, also
apply to any other relative treatment effects a researcher might seek
to estimate through 2SLS. An important case is \emph{ordered treatment
effects}\textemdash the effect of treatment $k$ relative to treatment
\emph{k-1}.\footnote{\citet{AngristImbens1995JASA} showed the conditions under which 2SLS
with the \emph{multivalued treatment indicator} $T$ as the endogenous
variable identifies a convex combination of the effect of treatment
$2$ relative to treatment \emph{$1$ }and the effect of\emph{ }treatment
$1$ relative to treatment $0$. In contrast, we seek to determine
the conditions under which 2SLS with two binary treatment indicators\textemdash $D_{1}=\mathbf{1}\left[T\geq1\right]$
and $D_{2}=\mathbf{1}\left[T=2\right]$\textemdash \emph{separately
}identifies the effect of each of the two treatment margins.} In the ordered case, 2SLS with one binary instrument per treatment
indicator assigns proper weights if and only if there exists a labeling
of the instruments such that instrument $k$ moves agents only from
treatment \emph{k-1} to treatment $k$. This condition also imposes
a particular restriction on agents' choice behavior: we show that
2SLS assigns proper weights in such ordered choice models only when
agents' preferences can be described as single-peaked over the treatments.
When treatments have a natural ordering\textemdash such as years of
schooling\textemdash the researcher might be able to make a strong
theoretical case in favor of such preferences.

We finally present another special case of ordered choice where our
conditions are satisfied: a classical threshold-crossing model applicable
when treatment assignment depends on a single index crossing multiple
thresholds. For instance, treatments can be grades and the latent
index the quality of the student's work, or treatments might be years
of prison and the latent index the severity of the committed crime.
Suppose the researcher has access to exogenous shocks to these thresholds,
for instance through random assignment to judges or graders that agree
on ranking but use different cutoffs. Then 2SLS assigns proper weights
provided that there is a linear relationship between the predicted
treatments\textemdash an easily testable condition.\footnote{More precisely, the conditional expectation of predicted treatment
$k$ must be a linear function of predicted treatment $l\neq k$.}

Our paper contributes to a growing literature on the use of instruments
to identify causal effects in settings with multiple treatments \citep{HeckmanEtal2008_Annales,kline2016evaluating,kirkeboen2016,HeckmanPinto2018,lee2018identifying,Galindo2020Empirical,pinto2022beyond,Humphries2022,heinesen2022instrumental,kamat2023identification}
or multiple instruments \citep{MogstadEtAl2019,Goff2020Vector,mogstad2020policy}.
Our main contribution is to provide the exact conditions under which
2SLS with multiple treatments assigns proper weights under arbitrary
treatment effect heterogeneity. We allow for any number of treatments,
any number of continuous or discrete instruments, any definition of
treatment indicators, and covariates. Moreover, we show how the conditions
can be refuted. By comparison, the existing literature provides only
sufficient conditions in the case with three treatments, three instrument
values, and no controls \citep{BehagheletAl2013,kirkeboen2016}.

In the case with one binary instrument per treatment indicator, we
show that the \emph{extended monotonicity} condition provided by \citet{BehagheletAl2013}
is not only sufficient but also \emph{necessary} for 2SLS to assign
proper weights, after a possible permutation of the instruments. This
non-trivial result gives rise to a new testable implication: For 2SLS
to assign proper weights in such models, each instrument can only
affect one treatment. Furthermore, we show that knowledge of agents'
next-best alternatives\textemdash as in \citet{kirkeboen2016}\textemdash is
implicitly assumed whenever estimates from just-identified 2SLS models
with multiple treatments are interpreted as a positively weighted
sum of individual treatment effects. We thus show that the assumption
that next-best alternatives are observed or can be inferred is not
only sufficient but also essentially \emph{necessary} for 2SLS to
identify a meaningful causal parameter.\footnote{The only exception being that next-best alternatives need not be observed
for always-takers.} We also provide new identification results for ordered treatments.
First, we show when 2SLS with multiple ordered treatments identifies
separate treatment effects in a standard threshold-crossing model
considered in the ordered choice literature (e.g., \citealt{carneiro2003estimating,cunha2007identification,Heckman2007EconometricII}).
While \citet{Heckman2007EconometricII} show that local IV identifies
ordered treatment effects in such a model, we show that 2SLS can also
identify the effect of each treatment transition under an easily testable
linearity condition. A similar result is found concurrently by \citet{Humphries2022}.\footnote{\citet{Humphries2022} also show that if treatment assignment depends
on several unobserved latent indices\textemdash instead of just one\textemdash 2SLS
does not, in general, assign proper weights.} We also show how the result of \citet{BehagheletAl2013} extends
to ordered treatment effects. Finally, we show that for 2SLS to assign
proper weights in versions of the \citet{BehagheletAl2013} model
with ordered treatment effects, it must be possible to describe agents'
preferences as single-peaked over the treatments.

In contrast to \citet{HeckmanPinto2018}, who provide general identification
results in a setting with multiple treatments and discrete instruments,
we focus specifically on the properties of 2SLS\textemdash a standard
and well-known estimator common in the applied literature. Other contributions
to the literature on the use of instrumental variables to separately
identify multiple treatments \citep{lee2018identifying,Galindo2020Empirical,lee2020filtered,Mountjoy2019,pinto2022beyond}
focus on developing new approaches to identification. We do not necessarily
recommend 2SLS over these alternative methods. For instance, the method
of \citet{HeckmanPinto2018} identifies causal effects under strictly
weaker assumptions than those required for 2SLS to assign proper weights
in models with two binary treatment indicators and three treatments
(see Section \ref{subsec:alternative}). Also, as argued by \citet{Heckman2007EconometricII},
the weighted average of treatment effects produced by 2SLS in overidentified
models is not necessarily an interesting parameter, even when the
weights are non-negative. The alternatives to 2SLS, discussed in Section
\ref{subsec:alternative}, arguably all target more policy-relevant
treatment effects. But given the popularity of 2SLS among practitioners,
we still see a need to clarify the exact conditions under which this
is a valid approach\textemdash contributing to a recent body of research
assessing the robustness of standard estimators to heterogeneous effects
(\emph{e.g.}, \citealt{de2020two,de2020two_several,callaway2020difference,goodman2021difference,sun2020estimating,borusyak2021revisiting,goldsmith2022contamination}).

In Section \ref{sec:general}, we develop the exact conditions for
the multivariate 2SLS to assign proper weights to agent-specific causal
effects and present testable implications. In Section \ref{sec:special},
we consider two special cases. Section \ref{sec:if_fail} discusses
implications when the conditions fail and alternatives to 2SLS. Section
\ref{sec:application} provides an illustration of our conditions
for the random judge IV design and tests these conditions using data
from \citet{BhullerEtAl2020}. We conclude in Section \ref{sec:Conclusion}.
Proofs and additional results are in the Appendix.

\section{Multivariate 2SLS with Heterogeneous Effects\label{sec:general}}

In this section, we develop sufficient and necessary conditions for
the multivariate 2SLS to identify a positively weighted sum of individual
treatment effects under heterogeneous effects and present testable
implications of these conditions.

\subsection{Definitions}

Fix a probability space where an outcome corresponds to a randomly
drawn \emph{agent} $i$.\footnote{All random variables thus correspond to a randomly drawn agent. We
omit $i$ subscripts.} Define the following random variables: A multi-valued treatment $T\in\mathcal{T}\equiv\left\{ 0,1,2\right\} $,
an outcome $Y\in\mathbb{R}$, and a vector-valued instrument $Z\in\mathcal{Z}\subseteq\mathbb{R}^{m}$
with $m\geq2$.\footnote{We focus on \emph{mutually exclusive }treatments. Treatments that
are not mutually exclusive can always be made into mutually exclusive
treatments. For instance, two not mutually exclusive treatments and
an excluded treatment can be thought of as four mutually exclusive
treatments: Receiving the excluded treatment, receiving only treatment
1, receiving only treatment 2, and receiving both treatments.} For expositional ease, we focus on the case with three treatments
and no control variables. In Section \ref{subsec:generalization},
we show that all our results generalize to an arbitrary number of
treatments, and in Section \ref{subsec:controls}, we show how our
results extend to the case with control variables.

Define $\mathcal{S}$ as all possible mappings $f:\mathcal{Z}\rightarrow\mathcal{\mathcal{T}}$
from instrument values to treatments\textemdash all possible ways
the instrument can affect treatment. Following \citet{HeckmanPinto2018},
we refer to the elements of $\mathcal{S}$ as \emph{response types}.
The random variable $S\in\mathcal{S}$ describes agents' \emph{potential
treatment choices}: If agent $i$ has $S=s$, then $s\left(z\right)$
for $z\in\mathcal{Z}$ indicates the treatment selected by agent $i$
if $Z$ is set to $z$. The response type of an agent describes how
the agent's choice of treatment reacts to changes in the instrument.
For example, in the case with a binary treatment and a binary instrument,
the possible response types are \emph{never-takers }($s\left(0\right)=s\left(1\right)=0$),\emph{
always-takers} ($s\left(0\right)=s\left(1\right)=1$),\emph{ compliers
}($s\left(0\right)=0$, $s\left(1\right)=1$),\emph{ }and\emph{ defiers
}($s\left(0\right)=1$, $s\left(1\right)=0$)\emph{.} Similarly, define
$Y\left(k\right)$ for $k\in\mathcal{\mathcal{T}}$ as the agent's
\emph{potential outcome} when $T$ is set to $k$.

Using 2SLS, a researcher can aim to estimate two out of three relative
treatment effects. We focus on two special cases: the \emph{unordered
}and \emph{ordered }case. In the unordered case, the researcher seeks
to estimate the effects of treatment 1 and 2 relative to treatment
0 by estimating 2SLS with treatment indicators $D_{1}^{\text{unordered}}\equiv\mathbf{1}\left[T=1\right]$
and $D_{2}^{\text{unordered}}\equiv\mathbf{1}\left[T=2\right]$.\emph{}\footnote{The more general case\textemdash where the researcher seeks to estimate
\emph{all} relative treatment effects\textemdash could be analyzed
by varying which treatment is considered to be the excluded treatment.
In that case, the researcher should discuss and test the conditions
in Section \ref{sec:general} for each choice of excluded treatment.
Often, however, the researcher will only be interested in \emph{some
}relative treatment effects.} In that case, the treatment effects of interest are represented by
the random vector $\beta^{\text{unordered}}\equiv\left(Y\left(1\right)-Y\left(0\right),Y\left(2\right)-Y\left(0\right)\right)^{T}$.
In the ordered case, the researcher is interested in $\beta^{\text{ordered}}\equiv\left(Y\left(1\right)-Y\left(0\right),Y\left(2\right)-Y\left(1\right)\right)^{T}$
and uses treatment indicators $D_{1}^{\text{ordered}}\equiv\mathbf{1}\left[T\geq1\right]$
and $D_{2}^{\text{ordered}}\equiv\mathbf{1}\left[T=2\right]$. In
general terms, we let $\beta\equiv\left(\beta_{1},\beta_{2}\right)^{T}$
denote the treatment effects of interest and $D\equiv\left(D_{1},D_{2}\right)^{T}$
the corresponding treatment indicators. Unless otherwise specified,
our results hold for any definition of $\beta$. But to ease exposition
we focus on the unordered case when interpreting our results. We maintain
the following standard IV assumptions throughout:
\begin{assumption}
\label{assu:iv}(Exogeneity and Exclusion). $\left\{ Y\left(0\right),Y\left(1\right),Y\left(2\right),S\right\} \perp Z$
\end{assumption}
\begin{assumption}
\label{assu:rank}(Rank). $\Cov\left(Z,D\right)$ has full rank.
\end{assumption}
For a response type $s\in\mathcal{S}$, let $s_{1}$ and $s_{2}$
be the induced mapping between instruments and treatment indicators.
For instance, $s_{k}\left(z\right)\equiv\mathbf{1}\left[s\left(z\right)=k\right]$
if we use unordered treatment indicators, and $s_{k}\left(z\right)\equiv\mathbf{1}\left[s\left(z\right)\geq k\right]$
if we use ordered treatment indicators.  Define the 2SLS estimand
$\beta^{\text{2SLS}}=\left(\beta_{1}^{\text{2SLS}},\beta_{2}^{\text{2SLS}}\right)^{T}$by
\[
\beta^{\text{2SLS}}\equiv\Var\left(P\right)^{-1}\Cov\left(P,Y\right)
\]
where $P$ is the linear projection of $D$ on $Z$
\[
P=\left(P_{1},P{}_{2}\right)^{T}\equiv\E\left[D\right]+\Var\left(Z\right)^{-1}\Cov\left(Z,D\right)\left(Z-\E\left[Z\right]\right).
\]
We refer to $P$ as the \emph{predicted treatments}\textemdash the
best linear prediction of the treatment indicators given the value
of the instruments. Similarly, we refer to $P_{k}$ for $k\in\left\{ 1,2\right\} $
as \emph{predicted treatment $k$.} Since 2SLS using predicted treatments
as instruments is numerically equivalent to using the original instruments,
we can think of $P$ as our instruments. In particular, $P$ is a
linear transformation of the original instruments $Z$ such that we
get one instrument corresponding to each treatment indicator. We will
occasionally refer to $P_{1}$ and $P_{2}$ as ``instrument $1$''
and ``instrument $2$''. Let $\tilde{P}_{1}\equiv P_{1}-\frac{\Cov\left(P_{1},P_{2}\right)}{\Var\left(P_{2}\right)}P_{2}$
be the residualized instrument $1$ after netting out any linear association
with instrument $2$.

\subsection{Identification Results\label{subsec:Identification}}

What does multivariate 2SLS identify under Assumptions \ref{assu:iv}
and \ref{assu:rank} when treatment effects are heterogeneous across
agents? The following proposition expresses the 2SLS estimand as a
weighted sum of average treatment effects across response types:\footnote{By symmetry, an analogous expression can be derived for $\beta_{2}^{\text{2SLS}}$.}
\begin{prop}
\label{prop:weights}Under Assumptions \ref{assu:iv} and \ref{assu:rank}

\[
\beta_{1}^{\text{2SLS}}=\E\left[w_{1}^{S}\beta_{1}^{S}+w_{2}^{S}\beta_{2}^{S}\right]
\]
where, for a realization $s\in\mathcal{S}$ of $S$ and $k\in\left\{ 1,2\right\} $
\begin{eqnarray*}
w_{k}^{s}\equiv\frac{\Cov\left(\tilde{P}_{1},s_{k}\left(Z\right)\right)}{\Var\left(\tilde{P}_{1}\right)}; &  & \beta_{k}^{s}\equiv\E\left[\beta_{k}\mid S=s\right].
\end{eqnarray*}

\end{prop}
In the case of a binary treatment and a binary instrument, $\beta^{\text{2SLS}}$
is a weighted sum of average treatment effects for compliers and defiers.
Proposition \ref{prop:weights} is a generalization of this result
to the case with three treatments and $m$ instruments. In this case,
$\beta_{1}^{\text{2SLS}}$ is a weighted sum of the average treatment
effects of all response types present in the population. The parameter
$\beta_{k}^{s}$ describes the average effects of treatment $k$ for
agents of response type $s$. The weight $w_{k}^{s}$ indicates how
the average effects of treatment $k$ of agents with response type
$s$ contribute to the estimated effect of treatment $1$. Without
further restrictions, these weights could be both positive and negative.
Thus, in general, both treatment effects for response type $s$ might
contribute, either positively or negatively, to the estimated effect
of treatment $1$.

In the canonical case of a binary treatment and a binary instrument,
identification is ensured when there are no defiers. Proposition \ref{prop:weights}
generates similar restrictions on the possible response types in the
case of multiple treatments. To become familiar with our notation
and the implications of Proposition \ref{prop:weights}, consider
the following example.
\begin{example}
Consider the case of three treatments and two mutually exclusive binary
instruments: $\left(Z_{1},Z_{2}\right)\in\left\{ \left(0,0\right),\left(1,0\right),\left(0,1\right)\right\} $.
Assume we are interested in unordered treatment effects, $\beta^{\text{unordered}}=\left(Y\left(1\right)-Y\left(0\right),Y\left(2\right)-Y\left(0\right)\right)^{T}$,
with corresponding treatment indicators $D_{1}=\mathbf{1}\left[T=1\right]$
and $D_{2}=\mathbf{1}\left[T=2\right]$. One possible response type
in this case is defined by
\[
s\left(z\right)=\begin{cases}
2 & \text{if }z=\left(0,0\right)\\
1 & \text{if }z=\left(1,0\right)\\
2 & \text{if }z=\left(0,1\right)
\end{cases}.
\]
Response type $s$ thus selects treatment 2 unless $Z_{1}$ is turned
on. When $Z_{1}=1$, $s$ selects treatment $1$. Assume the best
linear predictors of the treatments indicators are $P_{1}=0.1+0.4Z_{1}$
and $P_{2}=0.2+0.5Z_{2}$ and that all instrument values are equally
likely. By Proposition \ref{prop:weights}, this response type's contribution
to $\beta_{1}^{\text{2SLS}}$ would be $w_{1}^{s}\beta_{1}^{s}+w_{2}^{s}\beta_{2}^{s}=2\beta_{1}^{s}-2\beta_{2}^{s}$.\footnote{We have $\tilde{P}_{1}=\frac{1}{4}\left(1.05+2Z_{1}+Z_{2}\right)$,
$s_{1}\left(Z\right)=Z_{1}$, and $s_{2}\left(Z\right)=1-Z_{1}$,
which gives $w_{1}^{s}=\Cov\left(\tilde{P}_{1},s_{1}\left(Z\right)\right)/\Var\left(\tilde{P}_{1}\right)=\frac{1}{12}/\frac{1}{24}=2$
and $w_{2}^{s}=-2$.} The average effect of treatment $2$ for response type $s$ thus
contributes negatively to the estimated effect of treatment $1$.
The presence of this response type in the population would be problematic.
The response type
\[
s\left(z\right)=\begin{cases}
0 & \text{if }z=\left(0,0\right)\\
1 & \text{if }z=\left(1,0\right)\\
0 & \text{if }z=\left(0,1\right)
\end{cases}
\]
on the other hand, has weights $w_{1}^{s}=2$ and $w_{2}^{s}=0$.
This response type's effect of treatment $2$ does not contribute
to the estimated effect of treatment $1$. Two-stage least squares
assigns proper weights on this response type's average treatment effects.\footnote{The weights on $\beta_{1}^{s}$ and $\beta_{2}^{s}$ in $\beta_{2}^{\text{2SLS}}$
are both zero in this example.}\hfill{}$\square$
\end{example}
Under homogeneous effects, or homogenous responses to the instruments,
the weight 2SLS assigns on the treatment effects of a particular response
type is not a cause of concern. By the following corollary of Proposition
\ref{prop:weights}, the ``cross'' weights $w_{2}^{s}$ are zero
on average and the ``own'' weights $w_{1}^{s}$ are, on average,
positive:
\begin{cor}
\label{cor:weights-multiple}Under Assumptions \ref{assu:iv} and
\ref{assu:rank}, the 2SLS estimand $\beta_{1}^{\text{2SLS}}$ is
a weighted sum of $\beta_{1}^{s}$ and $\beta_{2}^{s}$ where the
weights on the first sum to one and the weights on the latter sum
to zero.
\end{cor}
Thus, if $\beta_{1}^{s}=\beta_{1}$ and $\beta_{2}^{s}=\beta_{2}$
for all $s\in\mathcal{S}$, we get $\beta_{1}^{\text{2SLS}}=\E\left[w_{1}^{S}\beta_{1}+w_{2}^{S}\beta_{2}\right]=\beta_{1}$.
Also, if the population consists of only one response type $s$, we
get $\beta_{1}^{\text{2SLS}}=\beta_{1}^{s}$.\footnote{One can allow for always-takers and never-takers.}
But under heterogeneous effects \emph{and} heterogeneous responses
to the instruments, the estimated effect of treatment $1$ might be
contaminated by the effect of treatment $2$. This makes interpreting
2SLS estimates hard. Ideally, we would like $w_{1}^{s}$ to be non-negative,
and $w_{2}^{s}$ to be zero for all $s$. Only in that case can we
interpret the 2SLS estimate of the effect of treatment $1$ as a positively
weighted average of the effect of treatment $1$ under heterogeneous
effects. Throughout the rest of the paper, we say that the 2SLS estimand
of the effect of treatment $1$ assigns proper weights if the following
holds:
\begin{defn}
The 2SLS estimand of the effect of treatment 1 assigns \emph{proper
weights }if for each $s\in\supp\left(S\right)$, the 2SLS estimand
$\beta_{1}^{\text{2SLS}}$ places non-negative weight on $\beta_{1}^{s}$
and zero weight on $\beta_{2}^{s}$.
\end{defn}
We say that 2SLS \emph{assigns proper weights} if it assigns proper
weights for both treatment effects.\footnote{The requirement that 2SLS assign proper weights is equivalent to 2SLS
being \emph{weakly causal} (\citealt{blandhol2022tsls})\textemdash providing
estimates with the correct sign\textemdash under arbitrary heterogeneous
effects (see Section \ref{subsec:WC}). Note that we do not require
the estimand to assign \emph{equal weights }to all response types.
Such a criterion would essentially rule out the use of 2SLS beyond
the case studied in Section \ref{sec:binary}: As noted by \citet{Heckman2007EconometricII},
2SLS assigns higher weights to response types that are more influenced
by the instruments, producing a weighted average that is not necessarily
of policy interest. See Section \ref{subsec:alternative} for alternatives
to 2SLS that seek to target more policy-relevant parameters.} When is the weight $w_{1}^{s}$ on $\beta_{1}^{s}$ non-negative
for all $s$? The formula for the weight $w_{1}^{s}$ equals the
coefficient in a hypothetical linear regression of $s_{1}\left(Z\right)$
on $\tilde{P}_{1}$ across realizations of $Z$.\footnote{Or, equivalently, the coefficient on $P_{1}$ in a hypothetical regression
of $s_{1}\left(Z\right)$ on $P_{1}$ and $P_{2}$. Such a regression
is ``hypothetical'' since, in practice, we only observe $s_{1}\left(Z\right)$
for one realization of $Z$.} Thus $\beta_{1}^{\text{2SLS}}$ puts more weight on $\beta_{1}^{s}$
the more $\tilde{P}_{1}$ tends to push response type $s$ into treatment
$1$. If $\tilde{P}_{1}$ tends to push the response type \emph{out
of }treatment $1$, the weight $w_{1}^{s}$ will be negative. The
condition that ensures that $\beta_{1}^{\text{2SLS}}$ assigns non-negative
weights on $\beta_{1}^{s}$ for all $s$ can be written succintly
as follows.
\begin{assumption}
\label{assumption:non_neg} (Average Conditional Monotonicity). For
all $s\in\mathcal{S}$ 
\[
\E\left[\tilde{P}_{1}\mid s\left(Z\right)=1\right]\ge\E\left[\tilde{P}_{1}\mid s\left(Z\right)\neq1\right].
\]
\end{assumption}
\begin{cor}
\label{cor:non_neg}Under Assumptions \ref{assu:iv} and \ref{assu:rank},
the weight on $\beta_{1}^{s}$ in $\beta_{1}^{\text{2SLS}}$ is non-negative
for all $s$ if and only if Assumption \ref{assumption:non_neg} holds.
\end{cor}
Assumption \ref{assumption:non_neg} requires that, after controlling
linearly for predicted treatment $2$, the average predicted treatment
$1$ can not be lower at instrument values where a given agent selects
treatment $1$ than at other instrument values. Intuitively, Assumption
\ref{assumption:non_neg} requires that, controlling linearly for
predicted treatment 2, there is a non-negative correlation between
predicted treatment $1$ and potential treatment $1$ for each agent
across values of the instruments. We refer to the condition as \emph{average
conditional monotonicity} since it generalizes the average monotonicity
condition defined by \citet{FrandsenEtAl2019} to multiple treatments.\footnote{With one treatment, Assumption \ref{assumption:non_neg} reduces to
$\E\left[P_{1}\mid s\left(Z\right)=1\right]\geq\E\left[P_{1}\right]\Leftrightarrow\Cov\left(P_{1},s_{1}\left(Z\right)\right)\geq0$
which coincides with the average monotonicity condition in \citet{FrandsenEtAl2019}.
Under Assumption \ref{assumption:zero}, Assumption \ref{assumption:non_neg}
is equivalent to the correlation between $P_{1}$ and $s_{1}\left(Z\right)$
being non-negative (without having to condition on $P_{2}$).} In particular, the positive relationship between predicted treatment
and potential treatment only needs to hold ``on average'' across
realizations of the instruments. Thus, an agent might be a ``defier''
for some pairs of instrument values as long as she is a ``complier''
for sufficiently many other pairs. Informally, we can think of Assumption
\ref{assumption:non_neg}, as requiring the partial effect of $P_{1}$\textemdash instrument
$1$\textemdash on treatment $1$ to be, on average, non-negative
for all agents.

The weight $w_{2}^{s}$ on $\beta_{2}^{s}$ in $\beta_{1}^{\text{2SLS}}$
equals the coefficient in a hypothetical linear regression of $s_{2}\left(Z\right)$
on $\tilde{P}_{1}$. Intuitively, if $P_{1}$ has a tendency to push
certain agents into or out of treatment $2$\textemdash even after
controlling linearly for $P_{2}$\textemdash the estimated effect
of treatment $1$ will be contaminated by the effect of treatment
$2$ on these agents. The following condition is necessary and sufficient
to avoid such contamination:
\begin{assumption}
\label{assumption:zero} (No Cross Effects). For all $s\in\mathcal{S}$

\[
\E\left[\tilde{P}_{1}\mid s\left(Z\right)=2\right]=\E\left[\tilde{P}_{1}\mid s\left(Z\right)\neq2\right].
\]

\end{assumption}
\medskip{}
\begin{cor}
\label{cor:zero}Under Assumptions \ref{assu:iv} and \ref{assu:rank},
the weight on $\beta_{2}^{s}$ in $\beta_{1}^{\text{2SLS}}$ is zero
for all $s$ if and only if Assumption \ref{assumption:zero} holds.
\end{cor}
Assumption \ref{assumption:zero} requires that, after controlling
linearly for predicted treatment $2$, the average predicted treatment
$1$ can not be different at instrument values where a given agent
selects treatment $2$ than at other instrument values. In other words,
after linearly controlling for predicted treatment $2$, there can
be no correlation between predicted treatment $1$ and potential treatment
$2$ for any agent. Informally, we can think of Assumption \ref{assumption:zero}
as requiring there to be no partial effect of instrument 1 on treatment
$2$ for any agent. The following restatement of Assumption \ref{assumption:zero}
provides further intuition:
\begin{prop}
\label{cor:zero_alt}Assumption \ref{assumption:zero} is equivalent
under Assumption \ref{assu:iv} to
\[
\Cov\left(P_{1},s_{2}\left(Z\right)\right)=\rho\Cov\left(P_{2},s_{2}\left(Z\right)\right)
\]
 for all $s\in\mathcal{S}$ and a constant $\rho$.
\end{prop}
Thus, Assumption \ref{assumption:zero} \emph{does not} require $\Cov\left(P_{1},s_{2}\left(Z\right)\right)=0$\textemdash that
take-up of treatment $2$ is (unconditionally) unaffected by instrument
$1$. Instead, the condition requires that the extent that instrument
$1$ affects treatment $2$ \emph{compared to} how instrument $2$
affects treatment $2$ is \emph{constant} across response types. Assumption
\ref{assumption:zero} thus imposes a certain homogeneity in agents'
responses to the instruments. In particular, it is not allowed that
some agents react strongly to instrument $1$ but weakly to instrument
$2$ in the take-up of treatment $2$, while the opposite is true
for other agents. It is important to note that Assumption \ref{assumption:zero}
is a knife-edge condition\textemdash requiring an \emph{exact zero}
partial effect. Small deviations from Assumption \ref{assumption:zero}
will, however\textemdash in combination with moderate heterogeneous
effects\textemdash lead only to a small asymptotic 2SLS bias (see
Section \ref{subsec:het}).

Assumptions \ref{assumption:non_neg} and \ref{assumption:zero} are
necessary and sufficient conditions for 2SLS to assign proper weights.
This is our main result:
\begin{cor}
\label{cor:avg_mono}The 2SLS estimand of the effect of treatment
1 assigns proper weights under Assumptions \ref{assu:iv} and \ref{assu:rank}
if and only if Assumptions \ref{assumption:non_neg} and \ref{assumption:zero}
hold.
\end{cor}
Informally, 2SLS assigns proper weights if and only if for all agents
and treatments $k$, (i) increasing instrument $k$ tends to weakly
increase adoption of treatment $k$ and (ii) conditional on instrument
$k$, increases in instrument $l\neq k$ do not tend to push the agent
into or out of treatment $k$. In Sections \ref{sec:IA-monotonicity}
and \ref{sec:U-monotonicity}, we show how our conditions relate to
two natural generalizations of the \citet{imbens1994identification}'s
monotonicity condition to multiple treatments. In particular, we show
that if either \citet{HeckmanPinto2018}'s \emph{unordered monotonicity}
or \citet{Kamat2023}'s \emph{joint monotonicity }holds, then 2SLS
assigns proper weights under an easily testable linearity condition.

\subsection{Testable Implications of the Identification Conditions\label{subsec:Testing}}

Assumptions \ref{assumption:non_neg} and \ref{assumption:zero} can
not be directly assessed since we only observe $s\left(z\right)$
for the observed instrument values. But the assumptions do have testable
implications. For simplicity, consider the case with unordered treatment
effects $D_{k}=\mathbf{1}\left[T=k\right]$.\footnote{See Section \ref{subsec:kitagawa-generalization} for how Proposition
\ref{prop:kitagawa} generalizes to other treatment indicators such
as $D_{k}=\mathbf{1}\left[T\geq k\right]$.} We then have the following result:\footnote{We thank the associate editor for suggesting these testable implications.
\citet{balke1997bounds} and \citet{heckman2005structural} showed
similar testable implications of the binary treatment IV model assumptions.
\citet{sun2020instrument} presents similar testable implications
of multiple-treatment IV assumptions under multivalued ordered monotonicity
and unordered monotonicity.}
\begin{prop}
\label{prop:kitagawa}Maintain Assumptions \ref{assu:iv} and \ref{assu:rank}
and let $y\leq y'$. If 2SLS with $D_{k}=\mathbf{1}\left[T=k\right]$
assigns proper weights then
\[
\Var\left(P\right)^{-1}\Cov\left(P,\mathbf{1}_{y\leq Y\leq y'}D\right)
\]
is a non-negative diagonal matrix.\footnote{In fact, $\Var\left(P\right)^{-1}\Cov\left(P,\mathbf{1}_{Y\in\mathcal{B}}D\right)$
must be non-negative diagonal for \emph{any }set $\mathcal{B}\subset\supp\left(Y\right)$.
But by the argument of \citet{kitagawa2015test}, Lemma B.7, it is
sufficient to consider sets of the form $\left\{ \left[y,y'\right]\mid y\leq y'\right\} $.}
\end{prop}
For instance, for a binary outcome variable $Y\in\left\{ 0,1\right\} $,
Proposition \ref{prop:kitagawa} implies that $\Var\left(P\right)^{-1}\Cov\left(P,YD\right)$
and $\Var\left(P\right)^{-1}\Cov\left(P,\left(1-Y\right)D\right)$
are non-negative diagonal matrices. That $\Var\left(P\right)^{-1}\Cov\left(P,YD\right)$
is non-negative diagonal can be tested by running the following regressions:\footnote{Whether $\Var\left(P\right)^{-1}\Cov\left(P,\left(1-Y\right)D\right)$
is non-negative diagonal can be tested in an analogous manner by using
$\left(1-Y\right)D_{1}$ and $\left(1-Y\right)D_{2}$ as outcome variables.} 
\[
YD_{1}=\varphi_{1}+\vartheta_{11}P_{1}+\vartheta_{12}P_{2}+\upsilon_{1}
\]
\[
YD_{2}=\varphi_{2}+\vartheta_{21}P_{1}+\vartheta_{22}P_{2}+\upsilon_{2}
\]
and test whether $\vartheta_{11}$ and $\vartheta_{22}$ are non-negative
and whether $\vartheta_{12}=\vartheta_{21}=0$. Intuitively, if no-cross-effects
is satisfied, $P_{2}$ can not increase nor decrease the share of
agents with $T=1$ and $Y\left(1\right)=1$, after controlling for
$P_{1}$. For a continuous outcome variable, the implications could
be tested using the method of \citet{mourifie2017testing}\textemdash first
transforming the Proposition \ref{prop:kitagawa} condition to conditional
moment inequalities and then applying \citet{chernozhukov2013intersection}.
In that case, the relevant conditional moment conditions are that
\[
\Var\left(P\right)^{-1}\E\left[\left(P-\E\left[P\right]\right)D\mid Y=y\right]
\]
is non-negative diagonal for all $y\in\supp\left(Y\right)$.\footnote{For $\mathcal{B}\subset\supp\left(Y\right)$, we have
\begin{eqnarray*}
\Cov\left(P,\mathbf{1}_{Y\in\mathcal{B}}D\right) & = & \E\left[P\mathbf{1}_{Y\in\mathcal{B}}D\right]-\E\left[P\right]\E\left[\mathbf{1}_{Y\in\mathcal{B}}D\right]\\
 & = & \left(\E\left[PD\mid Y\in\mathcal{B}\right]-\E\left[P\right]\E\left[D\mid Y\in\mathcal{B}\right]\right)\Pr\left[Y\in\mathcal{B}\right]\\
 & = & \E\left[\left(P-\E\left[P\right]\right)D\mid Y\in\mathcal{B}\right]\Pr\left[Y\in\mathcal{B}\right]
\end{eqnarray*}
} Note that this test is a joint test of Assumptions \ref{assu:iv},
\ref{assumption:non_neg}, and \ref{assumption:zero}. If the test
rejects, it could be due to a violation of exogeneity or the exclusion
restriction.

Further testable implications can be obtained if the researcher has
access to ``pre-determined'' covariates. In particular, let $X\in\left\{ 0,1\right\} $
be a random variable such that $Z\perp\left(X,S\right)$.\footnote{We have already assumed $Z\perp S$ (Assumption \ref{assu:iv}). The
condition $Z\perp\left(X,S\right)$ requires, in addition, that $Z$
is independent of the joint distribution of $X$ and $S$. If the
instrument $Z$ is truly random, then any variable $X$ that pre-dates
the randomization satisfies $Z\perp\left(X,S\right)$. For instance,
in a randomized control trial, $X$ can be any pre-determined characteristic
of the individuals in the experiment.} Informally, $X$ is a \emph{pre-determined} variable not influenced
by or correlated with the instrument. We then have the following testable
prediction.
\begin{prop}
\label{prop:test}Maintain Assumption \ref{assu:rank}, and assume
$X\in\left\{ 0,1\right\} $ with $Z\perp\left(X,S\right)$. If Assumptions
\ref{assumption:non_neg}\textendash \ref{assumption:zero} hold then
\[
\Var\left(P\mid X=1\right)^{-1}\Cov\left(P,D\mid X=1\right)
\]
is a diagonal non-negative matrix.
\end{prop}
This prediction can be tested by running the following regressions:
\[
D_{1}=\gamma_{1}+\eta_{11}P_{1}+\eta_{12}P_{2}+\varepsilon_{1}
\]
\[
D_{2}=\gamma_{2}+\eta_{21}P_{1}+\eta_{22}P_{2}+\varepsilon_{2}
\]
for the subsample $X=1$ and testing whether $\eta_{11}$ and $\eta_{22}$
are non-negative and whether $\eta_{12}=\eta_{21}=0$. The predicted
treatments $P$ can be estimated from a linear regression of the treatments
on the instruments on the whole sample\textemdash a standard first-stage
regression. This test thus assesses the relationship between predicted
treatments and selected treatments in subsamples.\footnote{The one-treatment version of this test is commonly applied in the
literature (e.g., \citealt{dobbie2018effects,BhullerEtAl2020}) and
formally justified by \citet{FrandsenEtAl2019}.} If Assumptions \ref{assumption:non_neg}\textendash \ref{assumption:zero}
hold, we should see a positive relationship between treatment $k$
and predicted treatment $k$ and no statistically significant relationship
between treatment $k$ and predicted treatment $l\neq k$ in (pre-determined)
subsamples.\footnote{If the test is applied on various subsamples, $p$-values need to
be adjusted to account for multiple testing. It is only meaningful
to apply the test on \emph{sub}samples. The condition is always mechanically
satisfied in the full sample.} Note that failing to reject that $\Var\left(P\mid X=1\right)^{-1}\Cov\left(P,D\mid X=1\right)$
is non-negative diagonal across all observable pre-determined $X$
does not \emph{prove} that 2SLS assigns proper weights, even in large
samples: There might always be \emph{unobserved} pre-determined characteristics
$X$ such that $\Var\left(P\mid X=1\right)^{-1}\Cov\left(P,D\mid X=1\right)$
is not non-negative diagonal.\footnote{Similarly, in randomized control trials, showing that treatment is
uncorrelated with \emph{observed} pre-determined covariates does not
prove that treatment is randomly assigned.} Also, note that the test is a joint test of Assumptions \ref{assumption:non_neg}\textendash \ref{assumption:zero}
\emph{and} the assumption that $X$ is pre-determined. Thus, if the
test rejects, the reason might be that $X$ is not pre-determined.

Finally, note that the Proposition \ref{prop:kitagawa} test can also
be applied on the subsample $X=1$ if the researcher is willing to
maintain that Assumptions \ref{assu:iv}\textendash \ref{assumption:zero}
hold conditional on $X$. In that case, the Proposition \ref{prop:test}
test can be seen as a special case of the Proposition \ref{prop:kitagawa}
test with $y=-\infty$ and $y'=\infty$.

\section{Special Cases\label{sec:special}}

In this section, we first provide general identification results in
a model with binary instruments and derive the implied restrictions
on choice behavior under ordered or unordered treatment effects. Then,
we provide identification results for a standard threshold-crossing
model\textemdash an example of an overidentified model with ordered
treatment effects.

\subsection{One Binary Instrument Per Treatment Indicator\label{sec:binary}}

\subsubsection{Identification Results}

The standard application of 2SLS involves one binary instrument and
one binary treatment. In this section, we apply the results in Section
\ref{sec:general} to show how this canonical case generalizes to
the case with three possible treatments and three possible values
of the instrument.\footnote{The results generalize to $n$ possible treatments and $n$ possible
values of the instruments; see Section \ref{subsec:generalization}.} We refer to this case as the\emph{ ``just-identified''} case:
the number of distinct values of the instruments equals the number
of treatments.\footnote{In models with homogeneous treatment effects, having the same number
of instrument values as treatments gives ``just enough'' instruments
to ensure identification of all model parameters. Note, however, that
this ``just-identified'' model does not identify all parameters
under heterogeneous effects, since the number of parameters is then
much larger than the number of treatments.} In particular, assume we have an instrument, $V\in\left\{ 0,1,2\right\} $,
from which we create two mutually exclusive binary instruments $Z=\left(Z_{1},Z_{2}\right)^{T}$with
$Z_{v}\equiv\mathbf{1}\left[V=v\right]$.\footnote{There are other ways of creating two instruments from $V$. For instance,
one could define $Z_{1}=\mathbf{1}\left[V\geq1\right]$ and $Z_{2}=\mathbf{1}\left[V=2\right]$.
Such a parameterization would give exactly the same 2SLS estimate.
We focus on the case of mutually exclusive binary instruments since
it allows for an easier interpretation of our results.} We maintain Assumptions \ref{assu:iv} and \ref{assu:rank}. This
setting is common in applications. For instance, $Z_{v}$ might be
an inducement to take up treatment $v$, as considered by \citet{BehagheletAl2013}.
Under which conditions does the multivariate 2SLS assign proper weights
in this setting? It turns out that proper identification is\emph{
}achieved only when each instrument affects only one treatment. For
simpler exposition, we represent in this section the response types
$s$ as functions $s:\left\{ 0,1,2\right\} \rightarrow\left\{ 0,1,2\right\} $
where $s\left(v\right)$ is the treatment selected by response type
$s$ when $V$ is set to $v$. Thus, $s\left(v\right)$ is the \emph{potential
treatment }when $Z_{v}=1$.
\begin{prop}
\label{prop:binary}2SLS assigns proper weights in the above model
if and only if there exists a one-to-one mapping $f:\left\{ 0,1,2\right\} \rightarrow\left\{ 0,1,2\right\} $
between instruments and treatments, such that for all $k\in\left\{ 1,2\right\} $
and $s\in\mathcal{S}$ either $s_{k}\left(v\right)=0$ for all $v\in\left\{ 0,1,2\right\} $,
$s_{k}\left(v\right)=1$ for all $v\in\left\{ 0,1,2\right\} $, or
$s_{k}\left(v\right)=1\Leftrightarrow f\left(v\right)=k$.
\end{prop}
In words, for each agent $i$ and treatment $k$, either $i$ never
takes up treatment $k$, always takes up treatment $k$, or takes
up treatment $k$ if and only if $f\left(V\right)=k$. Each instrument
$Z_{v}$ is thus associated with exactly one treatment $D_{f\left(v\right)}$.
For ease of exposition, we can thus, without loss of generality, assume
that the treatments and instruments are ordered in the same way:
\begin{assumption}
\label{assu:order}Assume instrument values are labeled such that
$f\left(k\right)=k$ for all $k\in\left\{ 0,1,2\right\} $ where $f$
is the unique mapping defined in Proposition \ref{prop:binary}.
\end{assumption}
To see the implications of Proposition \ref{prop:binary}, consider
the response types defined in Table \ref{tab:response_types}. It
turns out that, for 2SLS to assign proper weights, the population
can not consist of any other response type:
\begin{table}
\centering{}\caption{Allowed Response Types in the Just-Identified Case.\label{tab:response_types}}
\begin{tabular}{cc}
\hline 
Response Type & $\left(s\left(0\right),s\left(1\right),s\left(2\right)\right)$\tabularnewline
\hline 
\hline 
Never-taker & $\left(0,0,0\right)$\tabularnewline
Always-1-taker & $\left(1,1,1\right)$\tabularnewline
Always-2-taker & $\left(2,2,2\right)$\tabularnewline
1-complier & $\left(0,1,0\right)$\tabularnewline
2-complier & $\left(0,0,2\right)$\tabularnewline
Full complier & $\left(0,1,2\right)$\tabularnewline
\hline 
\end{tabular}
\end{table}

\begin{cor}
\label{cor:binary_2}2SLS with $D_{k}=\mathbf{1}\left[T=k\right]$
assigns proper weights under Assumption \ref{assu:order} if and only
if all agents are either never-takers, always-1-takers, always-2-takers,
1-compliers, 2-compliers, or full compliers.
\end{cor}
\citet{BehagheletAl2013} show that these assumptions are sufficient
to ensure that 2SLS assigns proper weights. See also \citet{kline2016evaluating}
for similar conditions in the case of multiple treatments and one
instrument.\footnote{Equation 1 in \citet{kline2016evaluating}, generalized to two instruments,
also describes the same response types: $s\left(1\right)\neq s\left(0\right)\Rightarrow s\left(1\right)=1$
and $s\left(2\right)\neq s\left(0\right)\Rightarrow s\left(2\right)=2$.
The requirement that the instruments can cause individuals to switch
only from ``no treatment'' (the excluded treatment) into some treatment
is shared by \citet{rose2021recoding}'s \emph{extensive margin compliers
only} assumption.} Proposition \ref{prop:binary} shows that these conditions are not
only sufficient but also \emph{necessary}, after a possible permutation
of the instruments. These response types are characterized by instrument
$k$ not affecting treatment $l\neq k$\textemdash no cross effects.
Under the response type restrictions of Table \ref{tab:response_types},
2SLS identifies the average treatment effect of treatment $1$ for
the combined population of 1-compliers and full compliers and the
average treatment effect of treatment $2$ for the combined population
of 2-compliers and full compliers.\footnote{Thus, in the just-identified case, whenever 2SLS assigns proper weights,
it also assigns \emph{equal }weight to all complier types. This result
does not generalize to the overidentified case.} These treatment effects coincide with the treatment effects identified
by the method of \citet{HeckmanPinto2018}.\footnote{The methods proposed by \citet{HeckmanPinto2018} can be applied to
identify further parameters of interest. In particular, if we denote
always-1-takers by $s_{A1}$, always-2-takers by $s_{A2}$, never-takers
by $s_{N}$, 1-compliers by $s_{C1}$, 2-compliers by $s_{C2}$, and
full compliers by $s_{F}$, their method allows to identify the counterfactuals
$\E\left[Y\left(1\right)\mid S=s_{A1}\right]$, $\E\left[Y\left(2\right)\mid S=s_{A2}\right]$,
$\E\left[Y\left(1\right)\mid S\in\left\{ s_{C1},s_{F}\right\} \right]$,
$\E\left[Y\left(0\right)\mid S\in\left\{ s_{C1},s_{F}\right\} \right]$,
$\E\left[Y\left(2\right)\mid S\in\left\{ s_{C2},s_{F}\right\} \right]$,
$\E\left[Y\left(0\right)\mid S\in\left\{ s_{C2},s_{F}\right\} \right]$,
$\E\left[Y\left(0\right)\mid S\in\left\{ s_{N},s_{C1}\right\} \right]$,
and $\E\left[Y\left(0\right)\mid S\in\left\{ s_{N},s_{C2}\right\} \right]$.
Moreover, their method allows for the characterization of always-1-takers,
always-2-takers, and the following combined populations: 1-compliers
and full compliers, 2-compliers and full compliers, 1-compliers and
never-takers, and 2-compliers and never-takers.}

\subsubsection{Testing For Proper Weighting in Just-Identified Models}

By Proposition \ref{prop:binary}, there must exist a permutation
of the instruments such that instrument $k$ affects only treatment
$k$. This result gives rise to more powerful versions of the Section
\ref{subsec:Testing} testable implications applicable in just-identified
models:\footnote{It is straightforward to see that the tests become more powerful when
$P$ is replaced by $Z$. For instance, if $Y\left(k\right)=0$ for
all $k$, $\Var\left(P\right)^{-1}\Cov\left(P,\mathbf{1}_{y\leq Y\leq y'}D\right)$
is always non-negative diagonal but $\Var\left(Z\right)^{-1}\Cov\left(Z,\mathbf{1}_{y\leq Y\leq y'}D\right)$
might not. Conversely, if $\Var\left(Z\right)^{-1}\Cov\left(Z,\mathbf{1}_{y\leq Y\leq y'}D\right)$
is non-negative diagonal for all $y$ and $y$', then this must also
be true for $\Var\left(Z\right)^{-1}\Cov\left(Z,D\right)$. Since
$P\equiv\E\left[D\right]+\Var\left(Z\right)^{-1}\Cov\left(Z,D\right)\left(Z-\E\left[Z\right]\right)$,
we must then have that $\Var\left(P\right)^{-1}\Cov\left(P,\mathbf{1}_{y\leq Y\leq y'}D\right)$
is also non-negative diagonal for all $y$ and $y$'.}
\begin{prop}
\label{prop:just-identified-test}Under Assumption \ref{assu:order},
Propositions \ref{prop:kitagawa} and \ref{prop:test} continue to
hold when $P$ is replaced by $Z$.
\end{prop}
The stronger version of Proposition \ref{prop:test} reads as follows:
If 2SLS assigns proper weights in just-identified models, there must
be a permutation of the instruments such that treatment $k$ is affected
only by instrument $k$ in first-stage regressions.\footnote{Typically, the researcher would have a clear hypothesis about which
instrument is supposed to affect each treatment, avoiding the need
to run the test for all possible permutations.} As in Section \ref{subsec:Testing}, this prediction can be tested
by running a first-stage regression\footnote{\citet{heinesen2022instrumental} show that the coefficients from
a first-stage regression on the \emph{full sample} can be used to
partially identify violations of ``irrelevance'' and ``next-best''
assumptions invoked in \citet{kirkeboen2016}. Their Proposition 4
implies that $\eta$ is non-negative diagonal if their monotonicity,
irrelevance, and next-best assumptions hold. We show that this test
is a valid test not only of their invoked assumptions, but also more
generally of whether 2SLS assigns proper weights. Moreover, we show
that the test can be applied on subsamples.}
\[
D=\gamma+\eta Z+\varepsilon
\]
in a ``pre-determined'' subsample and test whether $\eta$ is a
non-negative diagonal matrix. This test can also be applied in the
full sample. The stronger version of Proposition \ref{prop:kitagawa}
can be tested in a similar manner.

\subsubsection{Choice-Theoretic Characterization}

What does the condition in Proposition \ref{prop:binary} imply about
choice behavior? To analyze this, we use a random utility model.\footnote{In a seminal article, \citet{vytlacil2002independence} showed that
the \citet{imbens1994identification} monotonicity condition is equivalent
to assuming that agents' selection into treatment can be described
by a random utility model where agents select into treatment when
a latent index crosses a threshold. In this section, we seek to provide
similar characterizations of the condition in Proposition \ref{prop:binary}.} Assume that response type $s$'s indirect utility from choosing treatment
$k$ when $V=v$ is $I_{ks}\left(v\right)$ and that $s$ selects
treatment $k$ if $I_{ks}\left(v\right)>I_{ls}\left(v\right)$ for
all $l\neq k$.\footnote{Since all agents of the same response type have identical behavior,
it is without loss of generality to assume that all agents of a response
type have the same indirect utility function. We assume response types
are never indifferent between treatments.} The implicit assumptions about choice behavior differ according to
which treatment effects we seek to estimate. We here consider the
cases of ordered and unordered treatment effects.

\paragraph*{Unordered Treatment Effects.}

What are the implicit assumptions on choice behavior when $\beta=\left(Y\left(1\right)-Y\left(0\right),Y\left(2\right)-Y\left(0\right)\right)^{T}$?
In this case, instrument $k$ can affect only treatment $k$. Thus,
instrument $k$ can not impact the utility of treatment $l\neq k$
in any way that changes choice behavior. Also, for instrument $k$
to impact treatment $k$ without impacting treatment $l$, it must
be that instrument $k$ can change treatment status only from treatment
$0$ to treatment $k$ and vice versa. Thus, essentially, Proposition
\ref{prop:binary} requires that only instrument $k$ affects the
indirect utility of treatment $k$, and that the excluded treatment
always is at least the next-best alternative:\footnote{\citet{lee2020filtered} consider a similar random utility model.
Their Additive Random Utility Model in combination with \emph{strict
one-to-one targeting }and the assumption that all treatments except
$T=0$ are targeted gives $I_{ks}\left(v\right)=u_{ks}+\mu_{k}\mathbf{1}\left[v=k\right]$
for constants $u_{ks}$ and $\mu_{k}>0$. As shown by \citet{lee2020filtered},
these assumptions are generally not sufficient to point-identify local
average treatment effects. Our result shows that\textemdash in the
just-identified case\textemdash local average treatment effects can
be identified if we additionally assume that all agents have the same
next-best alternative.}
\begin{prop}
\label{prop:binary_roy}2SLS with $D_{k}=\mathbf{1}\left[T=k\right]$
assigns proper weights if and only if there exists $u_{ks}\in\mathbb{R}$
and $\mu_{ks}\geq0$ such that $s\left(v\right)=\arg\max_{k}I_{ks}\left(v\right)$
with\footnote{The assumption that $\arg\max_{k}I_{ks}\left(z\right)$ is a singleton
implies that preferences are strict.}
\[
I_{0s}\left(v\right)=0
\]
\[
I_{ks}\left(v\right)=u_{ks}+\mu_{ks}\mathbf{1}\left[v=k\right]
\]
\[
I_{ks}\left(v\right)>I_{0s}\left(v\right)\Rightarrow I_{ls}\left(v\right)<I_{0s}\left(v\right)
\]
for all $s\in\mathcal{S}$, $v\in\left\{ 0,1,2\right\} $, $k,l\in\left\{ 1,2\right\} $
with $l\neq k$.
\end{prop}
The excluded treatment is, thus, always ``in between'' the selected
treatment and all other treatments in this selection model. Since
next-best alternatives are not observed, there are selection models
consistent with 2SLS assigning proper weights where other alternatives
than the excluded treatment are occasionally next-best alternatives.
But when indirect utilities are given by $I_{0s}=0$ and $I_{ks}=u_{ks}+\mu_{ks}\mathbf{1}\left[v=k\right]$,
this can only happen when $s$ always selects the same treatment,
in which case the identity of the next-best alternative is irrelevant.\footnote{See the proof of Proposition \ref{prop:binary_roy}.}

\begin{figure}
\caption{\label{fig:school}Excluded Treatment Always the Best or the Next-Best
Alternative.}
\medskip{}

\begin{centering}
\includegraphics[width=0.7\textwidth]{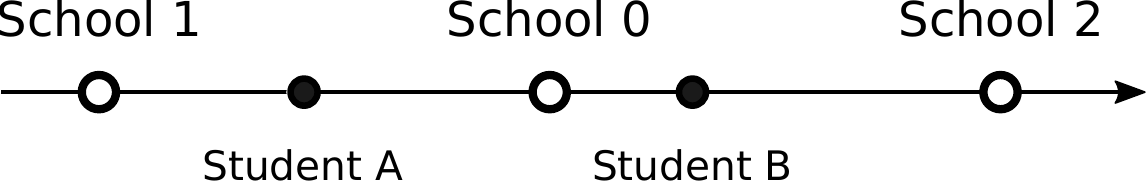}
\par\end{centering}
{\scriptsize{}\medskip{}
}{\scriptsize\par}

\emph{\scriptsize{}Note: }{\scriptsize{}Example of setting where the
excluded treatment could plausibly be argued to be the best or the
next-best alternative for all agents. Here the treatments are schools,
with School 0 being the excluded treatment. Agents are students choosing
which school to attend. The open dots indicate locations of schools,
and the closed dots indicate the location of two example students.
If students care sufficiently about travel distance, School 0 will
be either the best or the next-best alternative for all students.}{\scriptsize\par}
\end{figure}

In which settings can a researcher plausibly argue that the excluded
treatment is always the best or the next-best alternative? A natural
type of setting is when there are three treatments and the excluded
treatment is ``in the middle''. For example, consider estimating
the causal effect of attending ``School 1'' and ``School 2'' compared
to attending ``School 0'' on student outcomes. Here, School 0 is
the excluded treatment. Assume students are free to choose their preferred
school. If School 0 is geographically located in between School 1
and School 2, as depicted in Figure \ref{fig:school}, and students
care sufficiently about travel distance, it is plausible that School
0 is the best or the next-best alternative for all students. For instance,
student A in Figure \ref{fig:school}, who lives between School 1
and School 0, is unlikely to prefer School 2 over School 0. Similarly,
student B in Figure \ref{fig:school} is unlikely to have School 1
as her preferred school. If we believe no students have School 0 as
their least favorite alternative and we have access to random shocks
to the utility of attending Schools 1 and 2 multivariate 2SLS can
be safely applied.\footnote{Another example where the excluded treatment is always the best or
the next-best alternative is when agents are randomly encouraged to
take up one treatment and can not select into treatments they are
not offered. In that case, one might estimate the causal effects of
each treatment in separate 2SLS regressions on the subsamples not
receiving each of the treatments. But when control variables are included
in the regression, estimating a 2SLS model with multiple treatments
can improve precision.}

\citet{kirkeboen2016} and a literature that followed exploit knowledge
of next-best alternatives for identification in just-identified models.
In the Section \ref{subsec:kirkeboen}, we show that the conditions
invoked in this literature are not only sufficient for 2SLS to assign
proper weights but also essentially \emph{necessary}.\footnote{The \citet{kirkeboen2016} assumptions can be relaxed by allowing
for always-takers.} Thus, to apply 2SLS in just-identified models with arbitrary heterogeneous
effects, researchers have to either \emph{directly observe} next-best
alternatives or make assumptions about next-best alternatives based
on institutional and theoretical arguments.

\paragraph*{Ordered Treatment Effects.}

What are the implicit assumptions on choice behavior when $\beta=\left(Y\left(1\right)-Y\left(0\right),Y\left(2\right)-Y\left(1\right)\right)^{T}$?
In that case, instrument $k$ can influence only treatment indicator
$D_{k}=\mathbf{1}\left[T\geq k\right]$. Thus, instrument $k$ can
not impact relative utilities other than $I_{si}-I_{li}$ for $s\geq k$
and $l<k$ in a way that changes choice behavior. Thus, without loss
of generality, we can assume that instrument $k$ increases the utility
of treatments $t\geq k$ by the same amount while keeping the utility
of treatments $t<k$ constant. But this assumption is not sufficient
to prevent instrument $k$ from influencing treatment indicators other
than $D_{k}$. In addition, we need that preferences are \emph{single-peaked}:
\begin{defn}[Single-Peaked Preferences]
 \label{def:single-peaked}Preferences are \emph{single-peaked} if
for all $s\in\mathcal{S}$, $z\in\mathcal{Z}$, and $k,l,r\in\left\{ 1,\dots,n\right\} $
\[
k>l>r\text{ and }I_{ks}\left(v\right)>I_{ls}\left(v\right)\Rightarrow I_{ls}\left(v\right)\geq I_{rs}\left(v\right),
\]
\[
k<l<r\text{ and }I_{ls}\left(v\right)<I_{rs}\left(v\right)\Rightarrow I_{ks}\left(v\right)\geq I_{ls}\left(v\right).
\]
\end{defn}
\begin{figure}
\caption{\label{fig:single-peak}Single-Peaked Preferences and Ordered Treatment
Effects.}

\subfloat[\label{fig:not-single-peaked}Preferences not Single-Peaked.]{
\centering{}\includegraphics[width=0.45\textwidth]{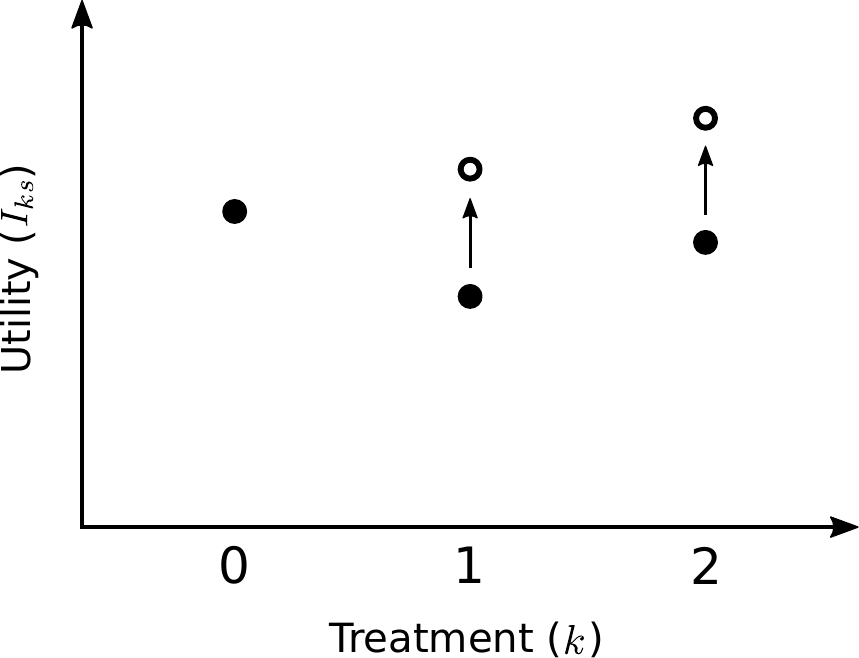}}\subfloat[\label{fig:single-peaked}Single-Peaked Preferences.]{
\centering{}\includegraphics[width=0.45\textwidth]{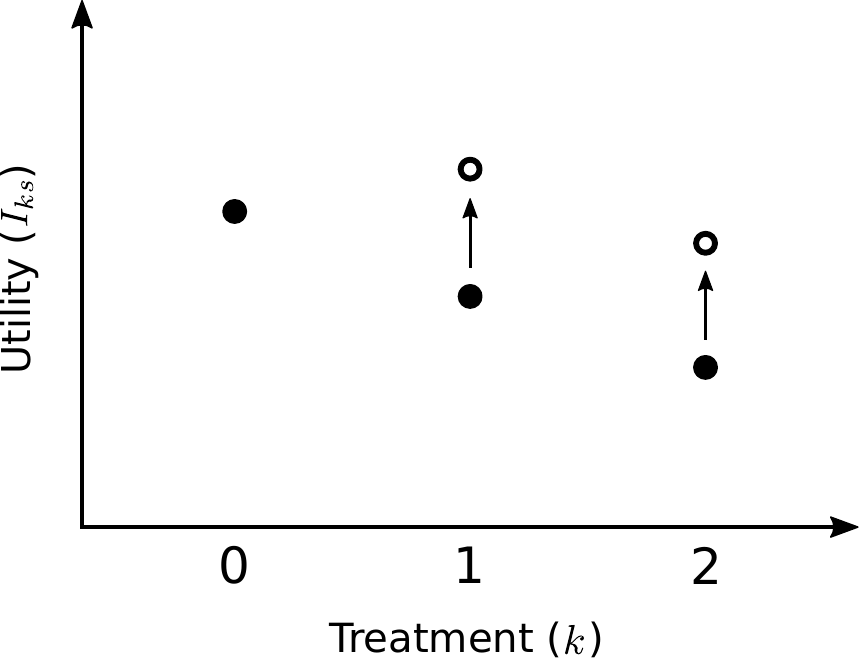}}

\emph{\scriptsize{}Note: }{\scriptsize{}Indirect utilities of response
type $s$ over treatments 0\textendash 2 when instrument 1 is turned
off (solid dots) and on (open dots). Figure (a) shows how instrument
1 might affect $D_{2}=\mathbf{1}\left[T\geq2\right]$ when preferences
are not single-peaked.}{\scriptsize\par}
\end{figure}

To see this, consider Figure \ref{fig:not-single-peaked}. The solid
dots indicate the indirect utilities of agent $i$ over treatments
0\textendash 2 when instrument 1 is turned off ($V=0$). In this case,
the agent's preferences has two peaks: one at $k=0$ and another at
$k=2$. If instrument $1$ increases the utilities of treatments $k\geq1$
by the same amount, the agent might change the selected treatment
from $T=0$ to $T=2$ as indicated by the open dots. This change impacts
both $D_{1}$ and $D_{2}$. In Figure \ref{fig:single-peaked}, however,
where preferences are single-peaked, a homogeneous increase in the
utility of treatments $k\geq1$ can never impact any other treatment
indicator than $D_{1}$. We thus have the following result.
\begin{prop}
\label{prop:binary_roy2}2SLS with $D_{k}=\mathbf{1}\left[T\geq k\right]$
assigns proper weights if and only if there exists $u_{ks}\in\mathbb{R}$
and $\mu_{ks}\geq0$ such that $s\left(v\right)=\arg\max_{k}I_{ks}\left(v\right)$
with
\[
I_{0s}=0
\]
\[
I_{ks}\left(v\right)=u_{ks}+\mu_{vs}\mathbf{1}\left[k\geq v\right]
\]
for all $s\in\mathcal{S},$ $v\in\left\{ 0,1,2\right\} $, and $k\in\left\{ 1,2\right\} $
and the preferences are single-peaked.
\end{prop}

\subsection{A Threshold-Crossing Model\label{subsec:threshold-crossing}}

In several settings, treatments have a clear ordering, and assignment
to treatment could be described by a latent index crossing multiple
thresholds. For instance, treatments could be grades and the latent
index the quality of the student's work. When estimating returns to
education, treatments might be different schools, thresholds the admission
criteria, and the latent index the quality of the applicant.\footnote{The model below applies in this setting if applicants agree on the
ranking of schools and schools agree on the ranking of candidates.} In the context of criminal justice, treatments might be conviction
and incarceration as in \citet{Humphries2022} or years of prison
as in \citet{rose2021does} and the latent index the severity of the
crime committed. If the researcher has access to random variation
in these thresholds, for instance through random assignment to graders
or judges who agree on ranking but use different cutoffs, then the
identification of the causal effect of each consecutive treatment
by 2SLS could be possible.

We here present a simple and easily testable condition under which
2SLS assigns proper weights in such a threshold-crossing model. Our
model is a simplified version of the model considered by \citet{Heckman2007EconometricII},
Section 7.2.\footnote{\citet{Heckman2007EconometricII} focus on identifying marginal treatment
effects and discuss what 2SLS with a multivalued treatment identifies
in this model, but do not consider 2SLS with multiple treatments.
\citet{lee2018identifying} consider a similar model with two thresholds
where agents are allowed to differ in two latent dimensions.} We assume agents differ by an unobserved latent index $U$ and that
treatment depends on whether $U$ crosses certain thresholds.\footnote{This specification is routinely applied to study ordered choices \citep{greene2010modeling}.}
In particular, assume there are thresholds $g_{1}\left(Z\right)<g_{2}\left(Z\right)$
such that
\[
T=\begin{cases}
0 & \text{if }U<g_{1}\left(Z\right)\\
1 & \text{if }g_{1}\left(Z\right)\leq U<g_{2}\left(Z\right)\\
2 & \text{if }U\geq g_{2}\left(Z\right)
\end{cases}
\]
We are interested in the ordered treatment effects $\beta=\left(Y\left(1\right)-Y\left(0\right),Y\left(2\right)-Y\left(1\right)\right)^{T}$,
using as treatment indicators $D_{k}=\mathbf{1}\left[U\geq g_{k}\left(Z\right)\right]$
for $k\in\left\{ 1,2\right\} $. Assume the first stage is correctly
specified:
\begin{assumption}
\label{assu:first-stage-correct}(First Stage Correctly Specified.)
Assume that $\E\left[D_{k}\mid Z\right]$ is a linear function of
$Z$ for all $k$.
\end{assumption}
This assumption is always true, for instance, when $Z$ is a set of
mutually exclusive binary instruments. We maintain Assumptions \ref{assu:iv}
and \ref{assu:rank}. We then have:\footnote{A similar result is found in a contemporaneous work by \citet{Humphries2022}
in the context of the random judge IV design. They suggest that the
required linearity condition between $P_{1}$ and $P_{2}$ can be
relaxed by running a 2SLS specification with $D_{1}$ as a single
treatment indicator and $P_{1}$ as the instrument while flexibly
controlling for $P_{2}$ (and vice versa). They further show that
if treatment assignment depends on several unobserved latent indices\textemdash instead
of just one\textemdash 2SLS does not, in general, assign proper weights.}
\begin{prop}
\label{prop:threshold-crossing}2SLS assigns proper weights in the
above model if $E\left[P_{k}\mid P_{l}\right]$ is linear in $P_{l}$
for all $k,l\in\left\{ 1,2\right\} $.
\end{prop}
Thus, when treatment can be described by a single index crossing multiple
thresholds, and we have access to random shocks to these thresholds,
multivariate 2SLS estimands of the effect of crossing each threshold
will be a positively weighted sum of individual treatment effects.
The required linearity condition between predicted treatments can
easily be tested empirically. In Proposition \ref{prop:linear-predicted},
we show that 2SLS \emph{does not} assign proper weights if $E\left[P_{k}\mid P_{l}\right]$
is non-linear for one $k$ and $l$ and there is a positive density
of agents at all values of $U\in\left[0,1\right]$. A linear relationship
between predicted treatments is thus close to \emph{necessary }for
2SLS to assign proper weights in the above model.

While an \emph{exact }linear relationship between $P_{1}$ and $P_{2}$
is unrealistic, small deviations from linearity are unlikely to lead
to a large 2SLS bias: Deviations from linearity will lead some agents
to be pushed out of treatment $2$ by $P_{1}$ and others to be pushed
into treatment $2$ by $P_{1}$. If the deviations from linearity
are ``local''\textemdash small irregularities in an otherwise linear
relationship\textemdash the agents pushed into treatment $2$ will
have similar $U$s as those pushed out of treatment $2$. If agents
with similar $U$s tend to have similar treatment effects, the bias
will be small. But large global deviations from linearity have the
potential to induce significant 2SLS bias in the presence of heterogeneous
effects. If such non-linearities are detected, one solution is to
follow the suggestion of \citet{Humphries2022}: Run 2SLS with one
endogenous variable $D_{1}$, $P_{1}$ as instrument, and control
non-linearly for $P_{2}$. In Section \ref{sec:application}, we discuss
what linearity between $P_{1}$ and $P_{2}$ means in a real application
and show how it can be tested using data from \citet{BhullerEtAl2020}.

\section{If the Conditions Fail\label{sec:if_fail}}

Our identification results show that if either Assumption \ref{assumption:non_neg}
or \ref{assumption:zero} is violated, then 2SLS does not assign proper
weights. If this turns out to be the case, the researcher has two
choices: Either impose assumptions on treatment effect heterogeneity
or select an alternative estimator.

\subsection{Assumptions on Treatment Effect Heterogeneity}

As we show in Section \ref{subsec:WC}, 2SLS assigning proper weights
is equivalent to 2SLS being \emph{weakly causal }\citep{blandhol2022tsls}\textemdash giving
estimates with the correct sign\textemdash under arbitrary heterogeneous
effects. But correctly signed estimates could also be ensured by imposing
assumptions on treatment effect heterogeneity. For instance, if treatment
effects do not systematically differ across response types or if the
amount of ``selection on gains'' and the violations of Assumptions
\ref{assumption:non_neg} and \ref{assumption:zero} are both moderate,
2SLS could still be weakly causal. See Section \ref{subsec:het} for
a formal analysis.

\subsection{Alternative Estimators\label{subsec:alternative}}

The econometrics literature has proposed several ways to identify
multiple treatment effects when 2SLS fails. The most general method
is provided by \citet{HeckmanPinto2018}. Their method allows the
researcher to learn which treatment effects are identified\textemdash and
\emph{how} they are identified\textemdash for any given restriction
on response types. The identification results apply to all settings
with discrete-valued instruments\textemdash also in cases where 2SLS
is not valid. For instance, if we allow for the presence of the response
type $\left(s\left(0\right),s\left(1\right),s\left(2\right)\right)=\left(2,1,2\right)$
in addition to the response types in Table \ref{tab:response_types}
in the Section \ref{sec:binary} model, 2SLS no longer assigns proper
weights, but the method of \citet{HeckmanPinto2018} still recovers
causal effects. \citet{HeckmanPinto2018} and \citet{pinto2022beyond}
discuss how to use revealed preference analysis to restrict the possible
response types.\footnote{For instance, \citet{pinto2022beyond} combines revealed preference
analysis with a functional form assumption to identify causal effects
in an RCT with multiple treatments and non-compliance.} In a related approach, \citet{lee2020filtered} show how assuming
that certain instrument values \emph{target }certain treatments can
lead to partial identification of treatment effects.

Another important strand of methods to identify multiple treatment
effects relies on continuous instruments and the marginal treatment
effects framework brought forward by \citet{heckman1999local,heckman2005structural}.
In the case of ordered treatment effects, \citet{Heckman2007EconometricII}
show how separate treatment effects can be identified if treatment
is determined by a single latent index crossing multiple thresholds
and the researcher has access to shocks to each threshold. Using this
method, separate treatment effects can be recovered in the Section
\ref{subsec:threshold-crossing} model also when predicted treatments
are not linearly related.\footnote{Intuitively, the causal effect of treatment $1$ can be obtained by
varying $P_{1}$ while keeping $P_{2}$ fixed. The 2SLS requires $E\left[P_{1}\mid P_{2}\right]$
to be linear since instead of keeping $P_{2}$ fixed it controls \emph{linearly
}for $P_{2}$.} Another advantage of the methods based on marginal treatment effects
is that they allow for the calculation of treatment effects that are
more policy-relevant than the weighted average produced by 2SLS. In
the context of unordered treatments, \citet{Heckman2007EconometricII}
and \citet{HeckmanEtal2008_Annales} show that analogous assumptions
can recover the causal effect of a given treatment versus the next-best
alternative.\footnote{A 2SLS version of this approach would be to run 2SLS with a single
binarized treatment indicator. The recovered treatment effect\textemdash the
effect of receiving a given treatment compared to a mix of alternative
treatments\textemdash might sometimes be a parameter of policy interest.
\citet{HeckmanPinto2018} extend this result to discrete instruments.
In particular, they show that the causal effect of a given treatment
versus the next-best alternative is identified under \emph{unordered
monotonicity}.} In their framework, causal effects between two specified treatments
can be identified by focusing on instrument realizations for which
the probability of taking up all other treatments is zero\textemdash if
such instrument realizations exist.\footnote{For instance, in the application considered in Section \ref{sec:application},
one could identify the causal effect of incarceration versus conviction
by studying cases assigned to judges that never acquit\textemdash if
such judges exist. The 2SLS equivalent of this approach is to run
2SLS on the subsample of judges that never acquit.} \citet{lee2018identifying} show how knowledge of the exact threshold
rules can be used to identify the effect of one treatment versus another
treatment without relying on such ``identification at infinity''
arguments.\footnote{\citet{kamat2023identification} show how treatment effects can be
partially identified in a similar model with discrete instruments.} \citet{Mountjoy2019} provides a method using continuous instruments
where knowledge of the exact threshold rules is not required either,
and identification is achieved by two relatively weak assumptions:
\emph{Partial unordered monotonicity} and \emph{comparative compliers}.\footnote{\citet{Mountjoy2019} considers a case with three treatments and two
continuous instruments. Partial unordered monotonicity requires that
instrument $k$ weakly increases up-take of treatment $k$ and weakly
reduces up-take of all other treatments for all agents. Partial unordered
monotonicity and our Assumptions \ref{assumption:non_neg} and \ref{assumption:zero}
do not nest each other. On the one hand, partial unordered monotonicity
allows for heterogeneity in how instrument $k$ induces agents out
of treatments $l\neq k$ in ways that would violate our no-cross-effect
condition. On the other hand, average conditional monotonicity allows
monotonicity to be violated for some instrument pairs in ways that
would violate partial unordered monotonicity. A main advantage of
partial unordered monotonicity, however, is that it has a clear economic
interpretation, which Assumptions \ref{assumption:non_neg} and \ref{assumption:zero}
lack. Comparative compliers requires that those induced from treatment
$l$ to treatment $k$ by a marginal increase in instrument $k$ are
similar to those induced from treatment $k$ to treatment $l$ by
a marginal increase in instrument $l$. This condition is satisfied
in a broad class of index models.}

A final alternative to 2SLS is to explicitly model the selection into
treatment using the method of \citet{heckman1979sample} (see, e.g.,
\citealt{kline2016evaluating}). This approach, however, requires
the researcher to make distributional assumptions about unobservables.

\section{Application: The Effects of Incarceration and Conviction\label{sec:application}}

In this section, we show how our results can be applied in practice.
Following \citet{Humphries2022} and \citet{Kamat2023}, we consider
the identification of the effects of conviction and incarceration
on defendant recidivism in a random judge IV design.\footnote{\citet{Humphries2022} and \citet{Kamat2023} also consider alternative
approaches beyond 2SLS.} We first discuss the conditions required for 2SLS to assign proper
weights in this setting and then test these conditions using data
from \citet{BhullerEtAl2020}.

\subsection{When Does 2SLS Assign Proper Weights?\label{subsec:application_illustration}}

Assume the possible treatments are 
\[
T\in\left\{ 0,1,2\right\} =\left\{ \text{acquittal},\text{non-incarceration conviction},\text{incarceration}\right\} 
\]
 We consider the treatment indicators $D_{1}=\mathbf{1}\left[T\geq1\right]$
(conviction) and $D_{2}=\mathbf{1}\left[T=2\right]$ (incarceration).
We thus seek to separately identify the effect of conviction versus
acquittal and the effect of incarceration versus conviction. As instruments
$Z$, we use randomly assigned judges.\footnote{Formally, $Z$ is a vector of binary judge indicators where $Z_{k}=1$
if the case is assigned judge $k$. In practice, following \citet{BhullerEtAl2020},
we use as our instruments the leave-one-out incarceration and conviction
rates of the assigned judge, calculated across all randomly assigned
cases to each judge, excluding the focal case.} The predicted treatments $P_{1}$ and $P_{2}$ then equal the rate
at which the randomly assigned judge convicts and incarcerates defendants,
respectively.

\begin{figure}[h]
\caption{Examples of Response Types in the Random Judge IV Design.\label{fig:example-response-types}}

\subfloat[Response Type Satisfying Assumptions \ref{assumption:non_neg}\textendash \ref{assumption:zero}\label{fig:satisfying}]{
\centering{}\includegraphics[width=0.45\textwidth]{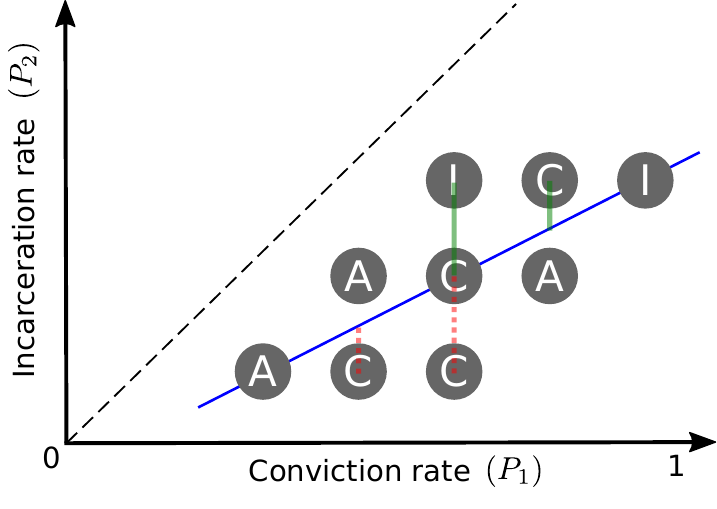}}\qquad{}\subfloat[Response Type Satisfying Assumptions \ref{assumption:non_neg}\textendash \ref{assumption:zero}\label{fig:satisfying2}]{
\centering{}\includegraphics[width=0.45\textwidth]{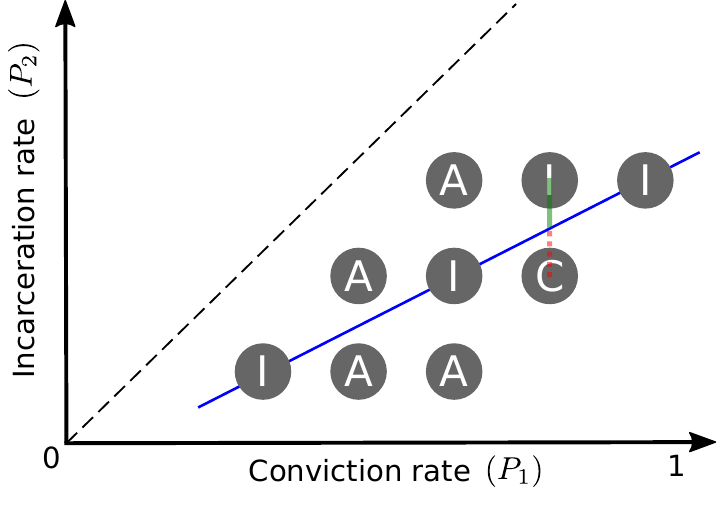}}

\subfloat[Response Type Violating Assumption \ref{assumption:zero}\label{fig:violating-nce}]{
\centering{}\includegraphics[width=0.45\textwidth]{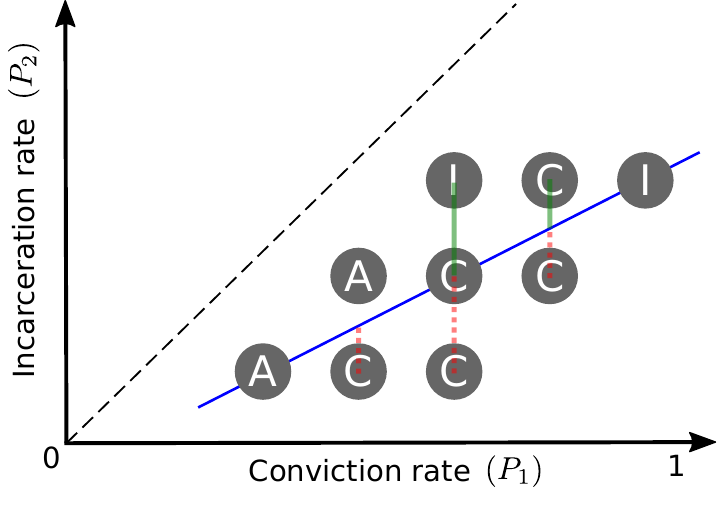}}\qquad{}\subfloat[Response Type Violating Assumption \ref{assumption:non_neg}\label{fig:violating-acm}]{
\centering{}\includegraphics[width=0.45\textwidth]{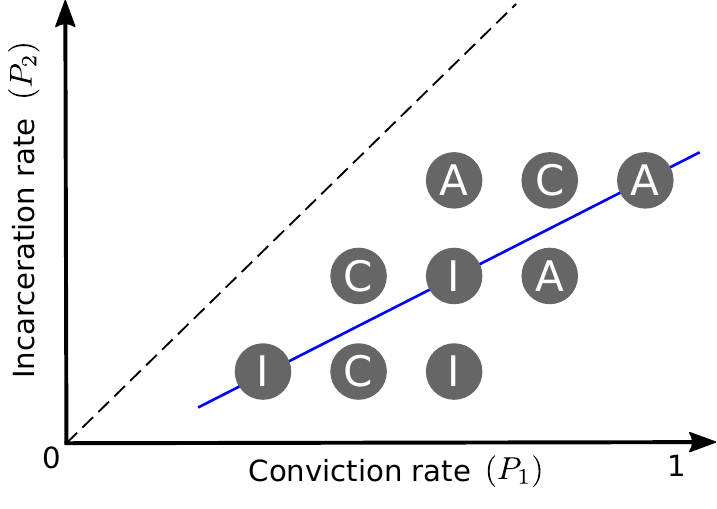}}

\emph{\footnotesize{}Note}{\footnotesize{}: Each dot represents a
judge. The letters indicate how the judge will decide if assigned
the case: aquit (A), convict without incarceration (C), or incarcerate
(I). The $x$-axis ($y$-axis) shows the judge's conviction (incarceration)
rate $P_{1}$ ($P_{2}$). The blue line is a linear regression of
$P_{2}$ on $P_{1}$. The green and the dotted red lines show the
distance to the blue line for the judges that convicts. ``No cross
effects'' requires that the sum of the green lines equals the sum
of the dotted red lines.}{\footnotesize\par}
\end{figure}

When does 2SLS assign proper weights in this setting? Applying the
Section \ref{subsec:threshold-crossing} model, we get that 2SLS assigns
proper weights if judges agree on how to rank cases but use different
cutoffs for conviction and incarceration and there is a linear relationship
between the judges' incarceration and conviction rates.\footnote{In the notation of Section \ref{subsec:threshold-crossing}, $U$
would represent the ``strength'' of the case, $g_{1}\left(Z\right)$
the conviction cutoff for the randomly chosen judge, and $g_{2}\left(Z\right)$
the incarceration cutoff.} In the Figure \ref{fig:example-response-types} example, the relationship
between judges' incarceration and conviction rates is linear.\footnote{In Figure \ref{fig:example-response-types}, the relationship between
$P_{1}$ and $P_{2}$ is \emph{exactly }linear. As noted in Section
\ref{subsec:threshold-crossing}, small local deviations from linearity
are unlikely to lead to a large 2SLS bias. But large global non-linearities
could cause significant 2SLS bias. For instance, it would be problematic
if judges with medium conviction rates tend to have a high incarceration
rate while judges with low or high conviction rates tend to have a
low incarceration rate. We do not find such non-linearities in Figure
\ref{fig:application_linearity_predictedT} for \citet{BhullerEtAl2020}.}

It is, however, not necessary to assume that judges agree on how to
rank cases for 2SLS to assign proper weights.\footnote{The single-index model has been criticized as an excessively restrictive
model of judge behavior \citep{Humphries2022,Kamat2023}. In particular,
\citet{Humphries2022} rejects the single-index model in their setting
by showing that the observable characteristics of those with $T=1$
change when holding $P_{1}$ constant and varying $P_{2}$. Under
the single-index model, this should not happen.} An example of a response type \emph{not} allowed by the single-index
model but satisfying Assumptions \ref{assumption:non_neg}\textendash \ref{assumption:zero}
is given in Figure \ref{fig:satisfying}. Here, there are nine judges
with their conviction rate ($P_{1}$) on the $x$-axis and their incarceration
rate ($P_{2}$) on the $y$-axis. After controlling linearly for the
conviction rate (the blue line), the judges convicting the defendant
do not, on average, have a higher nor lower incarceration rate than
the judges acquitting the defendant.\footnote{Graphically, the sum of the dotted red lines equals the sum of the
green lines.} Thus, by Corollary \ref{cor:zero}, the 2SLS estimand of the effect
of \emph{incarceration} places zero weight on this defendant's effect
of \emph{conviction}.\footnote{In Figure \ref{fig:satisfying}, the 2SLS estimand also places a positive
weight on the defendant's effect of incarceration: After controlling
linearly for the conviction rate, the judges incarcerating the defendant
tend to have an above-average incarceration rate. Thus, Assumption
\ref{assumption:non_neg} (``average conditional monotonicity'')
is satisfied. The 2SLS estimand of the effect of conviction versus
acquittal also assigns proper weights.} In fact, Assumptions \ref{assumption:non_neg}\textendash \ref{assumption:zero}
admit a wide range of response types beyond the Figure \ref{fig:satisfying}
response type. For instance, the Figure \ref{fig:satisfying2} response
type also satisfies Assumptions \ref{assumption:non_neg}\textendash \ref{assumption:zero}.

However, Assumption \ref{assumption:zero} (``no cross effects'')
is a knife-edge condition that can easily be violated by small deviations
from the allowed response types. For instance, the response type in
Figure \ref{fig:violating-nce} violates no cross effects.\footnote{Even after controlling linearly for the conviction rate, the judges
incarcerating the defendant have a lower-than-average conviction rate.
Graphically, the sum of the dotted red lines is higher than the sum
of the green lines. The 2SLS estimand of the effect of incarceration
will thus place a non-zero (negative) weight on the causal effect
of convicting this defendant.} One way to see why ``no cross effects'' must be violated for either
the Figure \ref{fig:satisfying} response type or the Figure \ref{fig:violating-nce}
response type is that the two response types are heterogeneous in
their relative responses to the instruments. Instrument $2$\textemdash the
judge's incarceration rate\textemdash has the same tendency to push
the two response types into conviction.\footnote{The covariance between $P_{2}$ and $s_{1}\left(Z\right)$ is exactly
the same.} But instrument $1$\textemdash the judge's conviction rate\textemdash has
a higher tendency to push a Figure \ref{fig:satisfying2} agent than
a Figure \ref{fig:satisfying} agent into conviction. Thus, the relative
responses to the two instruments are not homogeneous. By Proposition
\ref{cor:zero_alt}, ``no cross effects'' is violated for at least
one of the response types. Note, however, that if response types do
not deviate much from the allowed response types and heterogeneous
effects are moderate, the bias would still be small; see Section \ref{subsec:het}.

Finally, we note that ``average conditional monotonicity'' (Assumption
\ref{assumption:non_neg}) is satisfied in all of Figures \ref{fig:satisfying}\textendash \ref{fig:violating-nce}.
For this condition to be violated, it must be that the relationship
between the judge's incarceration rate and whether the agent is incarcerated
is \emph{flipped} for certain cases. The Figure \ref{fig:violating-acm}
response type violates average conditional monotonicity: After controlling
for the conviction rate, the judges incarcerating the agent have,
on average, a \emph{lower} incarceration rate than the other judges.
While unlikely to be widespread, such violations might happen. For
example, assume that\textemdash conditional on their conviction rate\textemdash judges
with higher incarceration rates are more ``conservative''. If conservative
judges tend to be \emph{less} likely to incarcerate defendants for
gun law violations, average conditional monotonicity might be violated
in gun law cases.

\subsection{Applying Our Tests\label{subsec:application_testing}}

To illustrate the application of the Section \ref{subsec:Testing}
tests, we use data from \citet{BhullerEtAl2020}. That paper used
a sample of 31,428 criminal cases that were quasi-randomly assigned
to judges in Norwegian courts from 2005\textendash 2009 and provided
an IV estimate of the effect of incarceration on the five-year recidivism
rate for defendants. Using exactly the same sample, we consider the
causal effects of conviction and incarceration on the same recidivism
outcome. Our 2SLS estimates are provided in Table \ref{tab:application_2sls_multipleT},
where Column (1) replicates the baseline single treatment 2SLS estimate
in \citet{BhullerEtAl2020} (for comparison, see Table 4, Column (3),
on p. 1297 in that paper), while Column (2) provides our multivariate
2SLS estimates of conviction and incarceration treatments.\footnote{While not their main focus, \citet{BhullerEtAl2020} also considered
multivariate 2SLS specifications with alternative non-mutually exclusive
treatments (see Appendix Tables B10 and B12 in that paper).}

We now test our conditions using this data. Let $Y$ be an indicator
for the defendant's five-year recidivism outcome, as used in the estimation
above. The Proposition \ref{prop:kitagawa} test, which we generalize
to ordered treatment indicators in Section \ref{subsec:kitagawa-generalization},
requires that $\vartheta_{11},\psi_{11},\vartheta_{22},\psi_{22}\geq0$,
$\vartheta_{22}+\vartheta_{12}=\psi_{22}+\psi_{12}=0$, and $\vartheta_{21}=\psi_{21}=0$
in the regressions
\begin{eqnarray*}
Y\left(D_{1}-D_{2}\right) & = & \varphi_{1X}+\vartheta_{11}P_{1}+\vartheta_{12}P_{2}+\upsilon_{1}\\
YD_{2} & = & \varphi_{2X}+\vartheta_{21}P_{1}+\vartheta_{22}P_{2}+\upsilon_{2}\\
\left(1-Y\right)\left(D_{1}-D_{2}\right) & = & \varphi_{3X}+\psi_{11}P_{1}+\psi_{12}P_{2}+\upsilon_{3}\\
\left(1-Y\right)D_{2} & = & \varphi_{4X}+\psi_{21}P_{1}+\psi_{22}P_{2}+\upsilon_{4}
\end{eqnarray*}
where $\left\{ \varphi_{1X},\dots,\varphi_{4X}\right\} $ are court-by-year
fixed effects.\footnote{In Proposition \ref{prop:kitagawax}, we show that the Proposition
\ref{prop:kitagawa} test extends to the case with control variables.} The results from these regressions are presented in Table \ref{tab:application_test_prop3_prop4_orderedT},
Panel A.\footnote{Note that there is a mechanical relationship between the coefficients
in Columns (1)\textendash (2) and Columns (3)\textendash (4) in Panel
A. We get $\psi_{11}=1-\vartheta_{11}$, $\psi_{12}=-1-\vartheta_{12}$,
$\psi_{21}=-\vartheta_{21}$, and $\psi_{22}=1-\vartheta_{22}$. The
additional tests in Columns (3)\textendash (4) are, however, not redundant:
The standard errors are different, and the conditions $\psi_{11}=1-\vartheta_{11}\geq0$
and $\psi_{22}=1-\vartheta_{22}\geq0$ are not tested in Columns (1)\textendash (2).} While we are unable to statistically reject the required conditions
at the 5\% level, the $\vartheta_{22}+\vartheta_{12}=0$ and $\psi_{22}+\psi_{12}=0$
conditions are close to rejection ($p$-values at 0.12 and 0.08).
Also, while we are unable to reject $\psi_{11}\geq0$, this estimate
has a negative sign. These results suggest that in a larger sample
we could have concluded that either Assumption \ref{assu:iv}, Assumption
\ref{assumption:non_neg}, or Assumption \ref{assumption:zero} is
violated.
\begin{center}
\begin{table}[H]
\begin{centering}
\caption{Empirical Application: 2SLS with Multiple Treatments.\label{tab:application_2sls_multipleT}}
\par\end{centering}
\begin{singlespace}
\noindent \begin{centering}
{\scriptsize{}}%
\begin{tabular}{l>{\centering}m{2.8cm}>{\centering}m{2.8cm}}
\multicolumn{3}{l}{\vspace{-0.75cm}
}\tabularnewline
\hline 
\hline 
\emph{\scriptsize{}Outcome:} & \multicolumn{2}{c}{\textbf{\scriptsize{}Recidivism}{\scriptsize{} $\left(Y\right)$}}\tabularnewline
\cline{2-3} \cline{3-3} 
\emph{\scriptsize{}Specification:} & {\scriptsize{}Single Treatment} & {\scriptsize{}Multiple Treatments}\tabularnewline
 & {\scriptsize{}(1)} & {\scriptsize{}(2)}\tabularnewline
\cline{2-3} \cline{3-3} 
\emph{\scriptsize{}Treatments:} &  & \tabularnewline
\textbf{\scriptsize{}Conviction}{\scriptsize{} $\left(D_{1}\right)$} & {\scriptsize{}\textendash{}} & {\scriptsize{}-0.222}\tabularnewline
 &  & {\scriptsize{}(0.819)}\tabularnewline
\textbf{\scriptsize{}Incarceration}{\scriptsize{} $\left(D_{2}\right)$} & {\scriptsize{}-0.293{*}{*}{*}} & {\scriptsize{}-0.281{*}{*}}\tabularnewline
 & {\scriptsize{}(0.106)} & {\scriptsize{}(0.126)}\tabularnewline
\hline 
{\scriptsize{}Dependent mean} & {\scriptsize{}0.70} & {\scriptsize{}0.70}\tabularnewline
{\scriptsize{}Number of cases} & {\scriptsize{}31,428} & {\scriptsize{}31,428}\tabularnewline
\hline 
\end{tabular}{\scriptsize\par}
\par\end{centering}
\end{singlespace}
\begin{singlespace}
\smallskip{}

\end{singlespace}

\textit{\scriptsize{}Note:}\textit{\emph{\scriptsize{} }}{\scriptsize{}This
table reports 2SLS estimates from model specification with a single
binary treatment indicator for incarceration (Column 1) and multiple
treatment indicators for conviction and incarceration (Column 2),
respectively, and randomly assigned judges as instruments using data
from \citet{BhullerEtAl2020}. The estimate reported in Column (1)
is identical to the IV estimate reported in the final row of Table
4, Column (3), on p. 1297 in \citet{BhullerEtAl2020}. The sample
consists of 31,428 non-confession criminal cases that were quasi-randomly
assigned to judges in Norwegian trial courts between 2005 and 2009.
All estimations control for court by court entry year fixed effects
and baseline controls listed in \citet{BhullerEtAl2020}, Table 1.
}\textit{\emph{\scriptsize{}{*}p<0.1, {*}{*}p<0.05, {*}{*}{*}p<0.01.}}{\scriptsize\par}
\end{table}
\vspace{-0.75cm}
\begin{sidewaystable}
\begin{centering}
\caption{Empirical Application: Testing Propositions \ref{prop:kitagawa} and
\ref{prop:test}.\label{tab:application_test_prop3_prop4_orderedT}}
\par\end{centering}
\begin{singlespace}
\noindent \begin{centering}
\renewcommand{\arraystretch}{.8}{\scriptsize{}}%
\begin{tabular}{l>{\centering}p{2.5cm}>{\centering}p{2.5cm}>{\centering}p{0.1cm}>{\centering}p{3.8cm}>{\centering}p{2.4cm}>{\centering}p{0.1cm}>{\centering}b{2.5cm}}
\multicolumn{8}{l}{\vspace{-0.75cm}
}\tabularnewline
\hline 
\hline 
\multicolumn{8}{l}{\textbf{\scriptsize{}A. Testing Proposition \ref{prop:kitagawa} for
Ordered Treatments: $\left\{ D_{1}=1\left[t\geq1\right],D_{2}=1\left[t=2\right]\right\} $}}\tabularnewline
\cline{2-8} \cline{3-8} \cline{4-8} \cline{5-8} \cline{6-8} \cline{7-8} \cline{8-8} 
 & \multicolumn{2}{c}{\textbf{\scriptsize{}Recidivism}{\scriptsize{} $\left(Y\right)$}} &  &  & \multicolumn{3}{c}{\textbf{\scriptsize{}No Recidivism}{\scriptsize{} $\left(1-Y\right)$}}\tabularnewline
\cline{2-3} \cline{3-3} \cline{6-8} \cline{7-8} \cline{8-8} 
\emph{\scriptsize{}Outcomes}{\scriptsize{}:} & {\scriptsize{}$Y\left(D_{1}-D_{2}\right)$} & {\scriptsize{}$YD_{2}$} &  & \noindent \raggedright{}\emph{\scriptsize{}Outcomes}{\scriptsize{}:} & \multicolumn{2}{c}{{\scriptsize{}$\left(1-Y\right)\left(D_{1}-D_{2}\right)$}} & {\scriptsize{}$\left(1-Y\right)D_{2}$}\tabularnewline
\emph{\scriptsize{}Regressors}{\scriptsize{}:} & \textbf{\scriptsize{}(1)} & \textbf{\scriptsize{}(2)} &  & \noindent \raggedright{}\emph{\scriptsize{}Regressors}{\scriptsize{}:} & \multicolumn{2}{c}{\textbf{\scriptsize{}(3)}} & \textbf{\scriptsize{}(4)}\tabularnewline
\cline{1-3} \cline{2-3} \cline{3-3} \cline{5-8} \cline{6-8} \cline{7-8} \cline{8-8} 
{\scriptsize{}Predicted Conviction $(P_{1})$} & {\scriptsize{}1.173 (1.005)} & {\scriptsize{}-0.689 (1.267)} &  & \noindent \raggedright{}{\scriptsize{}Predicted Conviction $(P_{1})$} & \multicolumn{2}{c}{{\scriptsize{}-0.173 (0.844)}} & {\scriptsize{}0.689 (0.659)}\tabularnewline
{\scriptsize{}Predicted Incarceration $(P_{2})$} & {\scriptsize{}-0.780{*}{*}{*} (0.169)} & {\scriptsize{}0.482{*}{*} (0.199)} &  & \noindent \raggedright{}{\scriptsize{}Predicted Incarceration $(P_{2})$} & \multicolumn{2}{c}{{\scriptsize{}-0.220{*} (0.124)}} & {\scriptsize{}0.518{*}{*}{*} (0.129)}\tabularnewline
\cline{1-3} \cline{2-3} \cline{3-3} \cline{5-8} \cline{6-8} \cline{7-8} \cline{8-8} 
{\scriptsize{}$H0:\vartheta_{11}\geq0,\vartheta_{21}=0$} & {\scriptsize{}{[}0.878{]}} & {\scriptsize{}{[}0.587{]}} &  & \noindent \raggedright{}{\scriptsize{}$H0:\psi_{11}\geq0,\psi_{21}=0$} & \multicolumn{2}{c}{{\scriptsize{}{[}0.419{]}}} & {\scriptsize{}{[}0.297{]}}\tabularnewline
{\scriptsize{}$H0:\vartheta_{22}\geq0$} & {\scriptsize{}\textendash{}} & {\scriptsize{}{[}0.992{]}} &  & \noindent \raggedright{}{\scriptsize{}$H0:\psi_{22}\geq0$} & \multicolumn{2}{c}{{\scriptsize{}\textendash{}}} & {\scriptsize{}{[}1.000{]}}\tabularnewline
{\scriptsize{}$H0:\vartheta_{22}+\vartheta_{12}=0$} & \multicolumn{2}{c}{{\scriptsize{}{[}0.123{]}}} &  & \noindent \raggedright{}{\scriptsize{}$H0:\psi_{22}+\psi_{12}=0$} & \multicolumn{3}{c}{{\scriptsize{}{[}0.080{]}}}\tabularnewline
{\scriptsize{}Observations:} & \multicolumn{2}{c}{{\scriptsize{}31,428}} &  & \noindent \raggedright{}{\scriptsize{}Observations:} & \multicolumn{3}{c}{{\scriptsize{}31,428}}\tabularnewline
\hline 
\hline 
\multicolumn{8}{l}{\textbf{\scriptsize{}B. Testing Proposition \ref{prop:test} for Ordered
Treatments: $\left\{ D_{1}=1\left[t\geq1\right],D_{2}=1\left[t=2\right]\right\} $}}\tabularnewline
\cline{2-8} \cline{3-8} \cline{4-8} \cline{5-8} \cline{6-8} \cline{7-8} \cline{8-8} 
\emph{\scriptsize{}Outcomes}{\scriptsize{}:} & \multicolumn{2}{c}{\textbf{\scriptsize{}Conviction}{\scriptsize{} $\left(D_{1}\right)$}} &  & \multicolumn{2}{c}{\textbf{\scriptsize{}Incarceration}{\scriptsize{} $\left(D_{2}\right)$}} &  & \tabularnewline
\cline{2-3} \cline{3-3} \cline{5-6} \cline{6-6} 
\emph{\scriptsize{}Regressors}{\scriptsize{}:} & {\scriptsize{}Predicted}{\scriptsize\par}

{\scriptsize{}Conviction $(P_{1})$} & {\scriptsize{}Predicted}{\scriptsize\par}

{\scriptsize{}Incarceration $(P_{2})$} &  & {\scriptsize{}Predicted}{\scriptsize\par}

{\scriptsize{}Conviction $(P_{1})$} & {\scriptsize{}Predicted Incarceration $(P_{2})$} &  & {\scriptsize{}Observations:}\tabularnewline
\emph{\scriptsize{}Sub-samples}{\scriptsize{}:} & \textbf{\scriptsize{}(1)} & \textbf{\scriptsize{}(2)} &  & \textbf{\scriptsize{}(3)} & \textbf{\scriptsize{}(4)} &  & \textbf{\scriptsize{}(5)}\tabularnewline
\hline 
{\scriptsize{}Previously Employed} & {\scriptsize{}1.275{*}{*} (0.523)} & {\scriptsize{}0.110 (0.085)} &  & -{\scriptsize{}0.127 (1.092)} & {\scriptsize{}1.195{*}{*}{*} (0.178)} &  & {\scriptsize{}16,547}\tabularnewline
{\scriptsize{}$H0:\eta_{11}\geq0,\eta_{12}=\eta_{21}=0,\eta_{22}\geq0$} & {\scriptsize{}{[}0.993{]}} & {\scriptsize{}{[}0.197{]}} &  & {\scriptsize{}{[}0.908{]}} & {\scriptsize{}{[}1.000{]}} &  & \tabularnewline
\cline{2-8} \cline{3-8} \cline{4-8} \cline{5-8} \cline{6-8} \cline{7-8} \cline{8-8} 
{\scriptsize{}Previously Non-Employed} & {\scriptsize{}0.641 (0.496)} & {\scriptsize{}-0.118 (0.079)} &  & {\scriptsize{}0.010 (1.162)} & {\scriptsize{}0.788{*}{*}{*} (0.186)} &  & {\scriptsize{}14,881}\tabularnewline
{\scriptsize{}$H0:\eta_{11}\geq0,\eta_{12}=\eta_{21}=0,\eta_{22}\geq0$} & {\scriptsize{}{[}0.902{]}} & {\scriptsize{}{[}0.137{]}} &  & {\scriptsize{}{[}0.920{]}} & {\scriptsize{}{[}1.000{]}} &  & \tabularnewline
\hline 
{\scriptsize{}Previously Incarcerated} & {\scriptsize{}1.239{*}{*}{*} (0.401)} & {\scriptsize{}0.062 (0.065)} &  & {\scriptsize{}-0.571 (1.084)} & {\scriptsize{}1.104{*}{*}{*} (0.175)} &  & {\scriptsize{}16,042}\tabularnewline
{\scriptsize{}$H0:\eta_{11}\geq0,\eta_{12}=\eta_{21}=0,\eta_{22}\geq0$} & {\scriptsize{}{[}0.999{]}} & {\scriptsize{}{[}0.339{]}} &  & {\scriptsize{}{[}0.598{]}} & {\scriptsize{}{[}1.000{]}} &  & \tabularnewline
\cline{2-8} \cline{3-8} \cline{4-8} \cline{5-8} \cline{6-8} \cline{7-8} \cline{8-8} 
{\scriptsize{}Previously Non-Incarcerated} & {\scriptsize{}0.883 (0.607)} & {\scriptsize{}-0.078 (0.098)} &  & {\scriptsize{}1.245 (1.116)} & {\scriptsize{}0.820{*}{*}{*} (0.180)} &  & {\scriptsize{}15,386}\tabularnewline
{\scriptsize{}$H0:\eta_{11}\geq0,\eta_{12}=\eta_{21}=0,\eta_{22}\geq0$} & {\scriptsize{}{[}0.927{]}} & {\scriptsize{}{[}0.424{]}} &  & {\scriptsize{}{[}0.265{]}} & {\scriptsize{}{[}1.000{]}} &  & \tabularnewline
\hline 
{\scriptsize{}Young (Age < 30)} & {\scriptsize{}1.498{*}{*}{*} (0.488)} & {\scriptsize{}-0.055 (0.079)} &  & {\scriptsize{}-1.049 (1.114)} & {\scriptsize{}1.131{*}{*}{*} (0.182)} &  & {\scriptsize{}15,956}\tabularnewline
{\scriptsize{}$H0:\eta_{11}\geq0,\eta_{12}=\eta_{21}=0,\eta_{22}\geq0$} & {\scriptsize{}{[}0.999{]}} & {\scriptsize{}{[}0.489{]}} &  & {\scriptsize{}{[}0.347{]}} & {\scriptsize{}{[}1.000{]}} &  & \tabularnewline
\cline{2-8} \cline{3-8} \cline{4-8} \cline{5-8} \cline{6-8} \cline{7-8} \cline{8-8} 
{\scriptsize{}Old (Age > 30)} & {\scriptsize{}0.483 (0.537)} & {\scriptsize{}0.055 (0.086)} &  & {\scriptsize{}1.088 (1.135)} & {\scriptsize{}0.870{*}{*}{*} (0.181)} &  & {\scriptsize{}15,472}\tabularnewline
{\scriptsize{}$H0:\eta_{11}\geq0,\eta_{12}=\eta_{21}=0,\eta_{22}\geq0$} & {\scriptsize{}{[}0.816{]}} & {\scriptsize{}{[}0.519{]}} &  & {\scriptsize{}{[}0.338{]}} & {\scriptsize{}{[}1.000{]}} &  & \tabularnewline
\hline 
{\scriptsize{}Below High School} & {\scriptsize{}1.238{*}{*}{*} (0.404)} & {\scriptsize{}-0.064 (0.065)} &  & {\scriptsize{}-0.766 (0.945)} & {\scriptsize{}1.025{*}{*}{*} (0.152)} &  & {\scriptsize{}22,651}\tabularnewline
{\scriptsize{}$H0:\eta_{11}\geq0,\eta_{12}=\eta_{21}=0,\eta_{22}\geq0$} & {\scriptsize{}{[}0.999{]}} & {\scriptsize{}{[}0.322{]}} &  & {\scriptsize{}{[}0.417{]}} & {\scriptsize{}{[}1.000{]}} &  & \tabularnewline
\cline{2-8} \cline{3-8} \cline{4-8} \cline{5-8} \cline{6-8} \cline{7-8} \cline{8-8} 
{\scriptsize{}High School or Above} & {\scriptsize{}0.403 (0.763)} & {\scriptsize{}0.169 (0.125)} &  & {\scriptsize{}1.855 (1.476)} & {\scriptsize{}0.949{*}{*}{*} (0.242)} &  & {\scriptsize{}8,777}\tabularnewline
{\scriptsize{}$H0:\eta_{11}\geq0,\eta_{12}=\eta_{21}=0,\eta_{22}\geq0$} & {\scriptsize{}{[}0.701{]}} & {\scriptsize{}{[}0.176{]}} &  & {\scriptsize{}{[}0.209{]}} & {\scriptsize{}{[}1.000{]}} &  & \tabularnewline
\hline 
{\scriptsize{}Had Children} & {\scriptsize{}1.248{*}{*} (0.631)} & {\scriptsize{}0.070 (0.100)} &  & {\scriptsize{}1.672 (1.331)} & {\scriptsize{}0.973{*}{*}{*} (0.211)} &  & {\scriptsize{}11,717}\tabularnewline
{\scriptsize{}$H0:\eta_{11}\geq0,\eta_{12}=\eta_{21}=0,\eta_{22}\geq0$} & {\scriptsize{}{[}0.976{]}} & {\scriptsize{}{[}0.484{]}} &  & {\scriptsize{}{[}0.209{]}} & {\scriptsize{}{[}1.000{]}} &  & \tabularnewline
\cline{2-8} \cline{3-8} \cline{4-8} \cline{5-8} \cline{6-8} \cline{7-8} \cline{8-8} 
{\scriptsize{}Had No Children} & {\scriptsize{}0.875{*}{*} (0.440)} & {\scriptsize{}-0.042 (0.072)} &  & {\scriptsize{}-0.931 (0.993)} & {\scriptsize{}1.013{*}{*}{*} (0.162)} &  & {\scriptsize{}19,711}\tabularnewline
{\scriptsize{}$H0:\eta_{11}\geq0,\eta_{12}=\eta_{21}=0,\eta_{22}\geq0$} & {\scriptsize{}{[}0.977{]}} & {\scriptsize{}{[}0.563{]}} &  & {\scriptsize{}{[}0.349{]}} & {\scriptsize{}{[}1.000{]}} &  & \tabularnewline
\hline 
{\scriptsize{}Low Incarceration Propensity} & {\scriptsize{}0.954{*} (0.548)} & {\scriptsize{}-0.040 (0.090)} &  & {\scriptsize{}-1.176 (1.101)} & {\scriptsize{}1.183{*}{*}{*} (0.183)} &  & {\scriptsize{}15,714}\tabularnewline
{\scriptsize{}$H0:\eta_{11}\geq0,\eta_{12}=\eta_{21}=0,\eta_{22}\geq0$} & {\scriptsize{}{[}0.959{]}} & {\scriptsize{}{[}0.657{]}} &  & {\scriptsize{}{[}0.285{]}} & {\scriptsize{}{[}1.000{]}} &  & \tabularnewline
\cline{2-8} \cline{3-8} \cline{4-8} \cline{5-8} \cline{6-8} \cline{7-8} \cline{8-8} 
{\scriptsize{}High Incarceration Propensity} & {\scriptsize{}1.058{*}{*} (0.470)} & {\scriptsize{}0.038 (0.075)} &  & {\scriptsize{}1.233 (1.120)} & {\scriptsize{}0.805{*}{*}{*} (0.179)} &  & {\scriptsize{}15,714}\tabularnewline
{\scriptsize{}$H0:\eta_{11}\geq0,\eta_{12}=\eta_{21}=0,\eta_{22}\geq0$} & {\scriptsize{}{[}0.988{]}} & {\scriptsize{}{[}0.615{]}} &  & {\scriptsize{}{[}0.271{]}} & {\scriptsize{}{[}1.000{]}} &  & \tabularnewline
\hline 
\end{tabular}{\scriptsize\par}
\par\end{centering}
\end{singlespace}
\begin{singlespace}
{\scriptsize{}\smallskip{}
}{\scriptsize\par}
\end{singlespace}

\textit{\scriptsize{}Note:}\textit{\emph{\scriptsize{} }}{\scriptsize{}This
table reports tests of Proposition \ref{prop:kitagawa} generalized
to ordered treatments in Section \ref{subsec:kitagawa-generalization}
(Panel A) and Proposition \ref{prop:test} (Panel B) using data from
\citet{BhullerEtAl2020}. See Section \ref{subsec:application_testing}
for the corresponding regression specifications. Proposition \ref{prop:kitagawax}
shows that Proposition \ref{prop:kitagawa} extends to the case with
control variables, while Proposition \ref{prop:testx} provides a
generalization of Proposition \ref{prop:test} under Assumption \ref{assu:covary_similarly}.
The sample consists of 31,428 non-confession criminal cases that were
quasi-randomly assigned to judges in Norwegian trial courts between
2005 and 2009. All estimations control for court by court entry year
fixed effects and baseline controls listed in \citet{BhullerEtAl2020},
Table 1. For each observation, we estimate an incarceration propensity
using pre-determined case and defendant characteristics listed in
\citet{BhullerEtAl2020}, Table 1, based on a logit model, and create
subsamples by low and high propensity based on the median predicted
propensity in the final two rows of Panel B. }\textit{\emph{\scriptsize{}{*}p<0.1,
{*}{*}p<0.05, {*}{*}{*}p<0.01.}}{\scriptsize\par}
\end{sidewaystable}
\vspace{-0.75cm}
\par\end{center}

In Table \ref{tab:application_test_prop3_prop4_orderedT}, Panel B,
we apply the subsample test of Proposition \ref{prop:test} on 12
different subsamples based on the defendant's past employment, past
incarceration, age, education, parental status, and the predicted
likelihood of incarceration based on pre-determined case and defendant
characteristics.\footnote{Ideally, one would select the subsamples using machine learning methods
as proposed by \citet{farbmacher2022instrument}. We leave how to
implement this in the multiple treatment setting for future research.} In each subsample, we run the regessions
\[
D_{1}=\gamma_{1X}+\eta_{11}P_{1}+\eta_{12}P_{2}+\varepsilon_{1}
\]
\[
D_{2}=\gamma_{2X}+\eta_{21}P_{1}+\eta_{22}P_{2}+\varepsilon_{2}
\]
where $\gamma_{1X}$ and $\gamma_{2X}$ are court-by-year fixed effects.\footnote{See Section \ref{subsec:controls} for how the Proposition \ref{prop:test}
generalizes to the inclusion of controls. Formally, these regressions
are joint tests of Assumption \ref{assumption:non_neg}\textendash \ref{assumption:zero}
and Assumption \ref{assu:covary_similarly}.} We can not reject $\eta_{11},\eta_{22}\geq0$ or $\eta_{12}=\eta_{21}=0$
in any subsample. With this test, we are unable to reject Assumptions
\ref{assumption:non_neg}\textendash \ref{assumption:zero}.

Finally, in Figure \ref{fig:application_linearity_predictedT}, we
assess whether the relationship between incarceration rates and conviction
rates among the judges in our data is linear, as required for 2SLS
to assign proper weights in the Section \ref{subsec:threshold-crossing}
model.\footnote{When the model includes court-by-year effects, such a linear relationship
must exist within each court-by-year cell (Proposition \ref{prop:threshold-crossing-x}).
We only consider the overall relationship in Figure \ref{fig:application_linearity_predictedT}.} While we can statistically reject that the relationship is \emph{exactly
}linear, the deviations from linearity are surprisingly small. Thus,
if judges agree on the ranking of cases but use different cutoffs
for conviction and incarceration, any bias in 2SLS due to heterogeneous
effects is likely to be negligible.
\begin{center}
\begin{figure}[H]
\caption{Empirical Application: Testing Proposition \ref{prop:threshold-crossing}.\label{fig:application_linearity_predictedT}}

\begin{centering}
\vspace{-0.25cm}
\includegraphics[width=0.7\textwidth]{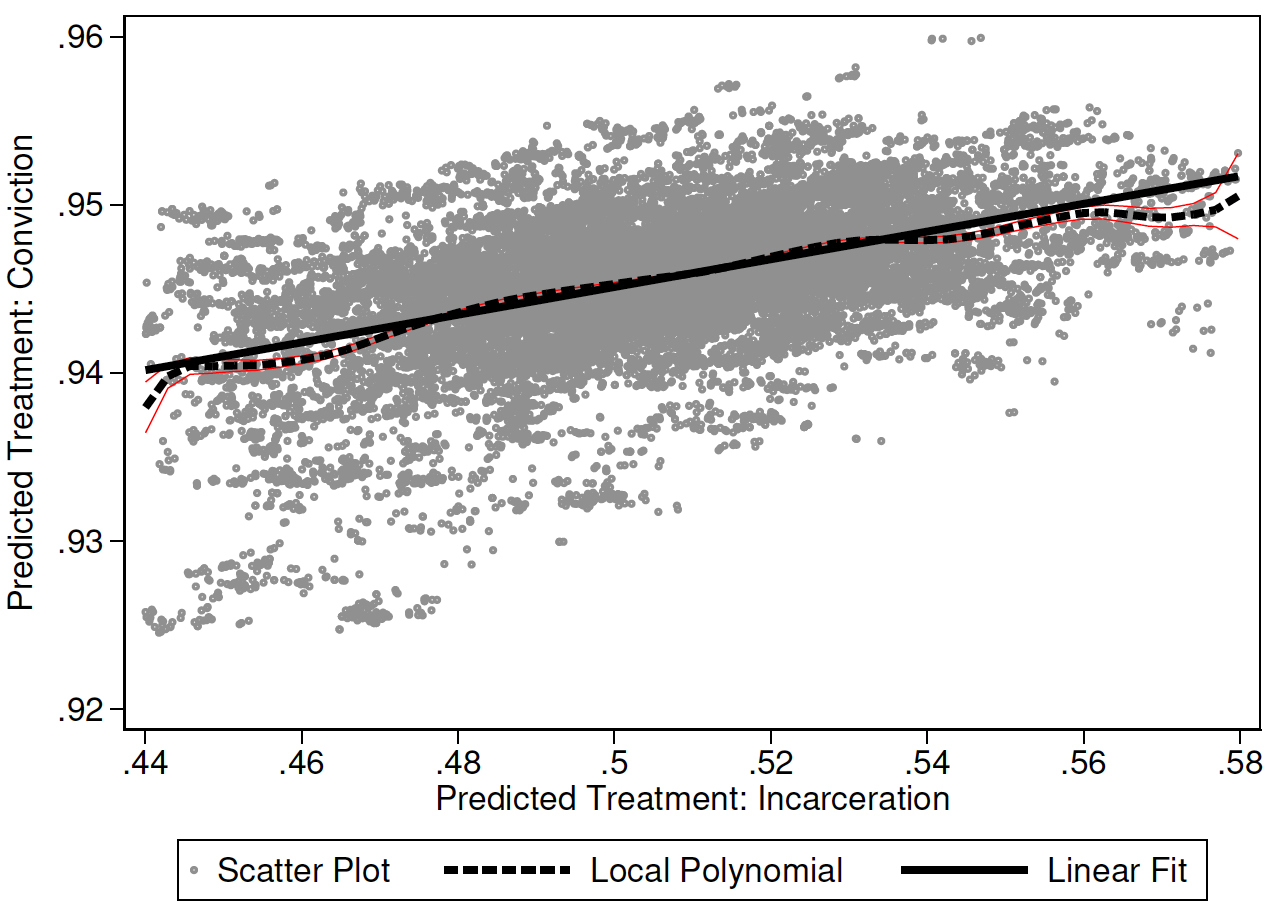}
\par\end{centering}
{\scriptsize{}\smallskip{}
}{\scriptsize\par}

\emph{\scriptsize{}Note:}{\scriptsize{} This figure shows a scatter
plot of predicted conviction treatment ($P_{1}$) and predicted incarceration
treatment ($P_{2}$) using data from \citet{BhullerEtAl2020}, along
with a linear regression fit and local polynomial regression fit.
The sample consists of 31,428 non-confession criminal cases that were
quasi-randomly assigned to judges in Norwegian trial courts between
2005 and 2009. The local polynomial regression uses an Epanechnikov
kernel with bandwidth of 0.01 and a third degree polynomial smoothing.
The red lines show the 99 percent confidence intervals corresponding
to the local polynomial regression. The linear regression of predicted
conviction treatment ($P_{1}$) on predicted incarceration treatment
($P_{2}$) has a coefficient of 0.083 (SE=0.001), and the Ramsey's
RESET test of a linear functional form is rejected with an F-statistic
of 445.1 (p-value=0.000). This provides a formal test of Proposition
\ref{prop:threshold-crossing} under a single-index threshold-crossing
model presented in Section \ref{subsec:threshold-crossing}. All estimations
control for court by court entry year fixed effects and baseline controls
listed in \citet{BhullerEtAl2020}, Table 1. Proposition \ref{prop:threshold-crossing-x}
shows that when the model includes court-by-year fixed effects, a
linear relationship between $P_{1}$ and $P_{2}$ must exist within
each court-by-year cell. For ease of illustration, we consider here
the overall relationship.}{\scriptsize\par}
\end{figure}
\par\end{center}

\begin{center}
\vspace{-0.75cm}
\par\end{center}

\section{Conclusion\label{sec:Conclusion}}

Two-stage least squares (2SLS) is a common approach to causal inference.
We have presented necessary and sufficient conditions for the 2SLS
to identify a properly weighted sum of individual treatment effects
when there are multiple treatments and arbitrary treatment effect
heterogeneity. The conditions require in just-identified models that
each instrument only affects one choice margin. In overidentified
models, 2SLS identifies ordered treatment effects in a general threshold-crossing
model conditional on an easily verifiable linearity condition. Whether
2SLS with multiple treatments should be used depends on the setting.
Justifying its use in the presence of heterogeneous effects requires
both running systematic empirical tests of the average conditional
monotonicity and no-cross-effects conditions and a careful discussion
of why the conditions are likely to hold.

\section*{Acknowledgments}

\begin{singlespace}
\noindent We thank Gaurab Aryal, Brigham Frandsen, Magne Mogstad,
Elie Tamer, the associate editor, and four anonymous referees for
useful comments. Manudeep Bhuller gratefully acknowledges support
from the European Research Council\textquoteright s Starting Grant
Project No. 757279. We thank Statistics Norway, Norwegian Courts Administration
and Norwegian Correctional Services for providing access to their
data.
\end{singlespace}

\bibliographystyle{authordate3}
\bibliography{references}
\pagebreak{}

\appendix
\setcounter{page}{1}\setcounter{footnote}{0}
\global\long\def\thepage{[APPENDIX-\arabic{page}]}%

\section{Appendix: Proofs}

\global\long\def\thetable{A.\arabic{table}}%
\setcounter{table}{0}
\global\long\def\thefigure{A.\arabic{figure}}%
\setcounter{figure}{0}
\global\long\def\theequation{A.\arabic{equation}}%

\setcounter{equation}{0}
\setcounter{thm}{0}
\setcounter{cor}{0}
\setcounter{prop}{0}
\setcounter{lem}{0}
\setcounter{assumption}{0}
\setcounter{example}{0}
\counterwithin{thm}{section}
\counterwithin{cor}{section}
\counterwithin{prop}{section}
\counterwithin{prop}{lem}
\counterwithin{assumption}{section}
\counterwithin{example}{section}
\renewcommand{\thethm}{\thesection.\arabic{thm}}
\renewcommand{\theprop}{\thesection.\arabic{prop}}
\renewcommand{\thecor}{\thesection.\arabic{cor}}
\renewcommand{\thelem}{\thesection.\arabic{lem}}
\renewcommand{\theassumption}{\thesection.\arabic{assumption}}
\renewcommand{\theexample}{\thesection.\arabic{example}}

\noindent In the proofs below we consider the general case with $n$
treatments: $T\in\left\{ 0,1,\dots,n\right\} $. We use the notation
$v_{s}\left(Z\right)\equiv\left[\begin{array}{ccc}
s_{1}\left(Z\right) & \dots & s_{n}\left(Z\right)\end{array}\right]^{\prime}$ to capture the vector of potential treatment indicator of response
type $s\in\mathcal{S}$. Our main results use the following Lemma.
\begin{lem}
\label{lem:weights}Under Assumptions \ref{assu:iv} and \ref{assu:rank},
$\beta^{\text{2SLS}}=\E\left[w^{S}\beta^{S}\right]$ where for $s\in\mathcal{S}$
\[
w^{s}\equiv\Var\left(P\right)^{-1}\Cov\left(P,v_{s}\left(Z\right)\right)\text{ and }\beta^{s}\equiv\E\left[\beta\mid S=s\right].
\]
\end{lem}
\begin{proof}
We have
\begin{eqnarray*}
\Cov\left(P,Y\right) & = & \E\left[\left(P-\E\left[P\right]\right)\left(Y-\E\left[Y\right]\right)\right]\\
 & = & \E\left[\left(P-\E\left[P\right]\right)Y\right]\\
 & = & \E\left[\E\left[\left(P-\E\left[P\right]\right)Y\mid Z,S\right]\right]\\
 & = & \E\left[\left(P-\E\left[P\right]\right)\E\left[Y\mid Z,S\right]\right]\\
 & = & \E\left[\left(P-\E\left[P\right]\right)\E\left[Y\left(0\right)+v_{S}\left(Z\right)\beta\mid Z,S\right]\right]\\
 & = & \E\left[\left(P-\E\left[P\right]\right)\left(\E\left[Y\left(0\right)\mid S\right]+v_{S}\left(Z\right)\E\left[\beta\mid S\right]\right)\right]\\
 & = & \E\left[\left(P-\E\left[P\right]\right)v_{S}\left(Z\right)\E\left[\beta\mid S\right]\right]\\
 & = & \E\left[\Cov\left(P,v_{S}\left(Z\right)\mid S\right)\E\left[\beta\mid S\right]\right]
\end{eqnarray*}
where $v_{s}\left(Z\right)\equiv\left[\begin{array}{ccc}
s_{1}\left(Z\right) & \dots & s_{n}\left(Z\right)\end{array}\right]^{\prime}$. The third equality uses the law of iterated expectations. The fourth
equality uses that $P$ is a deterministic function of $Z$. The sixth
and the seventh equality invokes Assumption \ref{assu:iv}.\footnote{The seventh equality also relies on the law of iterated expectations.
In particular
\begin{eqnarray*}
\E\left[\left(P-\E\left[P\right]\right)\E\left[Y\left(0\right)\mid S\right]\right] & = & \E\left[\E\left[\left(P-\E\left[P\right]\right)\E\left[Y\left(0\right)\mid S\right]\mid S\right]\right]\\
 & = & \E\left[\left(\E\left[P\mid S\right]-\E\left[P\right]\right)\E\left[Y\left(0\right)\mid S\right]\right]\\
 & = & \E\left[\left(\E\left[P\right]-\E\left[P\right]\right)\E\left[Y\left(0\right)\mid S\right]\right]\\
 & = & 0
\end{eqnarray*}
} Thus $\beta^{\text{2SLS}}=\E\left[w^{S}\beta^{S}\right]$ where $w^{s}\equiv\Var\left(P\right)^{-1}\Cov\left(P,v_{s}\left(Z\right)\right)$
and $\beta^{s}\equiv\E\left[\beta\mid S=s\right]$.
\end{proof}
\begin{proof}
(Proposition \ref{prop:weights}). By Lemma \ref{lem:weights}, $\beta^{\text{2SLS}}=\E\left[w^{S}\beta^{S}\right]$
where $w^{s}\equiv\Var\left(P\right)^{-1}\Cov\left(P,v_{s}\left(Z\right)\right)$.
Thus $\beta_{1}^{\text{2SLS}}=\E\left[w_{1}^{S}\beta_{1}^{S}+\dots+w_{n}^{S}\beta_{n}^{S}\right]$
where $w_{k}^{s}$ is the $k$th element in the first row of $w^{s}$.
The parameter $w_{k}^{s}$ equals the coefficient on $P_{1}$ in a
regression of $s_{k}\left(Z\right)$ on $P$. Applying the Frisch-Waugh-Lovell
theorem, we thus get
\[
w_{k}^{s}=\frac{\Cov\left(\tilde{P}_{1},s_{k}\left(Z\right)\right)}{\Var\left(\tilde{P}_{1}\right)}
\]
where $\tilde{P}_{1}$ is the residual in a regression of $P_{1}$
on $\left\{ P_{2},\dots,P_{n}\right\} $.
\end{proof}
\begin{proof}
(Corollary \ref{cor:weights-multiple}). We have
\begin{eqnarray*}
\E\left[\Cov\left(P,v_{S}\left(Z\right)\mid S\right)\right] & = & \Cov\left(P,\E\left[v_{S}\left(Z\right)\mid Z\right]\right)\\
 & = & \Cov\left(P,\E\left[D\mid Z\right]\right)\\
 & = & \Var\left(P\right)
\end{eqnarray*}
where $v_{s}\left(Z\right)\equiv\left[\begin{array}{ccc}
s_{1}\left(Z\right) & \dots & s_{n}\left(Z\right)\end{array}\right]^{\prime}$. The first equality uses the distributive property of covariances
of sums. The third equality uses that, by a standard property of
linear projections, $\Cov\left(Z,D-P\right)=0$. This property implies
$\Cov\left(Z,P\right)=\Cov\left(Z,D\right)=\Cov\left(Z,\E\left[D\mid Z\right]\right)$
which gives $\Cov\left(P,\E\left[D\mid Z\right]\right)=\Cov\left(P,P\right)=\Var\left(P\right)$
since $P$ is a linear function of $Z$.\footnote{By the law of iterated expectations
\begin{eqnarray*}
\Cov\left(Z,D\right) & = & \E\left[ZD\right]-\E\left[Z\right]\E\left[D\right]\\
 & = & \E\left[\E\left[ZD\mid Z\right]\right]-\E\left[Z\right]\E\left[\E\left[D\mid Z\right]\right]\\
 & = & \E\left[Z\E\left[D\mid Z\right]\right]-\E\left[Z\right]\E\left[\E\left[D\mid Z\right]\right]\\
 & = & \Cov\left(Z,\E\left[D\mid Z\right]\right)
\end{eqnarray*}
} Thus $\E\left[w^{S}\right]=I_{2}$ where $I_{2}$ is the $2\times2$
identity matrix.
\end{proof}
\begin{proof}
(Corollary \ref{cor:non_neg}). By Proposition \ref{prop:weights}
\[
w_{1}^{s}\geq0\Leftrightarrow\Cov\left(\tilde{P}_{1},s_{1}\left(Z\right)\right)\geq0\Leftrightarrow\E\left[\tilde{P}_{1}\mid s_{1}\left(Z\right)=1\right]\geq\E\left[\tilde{P}_{1}\mid s_{1}\left(Z\right)=0\right]
\]

\vspace{-0.25cm}
\end{proof}
\begin{proof}
(Corollary \ref{cor:zero}). By Proposition \ref{prop:weights}
\[
w_{2}^{s}=0\Leftrightarrow\Cov\left(\tilde{P}_{1},s_{2}\left(Z\right)\right)=0\Leftrightarrow\E\left[\tilde{P}_{1}\mid s_{2}\left(Z\right)=1\right]=\E\left[\tilde{P}_{1}\mid s_{2}\left(Z\right)=0\right]
\]
\vspace{-0.25cm}
\end{proof}
\begin{proof}
(Proposition \ref{cor:zero_alt}). Assumption \ref{assumption:zero}
is equivalent to $\Cov\left(\tilde{P}_{1},s_{2}\left(Z\right)\right)=0$
for all $s\in\mathcal{S}$. We have
\[
\Cov\left(\tilde{P}_{1},s_{2}\left(Z\right)\right)=\Cov\left(P_{1},s_{2}\left(Z\right)\right)-\frac{\Cov\left(P_{1},P_{2}\right)}{\Var\left(P_{2}\right)}\Cov\left(P_{2},s_{2}\left(Z\right)\right)
\]
Thus, Assumption \ref{assumption:zero} implies $\Cov\left(P_{1},s_{2}\left(Z\right)\right)=\rho\Cov\left(P_{2},s_{2}\left(Z\right)\right)$
for all $s\in\mathcal{S}$ with $\rho\equiv\Cov\left(P_{1},P_{2}\right)/\Var\left(P_{2}\right)$.
To prove the other direction of the equivalence, assume $\Cov\left(P_{1},s_{2}\left(Z\right)\right)=\rho\Cov\left(P_{2},s_{2}\left(Z\right)\right)$
for all $s\in\mathcal{S}$. Applying the law of iterated expectation
and using that $P\perp S$, we get $\Cov\left(P_{1},D_{2}\right)=\rho\Cov\left(P_{2},D_{2}\right)$.
By a standard property of linear projections, $\Cov\left(Z,P-D\right)=0$.
Since $P$ is a linear transformation of $Z$, we further get $\Cov\left(P,P-D\right)=0\Rightarrow\Var\left(P\right)=\Cov\left(P,D\right)$.
We thus have
\[
\rho=\frac{\Cov\left(P_{1},D_{2}\right)}{\Cov\left(P_{2},D_{2}\right)}=\frac{\Cov\left(P_{1},P_{2}\right)}{\Var\left(P_{2}\right)}
\]
which gives
\[
\Cov\left(\tilde{P}_{1},s_{2}\left(Z\right)\right)=\Cov\left(P_{1},s_{2}\left(Z\right)\right)-\frac{\Cov\left(P_{1},P_{2}\right)}{\Var\left(P_{2}\right)}\Cov\left(P_{2},s_{2}\left(Z\right)\right)=0
\]
\end{proof}
\begin{proof}
(Corollary \ref{cor:avg_mono}). This result follows immediately from
Corollaries \ref{cor:non_neg}\textendash \ref{cor:zero}.
\end{proof}
In our remaining proofs, we will also refer to the following corollary
of Lemma \ref{lem:weights}.
\begin{cor}
\label{cor:diagonal}2SLS assigns proper weights under Assumptions
\ref{assu:iv}\textendash \ref{assu:rank} if and only if $w^{s}\equiv\Var\left(P\right)^{-1}\Cov\left(P,v_{s}\left(Z\right)\right)$
is a non-negative diagonal matrix for all $s\in\mathcal{S}$.
\end{cor}
\begin{proof}
The diagonal elements of $w^{s}$ correspond to the weights on $\beta_{k}^{s}$
in $\beta_{k}^{\text{2SLS}}$ for $k\in\left\{ 1,\dots,n\right\} $.
Assumption \ref{assumption:non_neg} requires these weights to be
non-negative. The non-diagonal elements of $w^{s}$ correspond to
the weights on $\beta_{k}^{s}$ in $\beta_{l}^{\text{2SLS}}$ for
$l\neq k$. Assumption \ref{assumption:zero} requires these weights
to be zero.
\end{proof}
\begin{proof}
(Proposition \ref{prop:kitagawa}.) For $k\in\left\{ 1,\dots,n\right\} $,
we have
\begin{eqnarray*}
\Cov\left(P,\mathbf{1}_{y\leq Y\leq y'}D_{k}\right) & = & \Cov\left(P,\mathbf{1}_{y\leq Y\left(k\right)\leq y'}D_{k}\right)\\
 & = & \E\left[\Cov\left(P,\mathbf{1}_{y\leq Y\left(k\right)\leq y'}D_{k}\mid S\right)\right]\\
 & = & \E\left[\Cov\left(P,\mathbf{1}_{y\leq Y\left(k\right)\leq y'}S_{k}\left(Z\right)\mid S\right)\right]\\
 & = & \E\left[\Cov\left(P,S_{k}\left(Z\right)\mid S\right)\Pr\left[y\leq Y\left(k\right)\leq y'\mid S\right]\right]
\end{eqnarray*}
where the second equality uses the law of iterated expectations and
$P\perp S$. The fourth equality invokes Assumption \ref{assu:iv}.
Thus
\[
\Var\left(P\right)^{-1}\Cov\left(P,\mathbf{1}_{y\leq Y\leq y'}D\right)=\E\left[w^{S}c^{S}\right]
\]
where $w^{s}\equiv\Var\left(P\right)^{-1}\Cov\left(P,v_{s}\left(Z\right)\right)$
and $c^{s}$ is a non-negative matrix.\footnote{The $k$th column of the matrix $c_{s}$ is a vector with $n$ entries,
all equal to $\Pr\left[y\leq Y\left(k\right)\leq y'\mid S=s\right]\geq0$.} If 2SLS assigns proper weights, then $w^{s}$ is non-negative diagonal
for all $s\in\mathcal{S}$ by Corollary \ref{cor:diagonal}. We thus
must have that $\E\left[w^{S}c^{S}\right]$ is also a non-negative
diagonal matrix.
\end{proof}
\begin{proof}
(Proposition \ref{prop:test}). Assume Assumptions \ref{assumption:non_neg}
and \ref{assumption:zero} hold. Then $\Var\left(P\right)^{-1}\Cov\left(P,v_{s}\left(Z\right)\right)$
is a non-negative diagonal matrix for all $s\in\mathcal{S}$. Since
$Z\perp X\mid S$, we have 
\[
\Cov\left(P,v_{s}\left(Z\right)\mid X=1\right)=\Cov\left(P,v_{s}\left(Z\right)\right)
\]
 for each response type $s$ present in the subsample $X=1$ where
$v_{s}\left(Z\right)\equiv\left[\begin{array}{ccc}
s_{1}\left(Z\right) & \dots & s_{n}\left(Z\right)\end{array}\right]^{\prime}$. Furthermore, $\Var\left(P\mid X=1\right)=\Var\left(P\right)$.
Thus
\[
\Var\left(P\mid X=1\right)^{-1}\Cov\left(P,v_{s}\left(Z\right)\mid X=1\right)=\Var\left(P\right)^{-1}\Cov\left(P,v_{s}\left(Z\right)\right)
\]
is also non-negative diagonal. Noting that $\Cov\left(P,v_{s}\left(Z\right)\mid X=1\right)=\Cov\left(P,D\mid X=1,S=s\right)$,
we then have that
\[
\Var\left(P\mid X=1\right)^{-1}\Cov\left(P,D\mid X=1\right)=\E\left[\Var\left(P\mid X=1\right)^{-1}\Cov\left(P,D\mid X=1,S\right)\right]
\]
is a non-negative diagonal matrix.\footnote{\vspace{-0.75cm}
\begin{eqnarray*}
\E\left[\Cov\left(P,D\mid X=1,S\right)\right] & = & \E\left[\E\left[PD\mid X=1,S\right]\right]-\E\left[\E\left[P\mid X=1,S\right]\E\left[D\mid X=1,S\right]\right]\\
 & = & \E\left[PD\mid X=1\right]-\E\left[\E\left[P\mid X=1\right]\E\left[D\mid X=1,S\right]\right]\\
 & = & \E\left[PD\mid X=1\right]-\E\left[P\mid X=1\right]\E\left[D\mid X=1\right]\\
 & = & \Cov\left(P,D\mid X=1\right)
\end{eqnarray*}
where the second equality uses the law of iterated expectation and
that $P\perp S$.}
\end{proof}
To prove Proposition \ref{prop:binary}, the following lemma is useful.
\begin{lem}
\label{lem:n_equals_m}When n=m, 2SLS assigns proper weights if and
only if for all $s$, $k$, and $l$
\[
\frac{\Cov\left(Z_{k},s_{l}\left(Z\right)\right)}{\Cov\left(Z_{k},D_{l}\right)}=\frac{\Cov\left(Z_{l},s_{l}\left(Z\right)\right)}{\Cov\left(Z_{l},D_{l}\right)}\geq0
\]
\end{lem}
\begin{proof}
By Corollary \ref{cor:diagonal}, the 2SLS assigns proper weights
if and only if the matrix $w^{s}=\Var\left(P\right)^{-1}\Cov\left(P,v_{s}\left(Z\right)\right)$
is non-negative diagonal for all $s$ where $v_{s}\left(Z\right)\equiv\left[\begin{array}{ccc}
s_{1}\left(Z\right) & \dots & s_{n}\left(Z\right)\end{array}\right]^{\prime}$. When $n=m$, we can simplify to
\[
w^{s}=\Cov\left(Z,D\right)^{-1}\Cov\left(Z,v_{s}\left(Z\right)\right)\Rightarrow\Cov\left(Z,D\right)w^{s}=\Cov\left(Z,v_{s}\left(Z\right)\right)
\]
Thus, for all $s$, $k$, and $l$, $\Cov\left(Z_{k},D_{l}\right)w_{sll}=\Cov\left(Z_{k},s_{l}\left(Z\right)\right)$
where $w_{sll}\geq0$ is the $l$th diagonal element of $w^{s}$.
\end{proof}
\begin{proof}
(Proposition \ref{prop:binary}). Consider a treatment $k$. Defining
$\pi_{l}\equiv\Pr\left[V=l\right]$, we have
\begin{eqnarray*}
\Cov\left(Z_{l},s_{k}\left(V\right)\right) & = & \left(\Pr\left[s_{k}\left(V\right)=1\mid V=l\right]-\Pr\left[s_{k}\left(V\right)=1\mid V\neq l\right]\right)\Pr\left[V=l\right]\Pr\left[V\neq l\right]\\
 & = & \left(s_{k}\left(l\right)-\sum_{r\neq l}\frac{\pi_{r}s_{k}\left(r\right)}{1-\pi_{l}}\right)\pi_{l}\left(1-\pi_{l}\right)\\
 & = & \left(s_{k}\left(l\right)-\sum_{r}\pi_{r}s_{k}\left(r\right)\right)\pi_{l}\\
 & = & \left(s_{k}\left(l\right)-\E\left[s_{k}\left(V\right)\right]\right)\pi_{l}
\end{eqnarray*}
By Lemma \ref{lem:n_equals_m}, for all $l$ and $r$, we must have
that
\[
\frac{\left(s_{k}\left(l\right)-\E\left[s_{k}\left(V\right)\right]\right)\pi_{l}}{\Cov\left(Z_{l},D_{k}\right)}=\frac{\left(s_{k}\left(r\right)-\E\left[s_{k}\left(V\right)\right]\right)\pi_{r}}{\Cov\left(Z_{r},D_{k}\right)}\geq0
\]
is constant across $s$. These equations are satisfied for \emph{never-$k$-takers}\textemdash agents
with $s_{k}\left(l\right)=0$ for all $l$\textemdash and \emph{always-$k$-takers}\textemdash agents
with $s_{k}\left(l\right)=1$ for all $l$. Beyond never-$k$-takers
and always-$k$-takers, there can only be one response type for treatment
indicator $k$. To see this, assume response types $s$ and $s'$
are neither never-$k$-takers nor always-$k$-takers. We then must
have $s_{k}\left(l\right)=s_{k}\left(r\right)\Rightarrow\Cov\left(Z_{l},D_{k}\right)\pi_{r}=\Cov\left(Z_{r},D_{k}\right)\pi_{l}\Rightarrow s'_{k}\left(l\right)=s'_{k}\left(r\right)$.
Thus, either $s_{k}\left(l\right)=s'_{k}\left(l\right)$ or $s_{k}\left(l\right)=1-s_{k}'\left(l\right)$
for all $l$. The latter case is not possible since $s_{k}\left(l\right)-\E\left[s_{k}\left(V\right)\right]$
and $s_{k}'\left(l\right)-\E\left[s_{k}'\left(V\right)\right]$ will
then have opposite signs. Thus $s_{k}\left(l\right)=s'_{k}\left(l\right)$
for all $l$. Since, for all treatments $k$, there must exist at
least one response type that is neither an always-$k$-taker nor a
never-$k$-taker for the rank condition (Assumption \ref{assu:rank})
to hold and $n=m$, the response type for treatment $k$ must have
$s_{k}\left(l\right)=1$ for exactly one $l$. This defines a one-to-one
mapping $f:\left\{ 0,1,\dots,n\right\} \rightarrow\left\{ 0,1,\dots,n\right\} $
such that, for all $k$, the only non-trivial response type for treatment
indicator $k$ is defined by $s_{k}\left(v\right)=1\Leftrightarrow f\left(v\right)=k$.
\end{proof}
\begin{proof}
(Corollary \ref{cor:binary_2}). This result follows directly from
Proposition \ref{prop:binary}.
\end{proof}
\begin{proof}
(Proposition \ref{prop:just-identified-test}). Assume Assumptions
\ref{assumption:non_neg} and \ref{assumption:zero} hold. Let $y\leq y'$
and assume $X\in\left\{ 0,1\right\} $ with $Z\perp\left(X,S\right)$.
We want to show that $\Var\left(Z\right)^{-1}\Cov\left(Z,\mathbf{1}_{y\leq Y\leq y'}D\right)$
and $\Var\left(Z\mid X=1\right)^{-1}\Cov\left(Z,D\mid X=1\right)$
are non-negative diagonal. Let $a_{k}$ denote the share of always-$k$-takers
and $b_{k}$ the share of agents selecting treatment $k$ if and only
if $Z_{k}=1$ in the population. By Proposition \ref{prop:binary},
these are the only possible response types for treatment $k$ (when
we maintain Assumption \ref{assu:order}). Predicted treatments based
on the instruments $Z$ thus equals $P=a+bZ$ where $a\equiv\left(a_{1},\dots,a_{n}\right)'$
and $b$ is a diagonal matrix with diagonal elements $b_{k}$. By
Propositions \ref{prop:kitagawa} and \ref{prop:test}, the following
matrices must be non-negative diagonal
\begin{eqnarray*}
\Var\left(P\right)^{-1}\Cov\left(P,\mathbf{1}_{y\leq Y\leq y'}D\right) & = & \Var\left(bZ\right)^{-1}\Cov\left(bZ,\mathbf{1}_{y\leq Y\leq y'}D\right)\\
 & = & b^{-1}\Var\left(Z\right)^{-1}\Cov\left(Z,\mathbf{1}_{y\leq Y\leq y'}D\right)
\end{eqnarray*}
\begin{eqnarray*}
\Var\left(P\mid X=1\right)^{-1}\Cov\left(P,D\mid X=1\right) & = & \Var\left(bZ\mid X=1\right)^{-1}\Cov\left(bZ,D\mid X=1\right)\\
 & = & b^{-1}\Var\left(Z\mid X=1\right)^{-1}\Cov\left(Z,D\mid X=1\right)
\end{eqnarray*}

Since $b^{-1}$ is non-negative diagonal, we then get that $\Var\left(Z\right)^{-1}\Cov\left(Z,\mathbf{1}_{y\leq Y\leq y'}D\right)$
and $\Var\left(Z\mid X=1\right)^{-1}\Cov\left(Z,D\mid X=1\right)$
are also non-negative diagonal. 
\end{proof}
\begin{proof}
(Proposition \ref{prop:binary_roy}). We first show that indirect
utilities of the stated form satisfy the assumptions in Proposition
\ref{prop:binary} required for 2SLS to assign proper weights. We
need to show that (i) a response type never selects $k$ unless she
selects it at $V=k$ and (ii) a response type that selects $k$ when
$V=l\neq k$ always selects $k$. To show (i), assume $s$ selects
$l\neq k$ when $V=k$. Then $I_{ls}\left(k\right)=u_{ls}>I_{ks}\left(k\right)=u_{ks}+\mu_{ks}$.
Since $\mu_{ks}\geq0$, we must then have $I_{ls}\left(v\right)>I_{ks}\left(v\right)$
for all $v$\textemdash response type $s$ never selects $k$. To
show (ii), assume $s$ selects $k$ when $V=l\neq k$. Then $I_{ks}\left(l\right)=u_{ks}>0=I_{0s}\left(l\right)$.
Thus $I_{ks}\left(v\right)>I_{0s}\left(v\right)=0$ for all $v$.
By the last stated property of the indirect utilities in the proposition,
we must then have $I_{rs}\left(v\right)<I_{0s}\left(v\right)=0$ for
all $v$ and $r\notin\left\{ 0,k\right\} $. Thus $k$ is always selected.
To prove the other direction of the equivalence, assume that 2SLS
assigns proper weights. We want to show that choice behavior can be
described by indirect utilities of the form given in the proposition.
To do this, define indirect utilities $I_{ks}\left(v\right)=u_{ks}+\mu_{ks}\mathbf{1}\left[v=k\right]$
where

\[
\left(u_{ks},\mu_{ks}\right)=\begin{cases}
\left(1,0\right) & \text{if \ensuremath{s} always selects \ensuremath{k}}\\
\left(-1,0\right) & \text{if \ensuremath{s} never selects \ensuremath{k}}\\
\left(-1,2\right) & \text{if \ensuremath{s} selects \ensuremath{k} if and only if \ensuremath{V=k}}
\end{cases}
\]

By Proposition \ref{prop:binary}, these cases cover all possible
response types when 2SLS assigns proper weights. It is straightforward
to verify that $s\left(v\right)=\arg\max_{k}I_{ks}\left(v\right)$
with these values of $u_{ks}$ and $\mu_{ks}$. Note that no response
type is ever indifferent, so $\arg\max_{k}I_{ks}\left(v\right)$ is
indeed a singleton. The indirect utilities also satisfy $I_{ks}\left(v\right)>I_{0s}\left(v\right)\Rightarrow I_{ls}\left(v\right)<I_{0s}\left(v\right)$
for $l\notin\left\{ 0,k\right\} $: We only have $I_{ks}\left(v\right)>0=I_{0s}\left(v\right)$
when $s$ selects $k$ at $V=v$. In that case $l$ is not selected
at $V=v$ and $I_{ls}\left(v\right)=-1<0=I_{0s}\left(v\right)$.
\end{proof}
\begin{proof}
(Proposition \ref{prop:binary_roy2}). We first show that indirect
utilities of the stated form satisfy the assumptions in Proposition
\ref{prop:binary} required for 2SLS to assign proper weights. Under
ordered treatment effects, we have $s_{k}\left(v\right)=\mathbf{1}\left[s\left(v\right)\geq k\right]$.
The required conditions are thus that, for each $k$, a response type
$s$ either always selects $k$ or above, never selects $k$ or above,
or selects $k$ or above if and only if $V=k$. These conditions together
imply that $s$ either always select the same treatment or selects
treatment $k$ if $V=k$ and treatment $k-1$ if $V\neq k$ for some
$k$.

Assume $s$ selects $k-1$ when $V=0$. Thus $u_{k-1,s}>u_{ls}$ for
all $l\neq k-1$. We need to show that (i) $s$ always selects $k-1$
when $V\neq k$ and (ii) $s$ selects either $k$ or $k-1$ when $V=k$.
To show (i), consider first the case when $V=r<k$. Then $I_{k-1,s}\left(r\right)=u_{k-1,s}+\mu_{rs}$
is still the largest indirect utility, so $k-1$ is still selected.
Now, consider the case when $V=r>k$. Then, $I_{k-1,s}\left(r\right)=u_{k-1,s}>u_{ks}=I_{ks}\left(r\right)$
and $I_{k-1,s}\left(r\right)=u_{k-1,s}>u_{k-2,s}=I_{k-2,s}\left(r\right)$.
Informally, there is a ``peak'' at treatment $k-1$. By single-peakedness,
$I_{k-1,s}\left(r\right)$ must then be the largest indirect utility
when $V=r$. Thus $s$ always selects $k-1$ when $V\neq k$. To show
(ii), assume that $r\neq k$ is selected when $V=k$. Since $V=k$
only increases the utility of treatment $k$ and above we must have
$r\geq k-1$. If $r>k$, then $I_{rs}\left(k\right)=u_{rs}+\mu_{ks}>u_{ks}+\mu_{ks}=I_{ks}\left(k\right)\Rightarrow u_{rs}>u_{ks}$.
But then $I_{rs}\left(0\right)>I_{ks}\left(0\right)$ and $I_{k-1,s}\left(0\right)>I_{ks}\left(0\right)$,
violating single-peakedness at $V=0$. Thus, we must have $r=k-1$.

To prove the other direction of the equivalence, assume that 2SLS
assigns proper weights. We want to show that choice behavior can be
described by indirect utilities of the form given in the proposition.
To do this, define indirect utilities $I_{ks}\left(v\right)=u_{ks}+\mu_{vs}\mathbf{1}\left[k\geq v\right]$
where

\[
\left(u_{ks},\mu_{ks}\right)=\begin{cases}
\left(1,0\right) & \text{if \ensuremath{s} selects \ensuremath{k} when \ensuremath{V=0}}\\
\left(-1,0\right) & \text{if \ensuremath{s} never selects \ensuremath{k}}\\
\left(0,2\right) & \text{if \ensuremath{s} selects \ensuremath{k} if \ensuremath{V=k} and \ensuremath{k-1} otherwise}
\end{cases}
\]

These cases cover all possible response types when 2SLS assigns proper
weights: As argued above, under ordered treatment effects, the conditions
in Proposition \ref{prop:binary} require that each a response type
either always select the same treatment or selects treatment $k$
if $V=k$ and treatment $k-1$ if $V\neq k$ for some $k$.  It is
straightforward to verify that $s\left(v\right)=\arg\max_{k}I_{ks}\left(v\right)$
with these values of $u_{ks}$ and $\mu_{ks}$. Note that no response
type is ever indifferent, so $\arg\max_{k}I_{ks}\left(v\right)$ is
indeed a singleton.

We now verify that the indirect utilities given above are always single
peaked. First, consider the case when response type $s$ always selects
treatment $k$. Then $I_{ks}\left(v\right)=1$ and $I_{ls}\left(v\right)=-1$
for all $v$ and $l\neq k$ and the response type's preferences have
a single peak at $k$. Second, consider the case when response type
$s$ selects $k$ if $V=k$ and $k-1$ otherwise. We have $I_{ks}\left(k\right)=2$,
$I_{k-1,s}\left(k\right)=1$, $I_{ls}\left(k\right)=1$ for $l>k$,
and $I_{ls}$$\left(k\right)=-1$ for $l<k-1$. Thus preferences have
a single peak at treatment $k$ when $V=k$. Moreover, for $l\neq k$
we have $I_{k-1,s}$$\left(l\right)=1$, $I_{ks}\left(l\right)=0$,
and $I_{rs}\left(l\right)=-1$ for $r\notin\left\{ k,k-1\right\} $.
Thus preferences have a single peak (at treatment $k-1$) also when
$V=l\neq k$.

\end{proof}
\begin{proof}
(Proposition \ref{prop:threshold-crossing}). This result follows
directly from Proposition \ref{prop:linear-predicted2}.
\end{proof}
\begin{proof}
(Corollary \ref{cor:binary}). This result follows directly from Proposition
\ref{prop:binary}.
\end{proof}
\begin{proof}
(Proposition \ref{prop:weightsx}). We have
\begin{eqnarray*}
\Cov\left(\ddot{P},\ddot{Y}\right) & = & \E\left[\ddot{P}\ddot{Y}\right]-\E\left[\ddot{P}\right]\E\left[\ddot{Y}\right]\\
 & = & \E\left[\ddot{P}\ddot{Y}\right]\\
 & = & \E\left[\ddot{P}Y\right]\\
 & = & \E\left[\E\left[\ddot{P}Y\mid Z,X,S\right]\right]\\
 & = & \E\left[\ddot{P}\E\left[Y\mid Z,X,S\right]\right]\\
 & = & \E\left[\ddot{P}\left(\E\left[\alpha\mid X,S\right]+v_{S}\left(Z\right)\E\left[\beta\mid X,S\right]\right)\right]\\
 & = & \E\left[\ddot{P}v_{S}\left(Z\right)\E\left[\beta\mid X,S\right]\right]\\
 & = & \E\left[\Cov\left(\ddot{P},v_{S}\left(Z\right)\mid X,S\right)\E\left[\beta\mid X,S\right]\right]
\end{eqnarray*}
 where $v_{s}\left(Z\right)\equiv\left[\begin{array}{ccc}
s_{1}\left(Z\right) & \dots & s_{n}\left(Z\right)\end{array}\right]^{\prime}$. The second equality uses that $\E\left[\ddot{P}\right]=0$. The
third equality uses that 
\[
\E\left[\ddot{P}\ddot{Y}\right]=\E\left[\ddot{P}Y\right]-\E\left[\ddot{P}\E\left[Y\mid X\right]\right]
\]
 and, using $\E\left[\ddot{P}\mid X\right]=0$ and the law of iterated
expectations 
\begin{eqnarray*}
\E\left[\ddot{P}\E\left[Y\mid X\right]\right] & = & \E\left[\E\left[\ddot{P}\E\left[Y\mid X\right]\mid X\right]\right]\\
 & = & \E\left[\E\left[Y\mid X\right]\E\left[\ddot{P}\mid X\right]\right]=0
\end{eqnarray*}

The fourth equality applies the law of iterated expectations. The
sixth equality uses
\begin{eqnarray*}
\E\left[Y\mid Z,X,S\right] & = & \E\left[\alpha+D\beta\mid Z,X,S\right]\\
 & = & \E\left[\alpha+v_{S}\left(Z\right)\beta\mid Z,X,S\right]\\
 & = & \E\left[\alpha\mid X,S\right]+v_{S}\left(Z\right)\E\left[\beta\mid X,S\right]
\end{eqnarray*}
where the third equality invokes the conditional independence assumption.
The seventh equality uses $\E\left[\ddot{P}\mid X\right]=0$ combined
with the law of iterated expecations.\footnote{In particular,\vspace{-0.5cm}
\begin{eqnarray*}
\E\left[\ddot{P}\E\left[\alpha\mid X,S\right]\right] & = & \E\left[\E\left[\ddot{P}\E\left[\alpha\mid X,S\right]\mid X,S\right]\right]\\
 & = & \E\left[\E\left[\alpha\mid X,S\right]\E\left[\ddot{P}\mid X,S\right]\right]\\
 & = & \E\left[\E\left[\alpha\mid X,S\right]\E\left[\ddot{P}\mid X\right]\right]\\
 & = & 0
\end{eqnarray*}
where the third equality uses Assumption \ref{assu:ivx}.} We thus have
\[
\beta^{2SLS}=\Var\left(\ddot{P}\right)^{-1}\Cov\left(\ddot{P},\ddot{Y}\right)=\E\left[w^{S,X}\beta^{S,X}\right]
\]
where $w^{s,x}\equiv\Var\left(\ddot{P}\right)^{-1}\Cov\left(\ddot{P},v_{s}\left(Z\right)\mid X=x\right)$
and $\beta^{s,x}\equiv\left(\beta_{1}^{s,x},\dots,\beta_{n}^{s,x}\right)^{T}\equiv\E\left[\beta\mid S=s,X=x\right]$.
\end{proof}
\begin{proof}
(Proposition \ref{prop:kitagawax}). For $k\in\left\{ 1,\dots,n\right\} $,
we have
\begin{eqnarray*}
\Cov\left(\ddot{P},\mathbf{1}_{y\leq Y\leq y'}D_{k}\right) & = & \Cov\left(\ddot{P},\mathbf{1}_{y\leq Y\left(k\right)\leq y'}D_{k}\right)\\
 & = & \E\left[\ddot{P}\mathbf{1}_{y\leq Y\left(k\right)\leq y'}D_{k}\right]\\
 & = & \E\left[\E\left[\ddot{P}\mathbf{1}_{y\leq Y\left(k\right)\leq y'}D_{k}\mid S,X\right]\right]\\
 & = & \E\left[\E\left[\ddot{P}\mathbf{1}_{y\leq Y\left(k\right)\leq y'}S_{k}\left(Z\right)\mid S,X\right]\right]\\
 & = & \E\left[\E\left[\ddot{P}S_{k}\left(Z\right)\mid S,X\right]\Pr\left[y\leq Y\left(k\right)\leq y'\mid S,X\right]\right]\\
 & = & \E\left[\Cov\left(\ddot{P},S_{k}\left(Z\right)\mid S,X\right)\Pr\left[y\leq Y\left(k\right)\leq y'\mid S,X\right]\right]
\end{eqnarray*}
where the second equality uses $\E\left[\ddot{P}\right]=0$, the third
equality uses the law of iterated expectations, and the fifth equality
applies Assumption \ref{assu:ivx}. The last equality uses $\E\left[\ddot{P}\mid S,X\right]=\E\left[\ddot{P}\mid X\right]=0$.
Thus
\[
\Var\left(\ddot{P}\right)^{-1}\Cov\left(\ddot{P},\mathbf{1}_{y\leq Y\leq y'}D\right)=\E\left[w^{S,X}c^{S,X}\right]
\]
where $w^{s,x}\equiv\Var\left(\ddot{P}\right)^{-1}\Cov\left(\ddot{P},v_{s}\left(Z\right)\mid X=x\right)$
and $c^{s,x}$ is a non-negative matrix.\footnote{The $k$th column of $c^{s,x}$ is a vector with $n$ entries, all
equal to $\Pr\left[y\leq Y\left(k\right)\leq y'\mid S=s,X=x\right]$.} If 2SLS assigns proper weights then $w^{s,x}$ is non-negative diagonal
for all $s$ and $x$ by Proposition \ref{prop:weightsx}. We thus
have that $\E\left[w^{S,X}c^{S,X}\right]$ is also a non-negative
diagonal matrix.
\end{proof}
\begin{proof}
(Proposition \ref{prop:testx}). Assume 2SLS assigns proper weights.
We then have that
\begin{eqnarray*}
w^{s,x} & = & \Var\left(\ddot{P}\right)^{-1}\Cov\left(\ddot{P},v_{s}\left(Z\right)\mid X=x\right)\\
 & = & \Var\left(\ddot{P}\right)^{-1}\Cov\left(\ddot{P},v_{s}\left(Z\right)\mid W=1,X=x\right)\\
 & = & \Var\left(\ddot{P}\right)^{-1}\Cov\left(\ddot{P},D\mid W=1,X=x,S=s\right)
\end{eqnarray*}
is non-negative diagonal for all $x$ and $s$. The second equation
uses $Z\perp W\mid S,X$. Thus,
\begin{eqnarray*}
\E\left[w^{S,X}\right] & = & \Var\left(\ddot{P}\right)^{-1}\E\left[\Cov\left(\ddot{P},D\mid W=1,X,S\right)\right]\\
 & = & \Var\left(\ddot{P}\right)^{-1}\Cov\left(\ddot{P},D\mid W=1\right)\\
 & = & \Var\left(\ddot{P}\right)^{-1}\Cov\left(\ddot{P},\ddot{D}\mid W=1\right)
\end{eqnarray*}
is non-negative diagonal.\footnote{\vspace{-0.4cm}
\begin{eqnarray*}
\E\left[\Cov\left(\ddot{P},D\mid W=1,X,S\right)\right] & = & \E\left[\E\left[\ddot{P}D\mid W=1,X,S\right]\right]-\E\left[\E\left[\ddot{P}\mid W=1,X,S\right]\E\left[D\mid W=1,X,S\right]\right]\\
 & = & \E\left[\ddot{P}D\mid W=1\right]-\E\left[\E\left[\ddot{P}\mid X\right]\E\left[D\mid W=1,S\right]\right]\\
 & = & \E\left[\ddot{P}D\mid W=1\right]-0\\
 & = & \E\left[\ddot{P}D\mid W=1\right]-\E\left[\ddot{P}\mid W=1\right]\E\left[D\mid W=1\right]\\
 & = & \Cov\left(\ddot{P},D\mid W=1\right)
\end{eqnarray*}
where the second equality uses $\ddot{P}\perp S,W\mid X$ and the
third equality uses that $\E\left[\ddot{P}\mid X\right]=0$..}
\end{proof}
\begin{proof}
(Proposition \ref{prop:binaryx}). Under Assumptions \ref{assu:ivx}\textendash \ref{assu:flexible_interaction},
we have
\begin{eqnarray*}
w^{s,x} & = & \Var\left(\ddot{P}\right)^{-1}\Cov\left(\ddot{P},v_{s}\left(Z\right)\mid X=x\right)\\
 & = & a_{x}\Var\left(\ddot{P}\mid X=x\right)^{-1}\Cov\left(\ddot{P},v_{s}\left(Z\right)\mid X=x\right)\\
 & = & a_{x}\Var\left(P_{X}+E\left[D\mid X\right]\mid X=x\right)^{-1}\Cov\left(P_{X}+E\left[D\mid X\right],v_{s}\left(Z\right)\mid X=x\right)\\
 & = & a_{x}\Var\left(P_{X}\mid X=x\right)^{-1}\Cov\left(P_{X},v_{s}\left(Z\right)\mid X=x\right)
\end{eqnarray*}
where $v_{s}\left(Z\right)\equiv\left[\begin{array}{ccc}
s_{1}\left(Z\right) & \dots & s_{n}\left(Z\right)\end{array}\right]^{\prime}$. The result then follows by an application of Proposition \ref{prop:binary}
on each subsample $X=x$. The permutation $f$ must be the same across
values of $X$ since Assumption \ref{assu:flexible_interaction} requires
that $\E\left[D\mid Z,X\right]=\gamma Z+\phi_{X}$, i.e., the relationship
between $Z$ and $D$ can not differ across $X$.
\end{proof}
\begin{proof}
(Proposition \ref{prop:just-identified-test-x}). As shown in the
proof of Proposition \ref{prop:binaryx}, we have
\begin{eqnarray*}
w^{s,x} & = & a_{x}\Var\left(P_{X}\mid X=x\right)^{-1}\Cov\left(P_{X},v_{s}\left(Z\right)\mid X=x\right)
\end{eqnarray*}
under Assumptions \ref{assu:ivx}\textendash \ref{assu:flexible_interaction}
where $v_{s}\left(Z\right)\equiv\left[\begin{array}{ccc}
s_{1}\left(Z\right) & \dots & s_{n}\left(Z\right)\end{array}\right]^{\prime}$. The result then follows by applying Proposition \ref{prop:just-identified-test}
on each subsample $X=x$. The permutation $f$ must be the same across
values of $X$ since Assumption \ref{assu:flexible_interaction} requires
that $\E\left[D\mid Z,X\right]=\gamma Z+\phi_{X}$, i.e., the relationship
between $Z$ and $D$ can not differ across $X$.

\end{proof}
\begin{proof}
(Proposition \ref{prop:threshold-crossing-x}). As in the proof of
Proposition \ref{prop:binaryx}, Assumptions \ref{assu:ivx}\textendash \ref{assu:flexible_interaction}
imply
\begin{eqnarray*}
w^{s,x} & = & a_{x}\Var\left(P_{X}\mid X=x\right)^{-1}\Cov\left(P_{X},v_{s}\left(Z\right)\mid X=x\right)
\end{eqnarray*}
where $v_{s}\left(Z\right)\equiv\left[\begin{array}{ccc}
s_{1}\left(Z\right) & \dots & s_{n}\left(Z\right)\end{array}\right]^{\prime}$. The result then follows by applying Proposition \ref{prop:threshold-crossing}
on each subsample $X=x$.
\end{proof}
\begin{proof}
(Proposition \ref{prop:IA}). To see that joint monotonicity is not
sufficient for Assumption \ref{assumption:zero}, consider the following
example. In the model of Section \ref{sec:binary} with unordered
treatment effects, assume there are only two response types, $s$
and $s'$, defined by
\[
s\left(v\right)=\begin{cases}
0 & v=0\\
1 & v=1\\
2 & v=2
\end{cases}\text{ \,\,\,\,\,\,\,\,\,\,\,\,\,\,\,\,\,\,\,\,\, }s'\left(v\right)=\begin{cases}
0 & v=0\\
2 & v=1\\
2 & v=2
\end{cases}
\]
In a population composed of $s$ and $s'$, joint monotonicity is
trivially satisfied. But the no-cross effect condition (Assumption
\ref{assumption:zero}) is not satisfied. Instrument 2 affects both
treatment $1$ and treatment $2$ violating the conditions in Proposition
\ref{prop:binary}. In particular, response type $s'$ is not among
the allowed response types in Corollary \ref{cor:binary_2}.

To see that joint monotonicity does not imply Assumption \ref{assumption:non_neg},
assume $Z=\left\{ Z_{1},Z_{2}\right\} $ is composed of two continuous
instruments taking values between $0$ and $\bar{z}$. For simplicity,
assume only $Z_{1}$ affects treatment $1$ and $P_{1}\perp P_{2}$.
Assume the linear relationship between $Z_{1}$ and $D_{1}$ is increasing
but the highest propensity to take up treatment $1$ occurs at $Z_{1}=0$.
In other words, the linear relationship between the instrument and
treatment imposed by the first stage is misspecified. Then a response
type that selects $D_{1}$ if and only if $Z_{1}=0$ satisfies joint
monotonicity but violates average conditional monotonicity.\footnote{When $P_{1}\perp P_{2}$, average conditional monotonicity with respect
to treatment $1$ reduces to $\Cov\left(P_{1},s_{1}\left(Z\right)\right)\geq0$
which is violated in this case.}

To prove ii), consider the case when Assumption \ref{assu:first-stage-correct}
holds and $P_{1}\perp P_{2}$. Joint monotonicity then requires that
the selection of treatment $1$ is strictly non-decreasing in $P_{1}$
for all agents. But Assumption \ref{assumption:non_neg} is weaker
and only requires the selection of treatment $1$ to have a non-negative
\emph{correlation} with $P_{1}$ for all agents.
\end{proof}
\begin{proof}
(Proposition \ref{prop:linear-predicted}). When Assumption \ref{assu:first-stage-correct}
holds, $P_{k}$\textemdash the best linear prediction of $D_{k}$
given $Z$\textemdash equals $\E\left[D_{k}\mid Z\right]$, the propensity
of crossing threshold $k$ given $Z$. For each response type $s$
there must then exist a threshold $u_{ks}\in\left[0,1\right]$ such
that $P_{k}\geq u_{ks}\Leftrightarrow s_{k}\left(Z\right)=1$. By
Lemma \ref{lem:n_equals_m}, 2SLS assigns proper weights if and only
if
\[
\frac{\Cov\left(P_{k},s_{l}\left(Z\right)\right)}{\Cov\left(P_{k},D_{l}\right)}=\frac{\Cov\left(P_{l},s_{l}\left(Z\right)\right)}{\Cov\left(P_{l},D_{l}\right)}\geq0
\]
for all $l$ and $k$. Since 
\[
\Cov\left(P_{k},s_{l}\left(Z\right)\right)=\left(\E\left[P_{k}\mid P_{l}\geq u_{ls}\right]-\E\left[P_{k}\mid P_{l}<u_{ls}\right]\right)\Pr\left[s_{l}\left(Z\right)=1\right]\Pr\left[s_{l}\left(Z\right)=0\right]
\]
this is equivalent to
\begin{equation}
\frac{\E\left[P_{k}\mid P_{l}\geq u_{ls}\right]-\E\left[P_{k}\mid P_{l}<u_{ls}\right]}{\Cov\left(P_{k},D_{l}\right)}=\frac{\E\left[P_{l}\mid P_{l}\geq u_{ls}\right]-\E\left[P_{l}\mid P_{l}<u_{ls}\right]}{\Cov\left(P_{l},D_{l}\right)}\geq0\label{eq:proper}
\end{equation}

We want to show that this condition is equivalent to
\begin{equation}
\E\left[P_{k}\mid P_{l}\right]=\gamma_{kl}+\delta_{kl}P_{l}\label{eq:linear}
\end{equation}
for constants $\gamma_{kl}$ and $\delta_{kl}.$ First, assume Equation
\ref{eq:linear}. We then have\footnote{Using $\E\left[P_{k}\mid P_{l}\geq u_{ls}\right]=\gamma_{kl}+\delta_{kl}\E\left[P_{l}\mid P_{l}\geq u_{ls}\right]$
and $\E\left[P_{k}\mid P_{l}<u_{ls}\right]=\gamma_{kl}+\delta_{kl}\E\left[P_{l}\mid P_{l}<u_{ls}\right]$.}
\[
\E\left[P_{k}\mid P_{l}\geq u_{ls}\right]-\E\left[P_{k}\mid P_{l}<u_{ls}\right]=\delta_{kl}\left(\E\left[P_{l}\mid P_{l}\geq u_{ls}\right]-\E\left[P_{l}\mid P_{l}<u_{ls}\right]\right)
\]
and\footnote{\vspace{-0.5cm}
\begin{eqnarray*}
\Cov\left(P_{k},s_{l}\left(Z\right)\right) & = & \delta_{kl}\left(\E\left[P_{l}\mid P_{l}\geq u_{ls}\right]-\E\left[P_{l}\mid P_{l}<u_{ls}\right]\right)\Pr\left[s_{l}\left(Z\right)=1\right]\Pr\left[s_{l}\left(Z\right)=0\right]\\
 & = & \delta_{kl}\Cov\left(P_{l},s_{l}\left(Z\right)\right)
\end{eqnarray*}
}
\begin{eqnarray*}
\frac{\Cov\left(P_{k},D_{l}\right)}{\Cov\left(P_{l},D_{l}\right)} & = & \frac{\E\left[\Cov\left(P_{k},S_{l}\left(Z\right)\mid S\right)\right]}{\E\left[\Cov\left(P_{l},S_{l}\left(Z\right)\mid S\right)\right]}=\delta_{kl}
\end{eqnarray*}
Thus Equation \ref{eq:proper} holds. To prove the opposite direction
of the equivalence, assume Equation \ref{eq:proper} holds. Rearranging
Equation \ref{eq:proper} gives
\[
\E\left[P_{k}\mid P_{l}\geq u_{ls}\right]=\gamma+\delta\E\left[P_{l}\mid P_{l}\geq u_{ls}\right]
\]
where $\gamma\equiv\E\left[P_{k}\right]-\delta\E\left[P_{l}\right]$
and $\delta\equiv\frac{\Cov\left(P_{k},D_{l}\right)}{\Cov\left(P_{l},D_{l}\right)}$.
Thus, defining $f_{kl}\left(p_{k},p_{l}\right)$ as the joint distribution
of $P_{k}$ and $P_{l}$ and $f_{l}\left(p_{l}\right)$ as the distribution
of $P_{l}$, we get
\[
\frac{\int_{p_{l}\leq u_{ls}}p_{k}f_{kl}\left(p_{k},p_{l}\right)dp_{k}dp_{l}}{\int_{0}^{u_{ls}}f_{l}\left(p_{l}\right)dp_{l}}=\gamma-\delta\frac{\int_{0}^{u_{ls}}p_{l}f_{l}\left(p_{l}\right)dp_{l}}{\int_{0}^{u_{ls}}f_{l}\left(p_{l}\right)dp_{l}}
\]
\[
\Leftrightarrow\int_{p_{l}\leq u_{ls}}p_{k}f_{kl}\left(p_{k},p_{l}\right)dp_{k}dp_{l}=\gamma\int_{0}^{u_{ls}}f_{l}\left(p_{l}\right)dp_{l}-\delta\int_{0}^{u_{ls}}p_{l}f_{l}\left(p_{l}\right)dp_{l}
\]
Invoking Assumption \ref{assu:all-complier-types}, we can differentiate
this equation with respect to $u_{ls}$, giving $\E\left[P_{k}\mid P_{l}=u_{ls}\right]=\gamma+\delta u_{ls}$.
Thus, Equation \ref{eq:linear} is satisfied.
\end{proof}
\begin{proof}
(Proposition \ref{prop:linear-predicted2}). This follows from the
proof of Proposition \ref{prop:linear-predicted}: Assumption \ref{assu:all-complier-types}
is used in the proof only to prove that the linearity condition is
\emph{necessary}.
\end{proof}
\begin{proof}
(Corollary \ref{prop:linear-predicted-UM}.) Assumption \ref{assu:UM}
implies that Assumption \ref{assu:mono} holds for all possible choices
for the excluded treatment.  We can thus apply Proposition \ref{prop:linear-predicted2}
for each choice for the excluded treatment.
\end{proof}
\begin{proof}
(Proposition \ref{prop:kirkeboen}). First, assume the conditions
in the Proposition are true and fix a pair $\left\{ s,k\right\} $
for which neither $s_{k}\left(v\right)=0$ for all $v$ nor $s_{k}\left(v\right)=1$
for all $v$ ($s$ is neither an always-$k$-taker nor a never-$k$-taker).
We need to show that $s\left(k\right)=k$. By Condition 3, we must
have $s\left(0\right)=0$. If $s\left(k\right)=l$ with $l\neq k$
we must have, by Condition 2 (irrelevance), that $s_{l}\left(k\right)=s_{l}\left(0\right)=0\Rightarrow s\left(k\right)\neq l$,
a contradiction. Thus $s\left(k\right)=k$. The reverse direction
of the equivalence is straightforward to verify.
\end{proof}

\section{Appendix: Additional Results}

\global\long\def\thetable{B.\arabic{table}}%
\setcounter{table}{0}
\global\long\def\thefigure{B.\arabic{figure}}%
\setcounter{figure}{0}
\global\long\def\theequation{B.\arabic{equation}}%
\setcounter{equation}{0}
\setcounter{thm}{0}
\setcounter{cor}{0}
\setcounter{prop}{0}
\setcounter{lem}{0}
\setcounter{assumption}{0}
\setcounter{example}{0}
\counterwithin{thm}{section}
\counterwithin{cor}{section}
\counterwithin{prop}{section}
\counterwithin{prop}{lem}
\counterwithin{assumption}{section}
\counterwithin{example}{section}
\renewcommand{\thethm}{\thesection.\arabic{thm}}
\renewcommand{\theprop}{\thesection.\arabic{prop}}
\renewcommand{\thecor}{\thesection.\arabic{cor}}
\renewcommand{\thelem}{\thesection.\arabic{lem}}
\renewcommand{\theassumption}{\thesection.\arabic{assumption}}
\renewcommand{\theexample}{\thesection.\arabic{example}}

\subsection{Relationship to Weak Causality\label{subsec:WC}}

\citet{blandhol2022tsls} call an estimand \emph{weakly causal }if
it has the correct sign when the treatment effects have the same sign
across all agents. In our setting, 2SLS is weakly causal if $\beta_{k}^{2SLS}\geq0$
($\leq0$) whenever $\beta_{k}^{s}\geq0$ ($\leq0$) for all $s\in\mathcal{S}$
for $k\in\left\{ 1,2\right\} $. In this section, we show that 2SLS
assigns proper weights if and only if 2SLS is weakly causal for all
possible heterogeneous effects.\footnote{The proof is a straightforward generalization of Proposition 4 in
\citet{blandhol2022tsls}.}

First, assume 2SLS is weakly causal for all possible heterogeneous
effects. Assume the causal effect of treatment $1$ for response type
$s$ is $\beta_{1}^{s}=1$ and all other causal effects are $0$.
Weak causality then requires that $\beta_{1}^{\text{2SLS}}$ is non-negative.
Since, by Proposition \ref{prop:weights}, $\beta_{1}^{\text{2SLS}}=\E\left[w_{11}^{S}\beta_{1}^{S}+w_{12}^{S}\beta_{2}^{S}\right]=w_{11}^{s}\Pr\left[S=s\right]$
in this case, weak causality implies that $w_{11}^{s}$ is non-negative.
Similarly, weak causality requires that the 2SLS estimand for treatment
$2$ is zero.\footnote{Weak causality requires that the 2SLS estimand is non-negative (non-positive)
if the effect for all response types is non-negative (non-positive).
When the treatment effect is zero (both non-negative and non-positive)
for all response types, the 2SLS estimand is then required to be zero
(both non-negative and non-positive).} Since, by Proposition \ref{prop:weights} applied to $\beta_{2}^{\text{2SLS}}$,
we have $\beta_{2}^{\text{2SLS}}=\E\left[w_{21}^{S}\beta_{1}^{S}+w_{22}^{S}\beta_{2}^{S}\right]=w_{21}^{s}\Pr\left[S=s\right]$,
weak causality implies that $w_{21}^{s}$ is zero. The argument can
be repeated for all response types and for the remaining two elements
of the weight matrix. Thus, if 2SLS is weakly causal for all possible
heterogeneous effects, we must have that the weight matrix $w^{s}$
is non-negative diagonal for all response types $s$. The proof that
proper weights imply weak causality is trivial.

\subsection{Generalization to More Than Three Treatments\label{subsec:generalization}}

In this section, we show how our results generalize to the case with
more than three treatments. Assume there are $n+1$ treatments: $T\in\mathcal{T}\equiv\left\{ 0,1,\dots,n\right\} $.
Otherwise, the notation remains the same. In particular, the treatment
effect of treatment $k$ relative to treatment $l$ is given by $Y\left(k\right)-Y\left(l\right)$.
Using 2SLS, we can seek to estimate $n$ of these treatment effects
denoted by the random vector $\beta\equiv\left(\delta_{1},\delta_{2},\dots,\delta_{n}\right)^{T}$.
For instance, we might be interested in \emph{unordered }treatment
effects\textemdash the causal effect of each treatment $T>0$ compared
to the \emph{excluded treatment $T=0$. }In that case, we choose $\delta_{k}=Y\left(k\right)-Y\left(0\right)$.
Alternatively, we might be interested in estimating \emph{ordered}
treatment effects\textemdash the casual effect of treatment $T=k$
compared to treatment $T=k-1$\textemdash in which case we define
$\delta_{k}=Y\left(k\right)-Y\left(k-1\right)$.

\begin{figure}[H]
\caption{Treatment Effects Corresponding to Different 2SLS Specifications.}

\begin{centering}
\label{treatment_network}\subfloat[Unordered Treatment Effects.\bigskip{}
]{
\centering{}\includegraphics[width=0.27\textwidth]{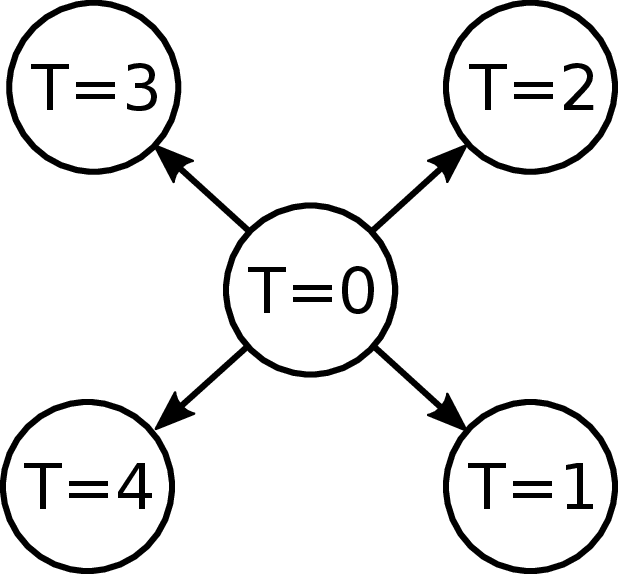}}
\par\end{centering}
\begin{centering}
\subfloat[Ordered Treatment Effects.\bigskip{}
]{
\centering{}\includegraphics[width=0.52\textwidth]{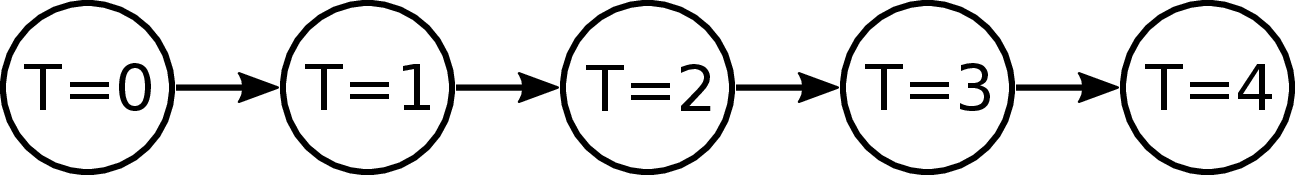}}
\par\end{centering}
\begin{centering}
\subfloat[Generic Example.\bigskip{}
]{
\centering{}\includegraphics[width=0.4\textwidth]{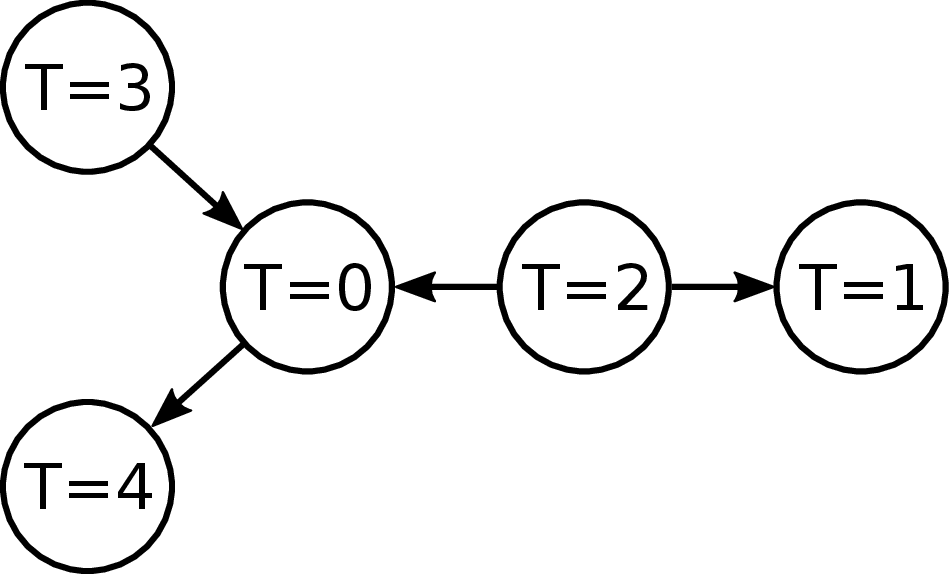}}
\par\end{centering}
\begin{singlespace}
{\scriptsize{}\medskip{}
}{\scriptsize\par}
\end{singlespace}

\emph{\scriptsize{}Note: }{\scriptsize{}Three possible choices for
the set of treatment effects of interest $\beta$. An arrow from $T=k$
to $T=l$ means that the researcher is interested in the treatment
effect $Y\left(k\right)-Y\left(l\right)$. In Figure (a), the treatment
effects of interest are $\delta_{k}=Y\left(k\right)-Y\left(0\right)$.
The corresponding treatment indicators are $D_{k}=\mathbf{1}\left[T=k\right]$.
In Figure (b), the treatment effects of interest are $\delta_{k}=Y\left(k\right)-Y\left(k-1\right)$
and the corresponding treatment indicators are $D_{k}=\mathbf{1}\left[T\geq k\right]$.
In Figure (c),
\[
\beta=\left[\begin{array}{c}
Y\left(1\right)-Y\left(2\right)\\
Y\left(0\right)-Y\left(2\right)\\
Y\left(0\right)-Y\left(3\right)\\
Y\left(4\right)-Y\left(0\right)
\end{array}\right]\text{ and }D=\left[\begin{array}{c}
\mathbf{1}\left[T=1\right]\\
\mathbf{1}\left[T\in\left\{ 0,3,4\right\} \right]\\
\mathbf{1}\left[T\in\left\{ 0,1,2,4\right\} \right]\\
\mathbf{1}\left[T=4\right]
\end{array}\right].
\]
}{\scriptsize\par}
\end{figure}

Once we have determined the treatment effects of interest, we define
corresponding treatment indicators $D\equiv\left\{ D_{1},\dots,D_{n}\right\} $
where $D_{k}=\mathbf{1}\left[T\in R_{k}\right]$ for a subset $R_{k}\subset\left\{ 0,\dots,n\right\} $.
The subsets $R_{k}$ are chosen by the researcher to correspond to
the treatment effects of interest. In particular, the sets $R_{k}$
are selected such that $Y=\alpha+D\beta$ where $\alpha$ is a function
of $\left\{ Y\left(0\right),Y\left(1\right),\dots,Y\left(n\right)\right\} $.
For instance, if $\delta_{k}=Y\left(k\right)-Y\left(0\right)$ we
define $D_{k}=\mathbf{1}\left[T=k\right]$ (i.e., $R_{k}=\left\{ k\right\} $)
such that $Y=Y\left(0\right)+D\beta$. Alternatively, if $\delta_{k}=Y\left(k\right)-Y\left(k-1\right)$
we define $D_{k}=\mathbf{1}\left[T\geq k\right]$ (e.g., $R_{k}=\left\{ k,k+1,\dots,n\right\} )$.
In Figure \ref{treatment_network}, we show these two special cases
and a third generic case as directed graphs where an arrow from $T=l$
to $T=k$ indicates that we are interested in estimating the treatment
effect $Y\left(k\right)-Y\left(l\right)$. By using different treatment
indicators one can estimate any subset of $n$ treatment effects that
connects the treatments in an acyclic graph.\footnote{Formally, when $\delta_{k}=Y\left(l\right)-Y\left(m\right)$, one
can define $D_{k}=\mathbf{1}\left[T\in R_{k}\right]$ where $R_{k}$
is the set of all treatments that are connected to $m$ through $l$
in the graph of treatment effects.}

For a response type $s\in\mathcal{S}$, define by $v_{s}$ the induced
mapping between instruments and treatment indicators: $v_{s}\left(z\right)\equiv\left(s_{1}\left(z\right),\dots,s_{n}\left(z\right)\right)^{T}$
for all $z\in\mathcal{Z}$ with $s_{k}\left(z\right)\equiv\mathbf{1}\left[s\left(z\right)\in R_{k}\right]$.
Proposition \ref{prop:weights} generalizes directly to this setting
with more than three treatments\textemdash the only required modification
to the proof is to replace $Y\left(0\right)$ with $\alpha$. Corollaries
\ref{cor:non_neg}\textendash \ref{cor:zero} generalize as follows:
The weight on $\beta_{sk}$ in $\beta_{k}^{2SLS}$ is non-negative
for all $s$ if and only if, for all $s$, the partial correlation
between $P_{k}$ and $s_{k}\left(Z\right)$ given $\left(P_{1},\dots,P_{k-1},P_{k+1},\dots,P_{n}\right)^{T}$
is non-negative. And the weight on $\beta_{sl}$ in $\beta_{k}^{2SLS}$
is zero for all $s$ and $l\neq k$ if and only if, for all $s$ and
$l\neq k$, the partial correlation between $P_{k}$ and $s_{l}\left(Z\right)$
given $\left(P_{1},\dots,P_{k-1},P_{k+1},\dots,P_{n}\right)^{T}$
is zero.

\subsubsection{Generalization of Proposition \ref{prop:kitagawa} to Alternative
Treatment Indicators\label{subsec:kitagawa-generalization}}

To see how Proposition \ref{prop:kitagawa} generalizes to treatment
indicators beyond $D_{k}=\mathbf{1}\left[T=k\right]$, consider the
case with four treatments and ordered treatment indicators: $D_{1}=\mathbf{1}\left[T\geq1\right]$,
$D_{2}=\mathbf{1}\left[T\geq2\right]$, and $D_{3}=\mathbf{1}\left[T=3\right]$.
From these indicators, we can \emph{reconstruct} the unordered treatment
indicators as follows: $\mathbf{1}\left[T=1\right]=D_{1}-D_{2}$,
$\mathbf{1}\left[T=2\right]=D_{2}-D_{3}$, and $\mathbf{1}\left[T=3\right]=D_{3}$.
We can then apply the steps of the Proposition \ref{prop:kitagawa}
proof and obtain, for instance
\[
\Cov\left(P,\mathbf{1}_{y\leq Y\leq y'}\left(D_{2}-D_{3}\right)\right)=\E\left[\Cov\left(P,S_{2}\left(Z\right)-S_{3}\left(Z\right)\mid S\right)\Pr\left[y\leq Y\left(2\right)\leq y'\mid S\right]\right]
\]

By Corollary \ref{cor:diagonal}, if 2SLS assigns proper weights,
we must have
\[
\Var\left(P\right)^{-1}\left[\begin{array}{ccc}
\Cov\left(P,s_{1}\left(Z\right)\right) & \Cov\left(P,s_{2}\left(Z\right)\right) & \Cov\left(P,s_{3}\left(Z\right)\right)\end{array}\right]=\left[\begin{array}{ccc}
a_{1s} & 0 & 0\\
0 & a_{2s} & 0\\
0 & 0 & a_{3s}
\end{array}\right]
\]
for all $s\in\mathcal{S}$ and some $a_{1s},a_{2s},a_{3s}\geq0$.
This equation implies
\[
\Var\left(P\right)^{-1}\left[\begin{array}{ccc}
\Cov\left(P,s_{1}\left(Z\right)-s_{2}\left(Z\right)\right) & \Cov\left(P,s_{2}\left(Z\right)-s_{3}\left(Z\right)\right) & \Cov\left(P,s_{3}\left(Z\right)\right)\end{array}\right]
\]
\[
=\left[\begin{array}{ccc}
a_{1s} & 0 & 0\\
-a_{2s} & a_{2s} & 0\\
0 & -a_{3s} & a_{3s}
\end{array}\right]
\]

Thus, if
\[
\Var\left(P\right)^{-1}\left[\begin{array}{ccc}
\Cov\left(P,\mathbf{1}_{y\leq Y\leq y'}\left(D_{1}-D_{2}\right)\right) & \Cov\left(P,\mathbf{1}_{y\leq Y\leq y'}\left(D_{2}-D_{3}\right)\right) & \Cov\left(P,\mathbf{1}_{y\leq Y\leq y'}D_{3}\right)\end{array}\right]
\]
\[
=\left[\begin{array}{ccc}
c_{11} & c_{12} & c_{13}\\
c_{21} & c_{22} & c_{23}\\
c_{31} & c_{32} & c_{33}
\end{array}\right]
\]
we get the following testable implications: $c_{11}\geq0$, $c_{12}=0$,
$c_{13}=0$, $c_{21}+c_{22}=0$, $c_{22}\geq0$, $c_{31}=0$, $c_{32}+c_{33}=0$,
and $c_{33}\geq0$.The Just-Identified Case with More than Three Treatments

The proof of Proposition \ref{prop:binary} is stated in terms of
$n$ treatments and the result thus trivially extends to more than
three treatments. To interpret Proposition \ref{prop:binary} with
more than three treatments, introduce the following terminology: For
each $k$, call response type $s$ an \emph{always-$k$-taker} if
$s_{k}\left(v\right)=1$ for all $v$, a \emph{never-$k$-taker} if
$s_{k}\left(v\right)=0$ for all $v$, and a \emph{$k$-complier}
if $s_{k}\left(v\right)=1\Leftrightarrow v=k$. A $k$-complier thus
selects treatment $k$ only when instrument $k$ is turned on.
\begin{cor}
\label{cor:binary}Under Assumption \ref{assu:order}, 2SLS assigns
proper weights if and only if, for all $k\neq0$, all agents are either
always-$k$-takers, never-$k$-takers, or $k$-compliers.
\end{cor}
\begin{table}[th]
\centering{}\caption{Allowed Response Types in the Just-Identified Case with Four Treatments.\label{tab:response_types4}}
\begin{tabular}{cc}
\hline 
Response Type & $\left(s\left(0\right),s\left(1\right),s\left(2\right),s\left(3\right)\right)$\tabularnewline
\hline 
\hline 
Never-taker & $\left(0,0,0,0\right)$\tabularnewline
Always-1-taker & $\left(1,1,1,1\right)$\tabularnewline
Always-2-taker & $\left(2,2,2,2\right)$\tabularnewline
Always-3-taker & $\left(3,3,3,3\right)$\tabularnewline
1-complier & $\left(0,1,0,0\right)$\tabularnewline
2-complier & $\left(0,0,2,0\right)$\tabularnewline
3-complier & $\left(0,0,0,3\right)$\tabularnewline
1+2-complier & $\left(0,1,2,0\right)$\tabularnewline
1+3-complier & $\left(0,1,0,3\right)$\tabularnewline
2+3-complier & $\left(0,0,2,3\right)$\tabularnewline
Full complier & $\left(0,1,2,3\right)$\tabularnewline
\hline 
\end{tabular}
\end{table}

In words, only instrument $k$ can influence treatment $k$. For instance,
in Table \ref{tab:response_types4} we show all allowed response types
in the case with four treatments. In the case of unordered treatments
effects, $D_{k}=\mathbf{1}\left[T=k\right]$, treatment $0$ must
play a special role: Unless response type $s$ always selects the
same treatment, we must have either $s\left(k\right)=k$ or $s\left(k\right)=0$,
for all $k$. If instrument $k$ is an inducement to take up treatment
$k$, an agent can thus never select any other treatment $l\notin\left\{ 0,k\right\} $
when induced to take up treatment $k$, unless the agent always selects
treatment $l$.

\subsection{Multivariate 2SLS with Covariates\label{subsec:controls}}

In this section, we show how our results generalize to 2SLS with multiple
treatments and control variables. In particular, assume we have access
to a vector of controls $X$ with finite support such that the instruments
$Z$ are exogenous only conditional on $X$:\footnote{We consider the case with $n+1$ treatments ($T\in\left\{ 0,\dots,n\right\} $)
also in this section.}
\begin{assumption}
\label{assu:ivx}(Conditional Exogeneity and Exclusion). $\left\{ Y\left(0\right),\dots,Y\left(n\right),S\right\} \perp Z\mid X$
\end{assumption}
We will consider the case of running 2SLS while flexibly controlling
for $X$ using fully saturated fixed effects.\footnote{ The results are also valid for the case where $X$ is only controlled
for linearly, provided that $\E\left[Y\mid X\right]$, $\E\left[D\mid X\right]$,
and $\E\left[Z\mid X\right]$ are indeed linear in $X$. If these
relationships are not linear, 2SLS with linear controls for $X$ might
fail to assign proper weights even with a binary treatment \citep{blandhol2022tsls}.} In particular, define the demeaned variables $\ddot{Y}=Y-\E\left[Y\mid X\right]$,
$\ddot{D}=D-\E\left[D\mid X\right]$, and $\ddot{Z}=Z-\E\left[Z\mid X\right]$.
The 2SLS estimand, controlling for $X$-specific fixed effects, is
then $\beta^{\text{2SLS}}\equiv\Var\left(\ddot{P}\right)^{-1}\Cov\left(\ddot{P},\ddot{Y}\right)$
where $\ddot{P}\equiv\Var\left(\ddot{Z}\right)^{-1}\Cov\left(\ddot{Z},\ddot{D}\right)\ddot{Z}$
is the linear projection of $\ddot{D}$ on $\ddot{Z}$. This estimand
is equivalent to the one obtained from 2SLS while controlling for
indicator variables for each possible value of $X$ in the first and
the second stage. Let $\mathcal{S}_{x}$ be the set of all response
types $s$ such that $\Pr\left[S=s\mid X=x\right]>0$. We maintain
the following assumption.
\begin{assumption}
\label{assu:rankx}(Rank). $\Cov\left(\ddot{Z},\ddot{D}\right)$ has
full rank.
\end{assumption}
Proposition \ref{prop:weights} then generalizes as follows:\footnote{The result is stated in the same way as Lemma \ref{lem:weights}.
One could obtain a statement more similar to Proposition \ref{prop:weights}
by applying the Frisch-Waugh-Lovell theorem.}
\begin{prop}
\label{prop:weightsx}Under Assumptions \ref{assu:ivx} and \ref{assu:rankx}
\[
\beta^{2SLS}=\E\left[w^{S,X}\beta^{S,X}\right]
\]
where, for $s\in\mathcal{S}_{x}$ and $x\in X$
\[
w^{s,x}\equiv\Var\left(\ddot{P}\right)^{-1}\Cov\left(\ddot{P},v_{s}\left(Z\right)\mid X=x\right)
\]
\[
\beta^{s,x}\equiv\left(\beta_{1}^{s,x},\dots,\beta_{n}^{s,x}\right)^{T}\equiv\E\left[\beta\mid S=s,X=x\right]
\]
with $v_{s}\left(Z\right)\equiv\left[\begin{array}{ccc}
s_{1}\left(Z\right) & \dots & s_{n}\left(Z\right)\end{array}\right]^{\prime}$.
\end{prop}
Thus, 2SLS with controls assigns proper weights if and only if $w^{s,x}$
is a non-negative diagonal matrix for all $s\in\mathcal{S}_{x}$ and
$x\in X$. This condition is, in general, hard to interpret. But the
condition has a straightforward interpretation under the following
assumption:
\begin{assumption}[Predicted treatments covary similarly across $X$]
\label{assu:covary_similarly} 
\[
\Var\left(\ddot{P}\mid X=x\right)=a_{x}\Var\left(\ddot{P}\right)\text{ for all \ensuremath{x\in\mathcal{X}} and constants \ensuremath{a_{x}>0}.}
\]
\end{assumption}
Then the condition reduces to Assumption \ref{assumption:non_neg}
and \ref{assumption:zero} holding conditional on $X$:\footnote{I.e., to $\Var\left(\ddot{P}\mid X=x\right)^{-1}\Cov\left(\ddot{P},v_{s}\left(Z\right)\mid X=x\right)$
being a non-negative diagonal matrix for all $s\in\mathcal{S}_{x}$
and $x\in X$.} The 2SLS estimand of the effect of treatment $k$ assigns proper
weights if and only if, for all agents, there is a non-negative (zero)
partial correlation between predicted treatment $k$ and potential
treatment $k$ ($l\neq k$) \emph{conditional on $X$}. For instance,
consider a random judge design where random assignment of cases to
judges holds only within courts, and we are interested in separately
estimating the effect of incarceration and conviction on defendants'
future outcomes. Assume incarceration rates and conviction rates covary
in a similar way across courts.\footnote{In this setting, $\Var\left(\ddot{P}\mid X=x\right)=a_{x}\Var\left(\ddot{P}\right)$
is equivalent to (i) the correlation between incarceration and conviction
rates between judges not varying across courts and (ii) the ratio
between the standard deviation of incarceration rates and the standard
deviation of conviction rates not varying across courts.} Then, the 2SLS estimand of the effect of incarceration assigns proper
weights if and only if the following holds for each case $i$ \emph{among
the judges sitting in the court where the case is filed} (i.e., conditional
on $X$):\footnote{Using the assigned judge as the instrument, the predicted treatments
are the judges' incarceration and conviction rates. In applied work,
these rates are typically estimated by leave-one-out estimates to
avoid small-sample bias.}
\begin{enumerate}
\item There is a non-negative correlation between incarerating in case $i$
and the judge's incarceration rate conditional on the judge's conviction
rate,
\item There is no correlation between convicting in case $i$ and the judge's
incarceration rate conditional on the judge's conviction rate.
\end{enumerate}
Note that 2SLS could assign proper weights even though Assumptions
\ref{assumption:non_neg} and \ref{assumption:zero} do not hold conditional
on $X$: It might be that the bias from such violations are counteracted
by bias coming from predicted treatments covarying differently across
values of $X$. This case, however, seems hard to characterize in
an interpretable way. Also, note that Assumption \ref{assu:covary_similarly}
only depends on population moments and can thus be easily tested.\footnote{For instance, in the Section \ref{sec:application} application, the
sample analogs of $\Var\left(\ddot{P}\right)$ and $\Var\left(\ddot{P}\mid X=x\right)$
for the court-by-year cell $x$ with the largest number of cases are
\[
\widehat{\Var}\left(\ddot{P}\right)=\left[\begin{array}{cc}
0.000017 & 0.000054\\
0.000054 & 0.000653
\end{array}\right]\,\,\text{and}\,\,\widehat{\Var}\left(\ddot{P}\mid X=x\right)=\left[\begin{array}{cc}
0.000043 & 0.00013\\
0.00013 & 0.0016
\end{array}\right]
\]
There are no $a_{x}$ such that $\widehat{\Var}\left(\ddot{P}\mid X=x\right)=a_{x}\widehat{\Var}\left(\ddot{P}\right)$.
A statistical test for contamination bias would, for instance, test
whether the non-diagonal elements of 
\[
\widehat{\Var}\left(\ddot{P}\right)^{-1}\widehat{\Var}\left(\ddot{P}\mid X=x\right)=\left[\begin{array}{cc}
2.6 & -0.58\\
-0.016 & 2.6
\end{array}\right]
\]
are zero using bootstrapped standard errors. As described below, the
estimated relative weight $\beta_{1}^{\text{2SLS}}$ assigns to $\beta_{2}^{s,x}$
compared to $\beta_{1}^{s,x}$ is given by $-0.016/2.6=0.006$, indicating
negligible contamination bias. On the other hand, the point estimates
suggest that $\beta_{2}^{\text{2SLS}}$ is $0.58/2.6=22\%$ ``contaminated''
by $\beta_{1}^{s,x}$ for this $X$-cell.}

It is useful to contrast 2SLS with multiple treatments and covariates
to ordinary least squares (OLS) with multiple treatments and covariates.
\citet{goldsmith2022contamination} show that OLS with multiple treatments
and controls is generally affected by ``contamination bias''\textemdash the
estimated effect of one treatment being contaminated by the effects
of other treatments. OLS can be seen as a special case of 2SLS with
$Z=D$. Proposition \ref{prop:weightsx} then implies that there is
no contamination bias in OLS whenever $\Var\left(\ddot{D}\mid X=x\right)=a_{x}\Var\left(\ddot{D}\right)$
for constants $a_{x}>0$.\footnote{In the OLS case, we get $w^{s,x}=\Var\left(\ddot{D}\right)^{-1}\Var\left(\ddot{D}\mid X=x\right)$.}
Since $D$ is restricted to be either zero or one, this condition
can only hold when $\E\left[D\mid X\right]=\E\left[D\right]$\textemdash when
treatment indicators are mean independent of $X$.\footnote{The condition implies that, for $x_{1},x_{2}\in\mathcal{X}$, we must
have $\Var\left(D\mid X=x_{1}\right)=a_{x_{1}x_{2}}\Var\left(D\mid X=x_{2}\right)$
for a constant $a_{x_{1}x_{2}}.$ In other words, we must have 
\[
\E\left[D_{k}\mid X=x_{1}\right]\left(1-\E\left[D_{k}\mid X=x_{1}\right]\right)=a_{x_{1}x_{2}}\E\left[D_{k}\mid X=x_{2}\right]\left(1-\E\left[D_{k}\mid X=x_{2}\right]\right)
\]
\[
\E\left[D_{k}\mid X=x_{1}\right]\E\left[D_{l}\mid X=x_{1}\right]=a_{x_{1}x_{2}}\E\left[D_{k}\mid X=x_{2}\right]\E\left[D_{l}\mid X=x_{2}\right]
\]
 for all $k\geq1$ and $l\notin\left\{ k,0\right\} $. Substracting
the latter equation from the first gives
\[
\E\left[D_{k}\mid X=x_{1}\right]\E\left[D_{0}\mid X=x_{1}\right]=a_{x_{1}x_{2}}\E\left[D_{k}\mid X=x_{2}\right]\E\left[D_{0}\mid X=x_{2}\right]
\]
where $D_{0}\equiv\mathbf{1}\left[T=0\right]$. Dividing the two last
equations be each other gives
\[
\frac{\E\left[D_{l}\mid X=x_{1}\right]}{\E\left[D_{0}\mid X=x_{1}\right]}=\frac{\E\left[D_{l}\mid X=x_{2}\right]}{\E\left[D_{0}\mid X=x_{2}\right]}
\]
for all $l$. Summing this equation across all $l$ gives $\E\left[D_{0}\mid X=x_{1}\right]=\E\left[D_{0}\mid X=x_{2}\right]$
which again implies $\E\left[D_{l}\mid X=x_{1}\right]=\E\left[D_{l}\mid X=x_{2}\right]$
for all $l$.

} \citet{goldsmith2022contamination} show that this is the only case
in which OLS does not suffer from contamination bias under arbitrary
treatment effect heterogeneity. This result, however, does not generalize
to 2SLS. In the case of 2SLS, the predicted treatments $P$ are not
restricted to be either zero or one. One can easily show that it is
possible that Assumption \ref{assu:covary_similarly} is satisfied
without instruments being unconditionally independent.\footnote{For instance, consider a random judge design with two courts, $X\in\left\{ 0,1\right\} $
with two judges each. In Court 0 ($X=0)$, the two judges have conviction
and incarceration rates of $\left(0.6,0.3\right)$ and $\left(0.7,0.2\right)$.
In Court 1 ($X=1$), the two judges have conviction and incarceration
rates of $\left(0.5,0.4\right)$ and $\left(0.8,0.1\right)$. Then
$\Var\left(\ddot{P}\mid X=1\right)=9\Var\left(\ddot{P}\mid X=0\right)$
where $\ddot{P}$ is the vector of the (demeaned) conviction and incarceration
propensities of the randomly assigned judge. Had incarceration and
conviction rates been forced to be either zero or one\textemdash as
in OLS\textemdash it would not have been possible to construct such
an example.} It is thus possible for 2SLS with multiple treatments and covariates
to not suffer from contamination bias even though the instruments
are independent only when conditioning on the controls.

If Assumptions \ref{assumption:non_neg} and \ref{assumption:zero}
hold conditional on $X$ but Assumption \ref{assu:covary_similarly}
is violated, 2SLS will suffer from contamination bias. In particular,
if Assumptions \ref{assumption:non_neg} and \ref{assumption:zero}
hold within the cell $X=x$ and 
\[
\Var\left(\ddot{P}\right)^{-1}\Var\left(\ddot{P}\mid X=x\right)=\left[\begin{array}{cc}
c_{11}^{x} & c_{21}^{x}\\
c_{12}^{x} & c_{22}^{x}
\end{array}\right]
\]
the extent that the estimated effect of $D_{1}$ is ``contaminated''
by the effect of $D_{2}$ in this cell is given by $c_{12}^{x}/c_{11}^{x}$.\footnote{Assume 
\[
\Var\left(\ddot{P}\mid X=x\right)^{-1}\Cov\left(\ddot{P},v_{s}\left(Z\right)\mid X=x\right)=\left[\begin{array}{cc}
a_{s} & 0\\
0 & b_{s}
\end{array}\right]
\]
 for $a_{s},b_{s}\geq0$ for all $s\in\mathcal{S}_{x}$. Then, the
2SLS weights are given by
\[
w^{s,x}=\left[\begin{array}{cc}
a_{s}c_{11}^{x} & b_{s}c_{21}^{x}\\
a_{s}c_{12}^{x} & b_{s}c_{22}^{x}
\end{array}\right]
\]
The relative weight $\beta_{1}^{\text{2SLS}}$ assigns to $\beta_{2}^{s,x}$
compared to $\beta_{1}^{s,x}$ is thus $a_{s}c_{12}^{x}/a_{s}c_{11}^{x}=c_{12}^{x}/c_{11}^{x}$.} The overall 2SLS bias due to non-zero contamination weights $c_{12}^{x}$
depends on the extent that heterogeneous effects across $X$-cells
are correlated with $c_{12}^{x}$.\footnote{Since $\E\left[\Var\left(\ddot{P}\mid X\right)\right]=\Var\left(\ddot{P}\right)$,
the ``contamination'' weights $c_{12}^{x}$ and $c_{21}^{x}$ are
zero on average. Thus, there is no ultimate bias unless the causal
effects correlate with the contamination weights.} One way to avoid contamination bias is to run 2SLS fully interacted
with $X$.\footnote{Such an approach is equivalent to running separate 2SLS regressions
for each value of $X$ and then aggregating the resulting estimates
across $X$. \citet{goldsmith2022contamination}'s solution to contamination
bias in OLS\textemdash running one-treatment-at-a-time regressions\textemdash is
not valid in the 2SLS case since it would entail running regressions
on samples selected based on an endogenous variable (received treatment).} Such an approach, however, requires a large number of observations
per value of $X$ to avoid issues of weak instruments.

In the case of a binary treatment, \citet{blandhol2022tsls} show
that 2SLS with flexible controls assigns proper weights under a \emph{monotonicity-correctness
}condition. The requirement that average conditional monotonicity
(Assumption \ref{assumption:non_neg}) needs to hold conditional on
each value of $X$ can be seen as a weakening of monotonicity-correctness.\footnote{Essentially, Assumption \ref{assumption:non_neg} only requires monotonicity-correctness
to hold on average across instrument values.}

Propositions \ref{prop:kitagawa} and \ref{prop:test} generalize
to 2SLS with covariates as follows:

\begin{prop}
\label{prop:kitagawax}Maintain Assumptions \ref{assu:ivx} and \ref{assu:rankx}
and let $y\leq y'$. If 2SLS assigns proper weights then
\[
\Var\left(\ddot{P}\right)^{-1}\Cov\left(\ddot{P},\mathbf{1}_{y\leq Y\leq y'}D\right)
\]
is a non-negative diagonal matrix.
\end{prop}
\begin{prop}
\label{prop:testx}Assume $W\in\left\{ 0,1\right\} $ with $Z\perp W\mid S,X$.
If 2SLS with flexible controls for $X$ assigns proper weights then
$\Var\left(\ddot{P}\right)^{-1}\Cov\left(\ddot{P},\ddot{D}\mid W=1\right)$
is a diagonal non-negative matrix.
\end{prop}
The Proposition \ref{prop:kitagawax} test can be implemented in the
same way as the Proposition \ref{prop:kitagawa} test.\footnote{For instance, if $Y$ is a binary variable, we should get $\vartheta_{1}\geq0$
and $\vartheta_{2}=0$ in the regression $YD_{1}=\varphi_{X}+\vartheta_{1}P_{1}+\vartheta_{2}P_{2}+\upsilon$
where $\varphi_{X}$ represent $X$-level fixed effects.} The prediction of Proposition \ref{prop:testx}, however, can not
be tested by a linear regression in the same way as the prediction
in Proposition \ref{prop:test} can (unless Assumption \ref{assu:covary_similarly}
holds).\footnote{If Assumption \ref{assu:covary_similarly} holds, $\Var\left(\ddot{P}\mid W=1\right)=h\Var\left(\ddot{P}\right)$
for a constant $h$. In that case, Proposition \ref{prop:testx} can
be tested as Proposition \ref{prop:test}, with the inclusion of $X$-specific
fixed effects. } The reason is that the covariance between $\ddot{P}$ and $\ddot{D}$
in the subpopulation $W=1$ is related to the variance of $\ddot{P}$
in the \emph{full} population. Instead, one can test this prediction
by, e.g., comparing the sample analog of $\Var\left(\ddot{P}\right)^{-1}\Cov\left(\ddot{P},\ddot{D}\mid W=1\right)$
to critical values obtained through resampling methods.

Due to the potential presence of contamination bias, Propositions
\ref{prop:binary}, \ref{prop:just-identified-test}, and \ref{prop:threshold-crossing}
do not immediately generalize to the case with covariates. The results,
however, do hold if one is willing to maintain Assumption \ref{assu:covary_similarly}
and that the first stage is sufficiently flexible to capture differences
in how agents respond to the instruments across $X$:
\begin{assumption}[First stage capturing heterogeneity across $X$]
\label{assu:flexible_interaction} 
\[
\ddot{P}=P_{X}-E\left[D\mid X\right]
\]
where $P_{X}\equiv\E\left[D\mid X\right]+\Var\left(Z\mid X\right)^{-1}\Cov\left(Z,D\mid X\right)\left(Z-\E\left[Z\mid X\right]\right)$.
\end{assumption}
This assumption requires that the predicted treatments estimated on
the full sample coincide with predicted treatments estimated separately
for each value of $X$. Assumption \ref{assu:flexible_interaction}
holds trivially when the instruments are flexibly interacted with
$X$. In the just-identified case, Assumption \ref{assu:flexible_interaction}
is equivalent to 
\[
\E\left[D\mid Z,X\right]=\gamma Z+\phi_{X}
\]
for a matrix of constants $\gamma$ and vectors of constants $\phi_{x}$.\footnote{By definiton, $\ddot{P}=\gamma\left(Z-\E\left[Z\mid X\right]\right)$
for a constant matrix $\gamma$. Furthermore, when $Z$ is a set of
indicators, we have $P_{X}=E\left[D\mid Z,X\right]$. The assumption
allows the baseline probabilities of receiving the various treatments
to vary across $X$ but requires the first-stage coefficients to be
homogenous across $X$.} We then have

\begin{prop}
\label{prop:binaryx} Proposition \ref{prop:binary} holds under Assumptions
\ref{assu:ivx}\textendash \ref{assu:flexible_interaction} when $X$
is being flexibly controlled for in the 2SLS specification.
\end{prop}
Again, if the conditions in Proposition \ref{prop:binary} hold but
Assumption \ref{assu:covary_similarly} or Assumption \ref{assu:flexible_interaction}
is violated, 2SLS might suffer from contamination bias. If one has
enough observations per value of $X$ this contamination bias can
be avoided by running 2SLS fully interacted with $X$.

Proposition \ref{prop:just-identified-test} generalizes to:
\begin{prop}
\label{prop:just-identified-test-x}Proposition \ref{prop:just-identified-test}
holds conditional on each value of $X$ under Assumptions \ref{assu:ivx}\textendash \ref{assu:flexible_interaction}
when $X$ is being flexibly controlled for in the 2SLS specification.
\end{prop}
 Finally, Proposition \ref{prop:threshold-crossing} generalizes
as follows:

\begin{prop}
\label{prop:threshold-crossing-x} Maintain Assumptions \ref{assu:ivx}\textendash \ref{assu:flexible_interaction}.
Then 2SLS assigns proper weights when controlling flexibly for $X$
in the Section \ref{subsec:threshold-crossing} model if $E\left[\ddot{P}_{k}\mid\ddot{P}_{l},X=x\right]$
is linear in $\ddot{P}_{l}$ for all $k$, $l$, and $x$.
\end{prop}

\subsection{Assumptions About Heterogeneous Effects\label{subsec:het}}

Average conditional monotonicity and no cross effects (Assumptions
\ref{assumption:non_neg} and \ref{assumption:zero}) make sure that
2SLS is weakly causal for arbitrary heterogeneous effects (see Section
\ref{subsec:WC}). But these assumptions can be relaxed if we are
willing to make some assumptions about heterogeneous effects.

To analyze this case, define \emph{compliers} (\emph{defiers}) as
the response types pushed into (out of) treatment $1$ by instrument
$1$. To simplify language, we here refer to $P_{1}$ as ``instrument
1'' and $P_{2}$ as ``instrument 2''.\footnote{All statements below are conditional on controlling linearly for the
other instrument. E.g., ``pushed into treatment $1$ by instrument
$1$''=''pushed into treatment $1$ by instrument $1$ when controlling
linearly for instrument $2$''.} Define the following parameters:
\begin{center}
\begin{tabular}{rlll}
$\beta_{1}^{\text{compliers}}$ & $\equiv$ & $\frac{\E\left[w_{11}^{S}\beta_{1}^{S}\mid w_{11}^{S}>0\right]}{\E\left[w_{11}^{S}\mid w_{11}^{S}>0\right]}$ & weighted average effect of treatment 1 for compliers\tabularnewline
 &  &  & \tabularnewline
$\beta_{1}^{\text{defiers}}$ & $\equiv$ & $\frac{\E\left[w_{11}^{S}\beta_{1}^{S}\mid w_{11}^{S}<0\right]}{\E\left[w_{11}^{S}\mid w_{11}^{S}<0\right]}$ & weighted average effect of treatment 1 for defiers\tabularnewline
 &  &  & \tabularnewline
$\beta_{2}^{\text{pushed in}}$ & $\equiv$ & $\frac{\E\left[w_{12}^{S}\beta_{2}^{S}\mid w_{12}^{S}>0\right]}{\E\left[w_{12}^{S}\mid w_{12}^{S}>0\right]}$ & weighted average effect of treatment $2$ for response types\tabularnewline
 &  &  & pushed out of treatment $2$ by instrument $1$\tabularnewline
$\beta_{2}^{\text{pushed out}}$ & $\equiv$ & $\frac{\E\left[w_{12}^{S}\beta_{2}^{S}\mid w_{12}^{S}<0\right]}{\E\left[w_{12}^{S}\mid w_{12}^{S}<0\right]}$ & weighted average effect of treatment $2$ for response types\tabularnewline
 &  &  & pushed out of treatment $2$ by instrument $1$\tabularnewline
$w^{\text{negative}}$ & $\equiv$ & $\frac{\E\left[w_{11}^{S}\mid w_{11}^{S}<0\right]}{\Pr\left[w_{11}^{S}<0\right]}$ & sum of the negative weights on treatment $1$ in $\beta_{1}^{\text{2SLS}}$\tabularnewline
 &  &  & \tabularnewline
$w^{\text{cross}}$ & $\equiv$ & $\frac{\E\left[w_{12}^{S}\mid w_{12}^{S}>0\right]}{\Pr\left[w_{12}^{S}>0\right]}$ & sum of the positive weights on treatment $2$ in $\beta_{1}^{\text{2SLS}}$\tabularnewline
\end{tabular}
\par\end{center}

\bigskip{}

We then obtain the following decomposition:\footnote{This identity follows directly from $\beta_{1}^{\text{2SLS}}=\E\left[w_{11}^{S}\beta_{1}^{S}+w_{12}^{S}\beta_{2}^{S}\right]$.}
\[
\beta_{1}^{\text{2SLS}}=\beta_{1}^{\text{compliers}}-\left(\beta_{1}^{\text{defiers}}-\beta_{1}^{\text{compliers}}\right)w^{\text{negative}}-\left(\beta_{2}^{\text{pushed out}}-\beta_{2}^{\text{pushed in}}\right)w^{\text{cross}}
\]

The 2SLS estimand thus equals the weighted average treatment effect
among the compliers and two ``bias terms'': The first bias term
is non-zero if there are defiers and defiers have a different weighted
average treatment effect than the compliers. The second bias term
is non-zero if there exist response types pushed into or out of treatment
$2$ by instrument $1$, and the weighted average effect of treatment
$2$ differ between those pushed into and out of treatment $2$. We
thus see that if we are willing to assume $\beta_{1}^{\text{defiers}}=\beta_{1}^{\text{compliers}}$
and $\beta_{2}^{\text{pushed out}}=\beta_{2}^{\text{pushed in}}$\textemdash ``no
selection on gains''\textemdash 2SLS is weakly causal without any
restrictions on response types.\footnote{A similar assumption is discussed by \citet{kolesar2013estimation}.}
This is, for instance, accomplished if $\beta\perp S$\textemdash treatment
effects do not systematically differ across response types.

The 2SLS estimand is also weakly causal if we are willing to impose
some restrictions on both response types and the amount of selection
on gains. For instance, if we maintain $w^{\text{cross}}=0$ (no cross
effects) then 2SLS is weakly causal if and only if\footnote{Remember that the 2SLS estimand is weakly causal if $\beta_{1}^{\text{2SLS}}\geq0$
($\beta_{1}^{\text{2SLS}}\leq0$) whenever $\beta\geq0$ ($\beta\leq0$)
for all agents. If $\beta\geq0$ for all agents, weak causality thus
requires $\beta_{1}^{\text{compliers}}-\left(\beta_{1}^{\text{defiers}}-\beta_{1}^{\text{compliers}}\right)w^{\text{negative}}\geq0\Rightarrow\frac{1+w^{\text{negative}}}{w^{\text{negative}}}\geq\frac{\beta_{1}^{\text{defiers}}}{\beta_{1}^{\text{compliers}}}$.
The same inequality is obtained when $\beta\leq0$ for all agents.}
\[
\frac{1+w^{\text{negative}}}{w^{\text{negative}}}\geq\frac{\beta_{1}^{\text{defiers}}}{\beta_{1}^{\text{compliers}}}
\]
Thus, if $w^{\text{negative}}=0.1$ (roughly 8\% defiers), 2SLS is
weakly causal unless the treatment effect for defiers is more than
11 times the treatment effect for compliers.\footnote{Since $w^{\text{positive}}-w^{\text{negative}}=1$ where $w^{\text{positive}}\equiv\frac{\E\left[w_{11}^{S}\mid w_{11}^{S}>0\right]}{\Pr\left[w_{11}^{S}>0\right]}$,
we get 
\[
w^{\text{negative}}=0.1\Leftrightarrow\frac{w^{\text{negative}}}{w^{\text{negative}}+w^{\text{positive}}}=\frac{1}{12}\approx0.83
\]
If the negative and positive weights are distributed in a similar
way, $w^{\text{negative}}=0.1$ thus corresponds to 8.3\% defiers
among agents affected by instrument $1$. } Similarly, if we maintain $w^{\text{negative}}=0$ (average conditional
monotonicity) then 2SLS is weakly causal if and only if\footnote{If $\beta\geq0$ for all agents, weak causality requires 
\[
\beta_{1}^{\text{compliers}}-\left(\beta_{2}^{\text{pushed out}}-\beta_{2}^{\text{pushed in}}\right)w^{\text{cross}}\geq0\Rightarrow\frac{1}{w^{\text{cross}}}\geq\frac{\beta_{2}^{\text{pushed out}}-\beta_{2}^{\text{pushed in}}}{\beta_{1}^{\text{compliers}}}
\]
The same inequality is obtained when $\beta\leq0$ for all agents.}
\[
\frac{1}{w^{\text{cross}}}\geq\frac{\beta_{2}^{\text{pushed out}}-\beta_{2}^{\text{pushed in}}}{\beta_{1}^{\text{compliers}}}
\]

Thus, if $w^{\text{cross}}=0.1$ (20\% are pushed into or out of treatment
2 by instrument 1), 2SLS is weakly causal unless the difference in
the treatment effects between those pushed into and out of treatment
2 is above ten times the treatment effect for compliers.\footnote{Since the weights on compliers sum to one when $w^{\text{negative}}=0$,
the magnitude of $2w^{\text{cross}}$ can be interpreted as the number
of agents pushed into or out of treatment $2$ by instrument $1$
as a share of the complier population.}

\subsection{Relationship to \citet{Kamat2023}'s Joint Monotonicity\label{sec:IA-monotonicity}}

In this section, we show how the conditions in Section \ref{subsec:Identification}
relate to one possible generalization of the classical \citet{imbens1994identification}
monotonicity condition to multiple treatments: \citet{Kamat2023}'s
\emph{joint monotonicity}. Joint monotonicity requires \citet{imbens1994identification}
monotonicity to hold for each treatment indicator $\left\{ D_{1},\dots,D_{k}\right\} $
separately.\footnote{\citet{Kamat2023} considered a setting with three ordered treatment
effects. Our definition encompasses any number of treatments and any
definition of treatment indicators.} Formally
\begin{assumption}
\label{assu:mono}(Joint Monotonicity.) \emph{Joint monotonicity }is
satisfied if for each treatment $k\in\left\{ 1,\dots,n\right\} $
and instrument values $z$ and $z'$ either $s_{k}\left(z\right)\geq s_{k}\left(z'\right)$
for all $s\in\mathcal{S}$ or $s_{k}\left(z\right)\leq s_{k}\left(z'\right)$
for all $s\in\mathcal{S}$.
\end{assumption}
It turns out that joint monotonicity is neither sufficient nor necessary
for 2SLS to assign proper weights under multiple treatments. On the
one hand, joint monotonicity is not sufficient to ensure proper weights
since it does not rule out cross effects (Assumption \ref{assumption:zero}).
Joint monotonicity might also fail to ensure average conditional monotonicity
(Assumption \ref{assumption:non_neg}) if the first stage is incorrectly
specified. On the other hand, joint monotonicity is not \emph{necessary}
for 2SLS to assign proper weights since Assumption \ref{assumption:non_neg}
only requires potential treatment $k$ to be non-decreasing, \emph{on
average}, in predicted treatment $k$.\footnote{Joint monotonicity requires potential treatment to be \emph{strictly
}non-decreasing in the probability of receiving treatment $k$ conditional
on the instruments. When there are many instrument values, as in,
for instance, random judge designs with many judges, joint monotonicity
is considerably more demanding than Assumption \ref{assumption:non_neg}.
In the just-identified case with $n$ treatments and $n$ mututally
exclusive binary instruments, however, Assumptions \ref{assumption:non_neg}
and \ref{assumption:zero} do imply joint monotonicity.} Thus, to summarize
\begin{prop}
\label{prop:IA}
\begin{description}
\item [{i)}] Joint monotonicity does not imply Assumption \ref{assumption:non_neg}
or Assumption \ref{assumption:zero}.
\item [{\enskip{}\qquad{}\qquad{}\qquad{}\quad{}ii)}] Assumptions
\ref{assumption:non_neg} and \ref{assumption:zero} do not imply
joint monotonicity.
\end{description}
\end{prop}
But we can ask under which additional conditions joint monotonicity
is sufficient to ensure that Assumptions \ref{assumption:non_neg}
and \ref{assumption:zero} hold. To answer this question, consider
the ideal case when Assumption \ref{assu:first-stage-correct} holds\textemdash the
first stage equation is correctly specified. Under Assumption \ref{assu:first-stage-correct},
joint monotonicity is equivalent to each agent selecting treatment
$k$ if predicted treatment $k$ is above a certain threshold. Thus,
for each response type $s$ in the population, there exists a value
$u\in\left[0,1\right]$ such that $P_{k}\geq u\Leftrightarrow s_{k}\left(Z\right)=1.$
It turns out that, even in this ideal case, joint monotonicity is
not sufficient to ensure that Assumptions \ref{assumption:non_neg}
and \ref{assumption:zero} hold\textemdash an additional linearity
assumption is needed. In particular, if all possible thresholds are
in use, multivariate 2SLS assigns proper weights under joint monotonicity
and Assumption \ref{assu:first-stage-correct} if and only if the
predicted treatments are linearly related:
\begin{assumption}
\label{assu:all-complier-types} (All Thresholds in Use.) For each
$k\in\left\{ 1,\dots,n\right\} $ and threshold $u\in\left[0,1\right]$
there exists a response type $s$ in the population with $P_{k}\geq u\Leftrightarrow s_{k}\left(Z\right)=1$.
\end{assumption}
\begin{prop}
\label{prop:linear-predicted}Under Assumptions \ref{assu:iv}, \ref{assu:rank},
\ref{assu:first-stage-correct}, \ref{assu:mono}, and \ref{assu:all-complier-types},
2SLS assigns proper weights if and only if $\E\left[P_{k}\mid P_{l}\right]=\gamma_{kl}+\delta_{kl}P_{l}$
for all $k,l\in\left\{ 1,\dots,n\right\} $ for constants $\gamma_{kl}$
and $\delta_{kl}$.
\end{prop}
To see why the predicted treatments need to be linearly related, consider
the following example. Assume there are three treatments and that
the relationship between $P_{2}$ and $P_{1}$ is given by $\E\left[P_{2}\mid P_{1}\right]=P_{1}-P_{1}^{2}$.
The probability of receiving treatment $2$ is highest when the probability
of receiving treatment $1$ is $50\%$ and equal to zero when the
probability of receiving treatment $1$ is either zero or one. Since
we only control linearly for $P_{1}$ in 2SLS, changes in $P_{2}$
will then push agents into or out of treatment $1$, violating the
no cross effects condition.

The required linearity condition is easy to test and ensures that
2SLS assigns proper weights also when Assumption \ref{assu:all-complier-types}
does not hold:

\begin{prop}
\label{prop:linear-predicted2}If Assumptions \ref{assu:iv}, \ref{assu:rank},
\ref{assu:first-stage-correct}, and \ref{assu:mono} hold and $\E\left[P_{k}\mid P_{l}\right]=\gamma_{kl}+\delta_{kl}P_{l}$
for all $k,l\in\left\{ 1,\dots,n\right\} $ and constants $\gamma_{kl}$
and $\delta_{kl}$, then 2SLS assigns proper weights.
\end{prop}
In Section \ref{subsec:threshold-crossing}, we apply Proposition
\ref{prop:linear-predicted2} to the case when treatment is characterized
by a latent index crossing multiple thresholds.

\subsection{Relationship to Unordered Monotonicity\label{sec:U-monotonicity}}

In this section, we show how the conditions in Section \ref{subsec:Identification}
relate to \citet{HeckmanPinto2018}'s unordered monotonicity condition.
Consider the case of unordered treatments ($D_{k}=\mathbf{1}\left[T=k\right]$)
and define $s_{0}\left(z\right)\equiv\mathbf{1}\left[s\left(z\right)=0\right]$.
\emph{Unordered monotonicity }is then defined by \citet{HeckmanPinto2018}
as:
\begin{assumption}
\label{assu:UM}(Unordered Monotonicity.) \emph{Unordered monotonicity
}is satisfied if for each treatment $k\in\left\{ 0,\dots,n\right\} $
and instrument values $z$ and $z'$ either $s_{k}\left(z\right)\geq s_{k}\left(z'\right)$
for all $s\in\mathcal{S}$ or $s_{k}\left(z\right)\leq s_{k}\left(z'\right)$
for all $s\in\mathcal{S}$.
\end{assumption}
Note that unordered monotonicity is a strictly stronger condition
than Assumption \ref{assu:mono}.\footnote{For instance, consider the just-identified case with three treatments
where all the response types in Table \ref{tab:response_types} are
present. In that case, Assumption \ref{assu:mono} is satisfied but
unordered monotonicity is violated: We have $s_{0}\left(2\right)>s_{0}\left(1\right)$
for ``1-compliers'' and $s_{0}\left(2\right)<s_{0}\left(1\right)$
for ``2-compliers''.} Assumption \ref{assu:mono} requires \citet{imbens1994identification}
monotonicity to hold for indicators for each treatment \emph{except
the excluded treatment. }Unordered monotonicity requires \citet{imbens1994identification}
monotonicity to hold also for an indicator for receiving the excluded
treatment ($k=0$). Define $P_{0}\equiv\Pr\left[T=0\mid Z\right]$.\footnote{Under Assumption \ref{assu:first-stage-correct}, $P_{k}\equiv\Pr\left[T=k\mid Z\right]$
for $k\in\left\{ 1,\dots,n\right\} $.} We then get the following corollary of Proposition \ref{prop:linear-predicted2}:\footnote{Proposition \ref{prop:linear-predicted} has a similar corollary.}
\begin{cor}
\label{prop:linear-predicted-UM}If Assumptions \ref{assu:iv}, \ref{assu:rank},
\ref{assu:first-stage-correct} and \ref{assu:UM} hold and $\E\left[P_{k}\mid P_{l}\right]=\gamma_{kl}+\delta_{kl}P_{l}$
for all $k,l\in\left\{ 0,\dots,n\right\} $ and constants $\gamma_{kl}$
and $\delta_{kl}$, then 2SLS assigns proper weights for all possible
choices for the excluded treatment.
\end{cor}
In other words, under unordered monotonicity and a linearity condition,
2SLS can identify \emph{all }possible relative treatment effects,
not just treatment effects relative to an excluded treatment.

\subsection{Knowledge of Next-Best Alternatives\label{subsec:kirkeboen}}

\citet{kirkeboen2016} and a following literature exploit data where
agents' next-best alternative\textemdash their treatment choice when
$V=0$\textemdash is plausibly observed. In such settings, 2SLS can
identify meaningful treatment effects under much weaker conditions
than those of Corollary \ref{cor:binary}. We show how our results
relate to the \citet{kirkeboen2016} approach in this section. Their
approach is to first identify all agents with treatment $k$ as their
next-best alternative and then run 2SLS on this subsample using treatment
$k$ as the excluded treatment. By varying $k$ one can obtain causal
estimates of \emph{all} relative treatments effects as opposed to
only treatment effects relative to \emph{one} excluded treatment.
\citet{kirkeboen2016} show that 2SLS assigns proper weights in this
setting under two relatively mild conditions: \emph{monotonicity }and
\emph{irrelevance}.\footnote{\citet{heinesen2022instrumental} study the properties of 2SLS when
irrelevance or the ``next-best'' assumption is violated and discuss
how these conditions can be tested.}

In our notation, their monotonicity condition is $s_{k}\left(k\right)\geq s_{k}\left(0\right)$
for all $s$ and $k$ and their irrelevance condition is $s_{k}\left(k\right)=s_{k}\left(0\right)\Rightarrow s_{l}\left(k\right)=s_{l}\left(0\right)$
for all $s$, $k$ and $l$. In words, irrelevance requires that if
instrument $k$ does not induce an agent into treatment $k$, it can
not induce her into treatment $l\neq k$ either. To see how our results
relate to the \citet{kirkeboen2016} approach, assume we run 2SLS
on the subsample of all agents we believe have treatment $0$ as their
next-best alternative using treatment $0$ as the excluded treatment.\footnote{The corresponding result for subsamples of agents selecting treatment
$k\neq0$ is analog.} It turns out that the conditions of \citet{kirkeboen2016} are almost
equivalent to the conditions in Corollary \ref{cor:binary} holding
in the subsample, with the only exception that our conditions allow
for the presence of always-takers:\footnote{The approach of \citet{kirkeboen2016} is to run 2SLS on the subsample
of agents having the same treatment as their next-best alternative.
Trivially, however, there is no problem if some agents with another
next-best alternative are included in the sample as long as they always
select this alternative\textemdash always-takers do not affect 2SLS
estimands.}
\begin{prop}
\label{prop:kirkeboen}Maintain Assumption \ref{assu:order} and assume
$D_{k}=\mathbf{1}\left[T=k\right]$. The conditions in Corollary \ref{cor:binary}
are then equivalent to the following conditions holding for all response
types $s$ in the population:

1. $s_{k}\left(k\right)\geq s_{k}\left(0\right)$ for all $k$ (Monotonicity)

2. $s_{k}\left(k\right)=s_{k}\left(0\right)\Rightarrow s_{l}\left(k\right)=s_{l}\left(0\right)$
for all $k$ and $l$ (Irrelevance)

3. Either $s\left(0\right)=0$ or there is a $k$ such that $s\left(v\right)=k$
for all $v$ (Next-best alternative or always-taker).
\end{prop}
In words, 2SLS applied on a subsample assigns proper weights if and
only if monotonicity and irrelevance holds and the subsample is indeed
composed only of (i) agents with the excluded treatment as their next-best
alternative and (ii) always-takers. The conditions of \citet{kirkeboen2016}
are thus not only sufficient for 2SLS to assign proper weights, but
also close to \emph{necessary}\textemdash they only way they can be
relaxed is by allowing for always-takers. Thus, knowing next-best
alternatives is \emph{necessary} for 2SLS to assign proper weights
in just-identified models with unordered treatment effects. In other
words, when interpreting estimates from a just-identified 2SLS model
with multiple treatments as a positively weighted sum of individual
treatment effects one implicitly assumes knowledge about agents' next-best
alternatives.
\end{document}